\documentclass[a4paper,fleqn,usenatbib]{mnras}
\usepackage[utf8]{inputenc}
\usepackage{gensymb}
\usepackage{natbib}
\usepackage{graphicx, color}
\usepackage{epsfig}
\usepackage{rotate}
\usepackage{latexsym}
\usepackage{amssymb}
\usepackage{psfrag}
\usepackage{caption}
\usepackage{booktabs}
\usepackage{amsmath}
\usepackage{hyperref}

\newcommand{\bfg}{\begin{figure*}}
\newcommand{\efg}{\end{figure*}}
\newcommand{\be}{\begin{equation}}
\newcommand{\ee}{\end{equation}}
\newcommand{\bdm}{\begin{displaymath}}
\newcommand{\edm}{\end{displaymath}}
\newcommand{\bea}{\begin{eqnarray}}
\newcommand{\eea}{\end{eqnarray}}
\newcommand{\nn}{\nonumber}
\newcommand{\bt}{\begin{tabular}}
\newcommand{\et}{\end{tabular}}
\newcommand{\lan}{\langle}
\newcommand{\ran}{\rangle}
\newcommand{\bi}{\begin{itemize}}
\newcommand{\ei}{\end{itemize}}
\newcommand{\g}{$\gamma$}

\def\BL{\Bigl}
\def\BR{\Bigr}

\def\Np{N_{\rm pix}\,}
\def\Nsub{N_{\rm sub}\,}
\def\Nr{N_{\rm r}\,}

\def\alm{a_{\ell m}}
\def\g{\gamma}
\def\Msolar{\rm M_\odot}
\def\M500{{\rm M_{500}}}
\def\heal{HEALPix\,}
\def\DeltaC2{\Delta \chi^2}

\def\flask{FLASK }
\def\fermi{\emph{Fermi}-LAT }

\usepackage{comment}

\title{Searching for gamma-ray emission from galaxy clusters at low redshift}

\author[M. Colavincenzo et al.]{Manuel Colavincenzo,$^{1,2}$\thanks{E-mail: colavincenzo.manuel@gmail.com}
    Xiuhui Tan$^{1,2,4,5,6}$,
    Simone Ammazzalorso$^{1,2}$,
    \newauthor  
    Stefano Camera$^{1,2,3}$,
    Marco Regis$^{1,2}$,
    Jun-Qing Xia$^{6}$,
   Nicolao Fornengo$^{1,2}$
  \\ \\
$^{1}$ Dipartimento di Fisica, Università di Torino, Via Pietro Giuria 1, I-10125, Torino, Italy\\
$^{2}$ INFN -- Istituto Nazionale di Fisica Nucleare, Sezione di Torino, via P. Giuria 1, I–10125 Torino, Italy\\
$^{3}$ INAF -- Istituto Nazionale di Astrofisica, Osservatorio Astrofisico di Torino, strada Osservatorio 20, I-10025 Pino Torinese, Italy\\
$^{4}$ Institute of High Energy Physics, Chinese Academy of Sciences, Beijing 100049, China\\
$^{5}$ School of Physical Sciences, University of Chinese Academy of Sciences, Beijing 100049, China\\
$^{6}$ Department of Astronomy, Beijing Normal University, Beijing 100875, China\\
}

\begin{document}

\label{firstpage}
\maketitle

\begin{abstract}
We report the identification of a positive cross-correlation signal between the unresolved $\g$-ray emission, measured by the \emph{Fermi} Large Area Telescope, and four different galaxy cluster catalogues.
The selected catalogues peak at low-redshift and span different frequency bands, including infrared, optical and X-rays. The signal-to-noise ratio of the detected cross-correlation amounts to 3.5 in the most significant case. We investigate and comment about its possible origin, in terms of compact $\g$-ray emission from AGNs inside clusters or diffuse emission from the intracluster medium.
The analysis has been performed by introducing an accurate estimation of the cross-correlation power-spectrum covariance matrix, built with mock realisations of the gamma and galaxy cluster maps. Different methods to produce the mock realizations starting from the data maps have been investigated and compared, identifying suitable techniques which can be generalized to other cross-correlation studies.
%We report the identification of a positive cross-correlation signal between the unresolved $\g$-ray emission, measured by the \emph{Fermi} Large Area Telescope, and  four different galaxy cluster catalogues: WHY18 (infrared galaxies), SDSSDR9 (optical galaxies), MCXCsub and HIFLUGCS (X-ray galaxies). The signal-to-noise ratio is 3.5 and 3.2 for MCXCsub
%and HIFLUGCS, respectively, while for WHY18 and SDSSDR9 is slightly in excess of 2. A deeper investigation of the most significant case of MCXCsub indicates that a fraction of the correlation signal seems to originate from large scales, a fact which could be related to emission from the intra-cluster medium. The statistical significance is nevertheless not enough to firmly assess its origin. 
%The analysis has been performed by introducing an accurate estimation of the cross-correlation power-spectrum covariance matrix, built with mock realisations of the gamma and galaxy cluster maps. Different methods to produce the mock realizations starting from the data maps have been investigated and compared, identifying the techniques which are more suitable for cross-correlation studies.
\end{abstract}

\begin{keywords}
     cosmology: observations – cosmology: theory – gamma-rays: diffuse backgrounds – large-scale structure of universe
\end{keywords}

\section{Introduction}
\label{sec:intro}

Galaxy clusters are one of the most important tracers of the Large Scale Structure (LSS) of the Universe, being the largest virialized objects formed by the gravitational instability. Because of their large dimension, mass and formation history, they represent a unique cosmological probe. They are also fundamental from an astrophysical point of view: they host galaxies, but also ionized hot gas thermalized via collisionless virial shocks, dark matter (DM) and relativistic cosmic rays (CRs) accelerated by the shocks present at the edge of the clusters. If we focus on $\gamma$-ray emission, the CRs can produce photons via inverse Compton, non-thermal bremsstrahlung and decay of $\pi^0$. DM particles can also induce $\g$-rays through the same mechanisms arising from the products of DM annihilation or decay. 

Within the hierarchical structure formation process, clusters form in the node of the cosmic web, where also different populations of astrophysical objects, as the Active Galactic Nuclei (AGN), are present. These astrophysical sources, that can be found within the clusters themselves, contribute to the total $\gamma$-ray flux that we observe. On top of this emission, clusters could host a spatially extended $\gamma$-ray contribution. The detection of a signal of $\gamma$-rays from CRs in clusters could help in understanding the origin of the radio halos (e.g. \cite{vanWeerenEtAl2019} for a review). The observation of $\gamma$-rays originated from annihilation or decay of DM particles would confirm their existence, whilst a non-detection can help to reduce the size of the bucket of DM particle models.

The study of the $\gamma$-rays from galaxy clusters requires a double effort from an observational point of view: we need accurate galaxy cluster catalogues (in terms of mass, dimension and position), in order to be able to distinguish between the DM halo, the Intra-Cluster Medium (ICM) and the sources (such as AGN) inside the cluster, and precise measurements of the $\gamma$-ray photon flux. From the cluster catalogues side there are several surveys in the literature, obtained with different telescopes and at different wavelengths, that can be used for this purpose, while from the $\gamma$-ray side the most detailed sky recognition comes from the \emph{Fermi} Large Area Telescope (LAT) \cite{AckermanEtAl2010}. Most recent analyses of the \fermi\ data looking for a $\gamma$-ray signal from galaxy clusters using different techniques include \citep{BranchiniEtAl2017,ReissEtAl2017,BrunettiEtAl2017,LisantiEtAl2017,HashimotoEtAl2019}.

In this work we focus on the information we can derive from the joint analysis of the two observables, the $\gamma$-ray photon flux and the galaxy clusters distribution; in other words we proceed with a cross-correlation analysis. A similar approach was undertaken by \cite{BranchiniEtAl2017} and \cite{HashimotoEtAl2019}, but using different cluster samples. Several analyses in the literature have been studying the cross-correlation of $\gamma$-ray with other LSS tracers  \citep{AndoEtAl2014,ShirasakiEtAl2014,FornengoEtAl2015,RegisEtAl2015,CuocoEtAl2015,ShirasakiEtAl2015,AndoAndIshiwata2016,ShirasakiEtAl2016,FengEtAl2017}.

Here we study the cross-correlation angular power spectrum (APS) between four selected galaxy cluster catalogues obtained in different bands (optical, infrared and X-ray) and $\gamma$-rays from the \fermi in different energy bins. 
We focus on low redshift catalogues and high-mass clusters, since our main goal is to disentangle a possible (and long-sought after) extended $\gamma$-ray emission from clusters. Such a signal would be originated from ICM or DM, since AGNs and galaxies have much more compact emissions.
Improving from the previous works listed above, we introduce an accurate estimation of the power spectrum covariance matrix. This is built with mock realisations of the gamma and galaxy cluster maps and allows a precise statistical evaluation of the significance of the measured APS.

\section{Data}
\label{sec:data}

The cross-correlation analyses we have carried out in this paper are based on: (i) The full-sky $\g$-ray intensity emission measured by the \fermi, for which we consider 9-years of data in the energy range between 630 MeV and 1 TeV; (ii) A series of low-redshift galaxy cluster catalogues built in different electromagnetic bands.

\subsection{Fermi-LAT $\g$-rays maps}
\label{subsec:fermi}

\begin{figure*}
    \centering
    \includegraphics[width=0.45\textwidth]{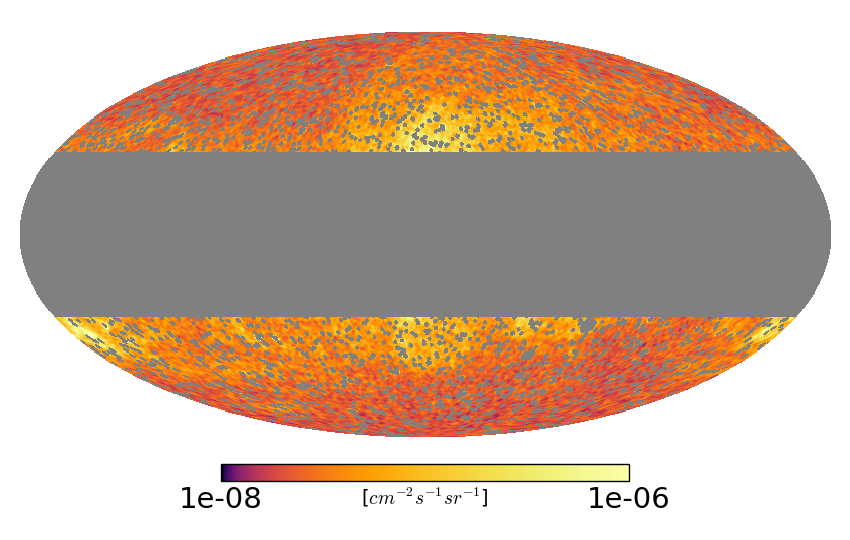}
    \includegraphics[width=0.45\textwidth]{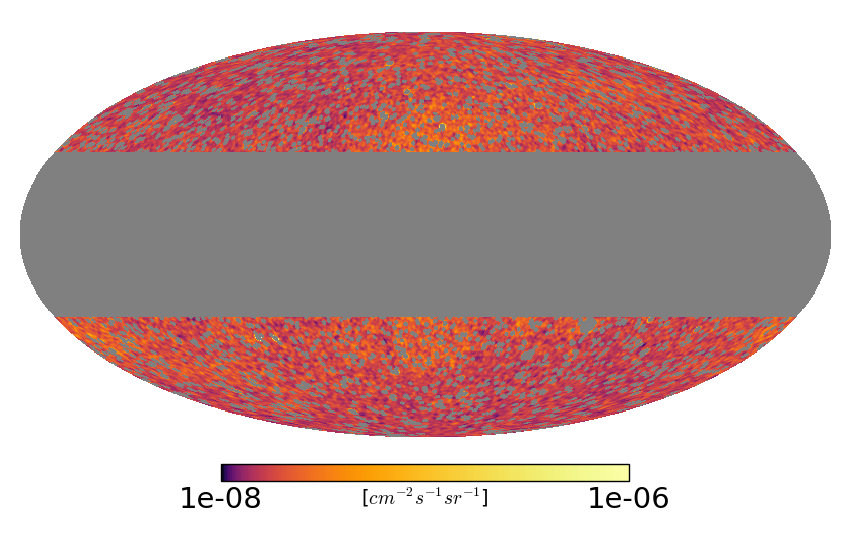}
    \caption{$\g$-ray maps in the $(2.3-4.8)$ GeV energy bin, masked with the procedure described in the text. The panels show the flux map before (left) and after (right) galactic-foreground subtraction. For illustration purposes, the maps have been rescaled to $N_{\rm side} = 128$ and smoothed with a Gaussian beam of size $\sigma = 0.4^\circ$.}
    \label{fig:fermimap}
\end{figure*}

The Fermi-LAT pair-conversion telescope achieved remarkable results for 
$\g$-ray astronomy during its 10 years of operations. 
Its large energy coverage (20 MeV - 1 TeV) and the capability of rejecting the contamination from charged cosmic rays make the instrument particularly suitable to investigate the nature of the unresolved extra-galactic $\g$-ray background (UGRB).
The angular resolution of the instrument is energy dependent and reaches $\sim$ 0.1$^\circ$ above 10 GeV. The photon and exposure maps adopted in the present analysis are produced with the Fermi Tools\footnote{We used the LAT Science Tools version \href{v10r0p5.}{https://fermi.gsfc.nasa.gov/ssc/data/analysis/software/}}. The current version of the Fermi Tools subdivide the photon events into quartiles of angular resolution, from PSF0 to PSF3, corresponding to a progress from the worst to the best Point Spread Function (PSF). We find a trade-off between photon statistics and angular resolution, by selecting the best quartile PSF3 for energies below 1.2 GeV (where we have the highest photon counts, but worst PSF) and PSF1+2+3 for higher energies.
In order to have the lowest contamination from cosmic-rays, we selected the Pass8 ULTRACLEANVETO event class\footnote{See
  \href{http://www.slac.stanford.edu/exp/glast/groups/canda/lat_Performance.htm}{http://www.slac.stanford.edu/exp/glast/groups/canda/lat\_
    Performance.htm} for further details on photon event classes.}, that is recommended for diffuse emission analysis.  

In this work we used 108 months of data (from mission week 9 to week
476). The analyses are performed on photon intensity maps, obtained by dividing the count maps by the exposure maps and the pixel area $\Omega_{\rm pix} = 4\pi/\rm N_{\rm pix}$. We adopt a HEALPix \citep{Healpix2005} pixelation format with resolution parameter
$\rm N_{\rm side} =$ 1,024, which corresponds to a total number of pixel $\rm N_{\rm pix}$ = 12,582,912 and a mean spacing of $\sim 0.06\degree$, similar to the best angular resolution of the $\g$-ray data.
We derived the intensity maps in 100 energy bins, evenly spaced in logarithmic scale
between 100 MeV and 1 TeV. These micro-bins were subsequently re-binned into 9 larger energy bins, from 631 MeV to 1 TeV, which are listed in table \ref{tab:Ebins}. The data selection and the pre-processing steps follow the same procedure described in \cite{AmmazzalorsoEtAl2018}.

%\subsubsection{Masking $\g$-ray data}
%\label{subsec:MaskGamma}

Since there is no clear sign of resolved extended emission from clusters, we adopt a cross-correlation technique which focusses on the UGRB component. To this aim, we need to mask resolved point sources. Moreover, we mask the
Galactic emission, which acts as a foreground and, while not correlating with the extra-galactic cluster distribution, nevertheless contributes a sizeable source of noise to the error budget. We therefore build a set of masks for the 
$\gamma$-ray maps by adopting the following criteria: 
\bi
\item We mask resolved sources by adopting to the 4FGL catalogue \citep{4FGL} that contains 5523 sources. Above 10 GeV, we include also additional sources present in the 3FHL catalogue \citep{AjeloEtAll2017}, that is specific to high energies.  Each source is masked taking into account both the source brightness and the PSF resolution in the specific bin, as done in \cite{AmmazzalorsoEtAl2018}.
\item The galactic disk emission is masked by means of a latitude cut that excludes the portion of the sky with $|b|<30^\circ$.
\ei 
For point sources, the masking radius around each catalogues sources is defined as:

\be
F_{\Delta E}^\gamma \, \exp{\left(-\frac{R^2}{2 \theta_{\Delta E}^2}\right)} > \frac{F_{\Delta E,\rm faintest}^\gamma}{5}\,,
\label{eq:radius}
\ee
where $F_{\Delta E}^\gamma$ is integral flux of the source in the specific
energy bin $\Delta E$ under consideration, $F_{\Delta E,\rm faintest}^\gamma$ is the
flux of the faintest source in the same energy bin, and
$\theta_{\Delta E}$ is the 68\% containment angle in that energy bin,
as provided by the Fermi-LAT PSF. The resulting energy-dependent masks aim at properly covering resolved sources and avoiding artefacts due to source flux leakage outside the mask, but at the same time maintaining a good sky coverage. For further details and impact of the mask, see also \cite{AckermannEtAl2018}. 

\begin{table*}
    \centering
    \begin{tabular}{cccccc}
        \toprule
        Bin & $E_{\rm min}$ [GeV] & $E_{\rm max}$  [GeV]  & $\ell_{\mathrm{min}}$ & $\ell_{\mathrm{max}}$ & $\theta_{\rm cont} $(deg) \\% & $f_{\rm sky}$ \\
        \midrule
         1 & 0.631 & 1.202 & 40 & 251  & 0.50 \\% & 0.315   \\ 
         2 & 1.202 & 2.290 & 40 & 316  & 0.58 \\% &  0.289   \\ 
         3 & 2.290 & 4.786 & 40 & 501  & 0.36 \\% &  0.404   \\
         4 & 4.786 & 9.120 & 40 & 794  & 0.22 \\% &  0.449   \\
         5 & 9.120 & 17.38 & 40 & 1000 & 0.15 \\% &  0.468  \\
         6 & 17.38 & 36.31 & 40 & 1000 & 0.12 \\% &  0.475  \\
         7 & 36.31 & 69.18 & 40 & 1000 & 0.11 \\% &  0.476  \\
         8 & 69.18 & 131.8 & 40 & 1000 & 0.10 \\% &  0.475  \\
         9 & 131.8 & 1000  & 40 & 1000 & 0.10 \\% &  0.473  \\
         \bottomrule
    \end{tabular}
    \caption{Energy bins in GeV used in our analysis. $E_{\rm min}$ and $E_{\rm max}$ denote the lower and upper bound of the bins, while $\ell_{\mathrm{min}}$ and $\ell_{\mathrm{max}}$ show the interval in multipole $\ell$ over which the fit of the angular power spectrum is performed: the lower bound is chosen in order to exclude possible galactic-foreground residual contamination, the upper limit is driven by the \fermi  PSF, whose 68\% containment angle $\theta_{\rm cont}$ is reported.} 
    \label{tab:Ebins}
\end{table*}

%\subsubsection{Foreground removal}
%\label{subsec:ForegroundGamma}

Even though we adopt a mask to cover the brightest part of the galactic plane emission, we nevertheless additionally adopt a procedure of foreground removal at higher latitudes, in order to reduce the contribution of this component to the error budget. This foreground removal is obtained by subtracting a model of the galactic foreground contribution, for which we use the template maps provided by the \fermi\ Collaboration with the Galactic emission model {\tt
  gll\_iem\_v06.fits}\footnote{\href{https://fermi.gsfc.nasa.gov/ssc/data/access/lat/BackgroundModels.html}{https://fermi.gsfc.nasa.gov/ssc/data/access/lat/BackgroundModels.html}}. The foreground template is projected in HEALPix maps keeping the same
$N_{\rm side}$ of the intensity maps and the same micro binning. A free normalisation is assigned to the template map and a
free constant is added, the latter representing the average UGRB emission and a possible cosmic-ray contamination of the $\gamma$-ray maps. The resulting foreground model is fitted with a Poissonian likelihood to the masked intensity maps. The best-fit normalisations obtained with this procedure are all well compatible with unity, showing a successful description of the foreground emission. We normalise the foreground templates with the obtained parameters and then subtract the foreground maps from the corresponding intensity maps: this procedure is performed after having re-binned them in the macro energy-bins then used in the cross-correlation analysis. For additional information about the foreground removal, see \cite{AckermannEtAl2018} and \cite{AmmazzalorsoEtAl2018}, where the same procedure is adopted.
In figure \ref{fig:fermimap}, we show as an example the $\g$-ray masked map in the ($2.3-4.8$) GeV energy bin before (left panel) and after (right panel) the galactic foreground removal.

\begin{table}
  \centering
  \begin{tabular}{ccccccc}
      \toprule
%      Catalogue & Type & Reference & Total & GT & LT & $\bar{\theta}_{500}$ (DEG) & SNR & $\DeltaC2_{\mathrm{FLAT-AGN}}$\\ 
      Catalogue & Type & Reference & Total &  SNR & $\DeltaC2_{\mathrm{FLAT-AGN}}$\\ 
      \toprule
%      WHY18  & infrared & \cite{WHY182018}  & 8999 & 5642 & 3357 & 0.068 & 2.1 & 0.6            \\
%      SDSSDR9 & optical & \cite{SDSSDR9_2018}  & 7384 & 4802 & 2582 & 0.104 & 2.3 & 0.2             \\
%      MCXCsub  & X-ray & \cite{mcxcsub2018} & 109 & 72 & 37 & 0.267 & 3.5 & 4.5            \\
%      HIFLUGCS & X-ray & \cite{HIFLUGCS2002} & 105 & 77 & 28 & 0.416  & 3.2 & 0.6           \\
      WHY18  & infrared & \cite{WHY182018}  & 8999  & 2.1 & 0.6            \\
      SDSSDR9 & optical & \cite{SDSSDR9_2018}   & 2582 & 2.3 & 0.2             \\
      MCXCsub  & X-ray & \cite{mcxcsub2018} & 109  & 3.5 & 4.5            \\
      HIFLUGCS & X-ray & \cite{HIFLUGCS2002} & 105  & 3.2 & 0.6           \\
      \bottomrule
  \end{tabular}
%  \caption{Cluster catalogues used in our analysis. For each catalog, we show the
%total number of clusters after the selection discussed in the text, the number of clusters that have an angular diameter (measured in terms of $\theta_{500}$) larger (GT) or smaller (LT) than the average angular size $\bar{\theta}_{500}$ of the corresponding catalog, the average angular size $\bar{\theta}_{500}$,  the signal-to-noise ratio for cross-correlation with the {\it Fermi}-LAT maps and the difference in the best-fit $\chi^2$'s of a flat angular power spectrum in multipole and the physical AGN model.}
   \caption{Cluster catalogues used in our analysis. For each catalogue, we show the
total number of clusters after the selection discussed in the text,  the signal-to-noise ratio for cross-correlation with the {\it Fermi}-LAT maps and the difference in the best-fit $\chi^2$'s of a flat angular power spectrum in multipole and the physical AGN model.}
 \label{tab:deltaChi2FinalFull}
\end{table}

\subsection{Galaxy cluster catalogues}
\label{subsec:ClusterCat}

We base our analysis on four galaxy cluster catalogues obtained from observations in different frequency bands: one in the infrared (WHY18), one in the optical (SDSSDR9) and two in the X-ray band (MCXCsub and HIFLUGCS). We use these catalogues because they contain clusters which are massive and located at relatively low-redshift, thus offering an expected enhanced cross-correlation signal. Being massive and relatively close, some of them might also extend beyond the angular resolution of the {\it Fermi}-LAT detector, allowing us to investigate if a cross-correlation signal due to diffuse emission in clusters is present on large scales.

WHY18 is obtained by combining photometric galaxies from 2MASS, the Wide-field Infrared Survey Explorer \cite[WISE,][]{WISE2010} and the SuperCOSMOS Sky Survey \citep{SUPERCosmos2001}. The main selection applied to the sources is such to include clusters with $\M500>3\times 10^{14}\Msolar$ \footnote{In our analysis we consider $\M500$ corresponding to an overdensity $\Delta_c = 500$. The mass in this case is defined as $\M500 = 4\pi\, r_{500}^3 \, 500\, \rho_c/3 = 250\, r_{500}^3\, H(z)^2/G$, where $\rho_c$ is the critical density, $r_{500}$ is the virial radius, $H(z)$ the Hubble parameter and $G$ the gravitational constant. \label{foot:virial}}  and redshift between 0.025 and 0.3. The final number of clusters in this catalogue is 47,500~\citep{WHY182018}.

SDSSDR9 is part of the full Sloan Digital Sky Survey \cite[SDSS,][]{SDSS2000} catalogue, obtained using an Adaptive Matched Filter \cite[AMF,][]{AMF1999} technique, based on a model of galaxy distribution. This filter is built using the cluster density radial profile, the galaxy luminosity function and the redshift. After applying the filter, the catalogues contains 49,479 clusters with redshift between 0.045 and 0.691 \citep{SDSSDR9_2018};

MCXC is an X-ray catalogue \citep{MCXC2011} obtained by collecting 1,743 clusters from two main types of X-ray observations: (i) contiguous area survey ROSAT \citep{ROSAT1999}; (ii) deeper pointer X-ray observations. MCXCsub is built from the full MCXC catalogue by selecting a sub-set of 112 clusters with $\M500 > 10^{13} \Msolar$, angular diameter larger than 0.2\degree, latitude larger than 20\degree and in positions of the sky to avoid contamination from bright $\g$-ray point-sources \citep{ReissEtAl2017}.

HIFLUGCS contains 63 clusters selected from ROSAT with latitude larger than 20\degree and flux between 0.1 and 2.4 keV larger than $2 \times 10^{-11} \rm ergs \ s^{-1} \ cm^{-2}$ \citep{2002ApJ...567..716R}.

\begin{figure*}
    \centering
    \includegraphics[width=0.5\textwidth]{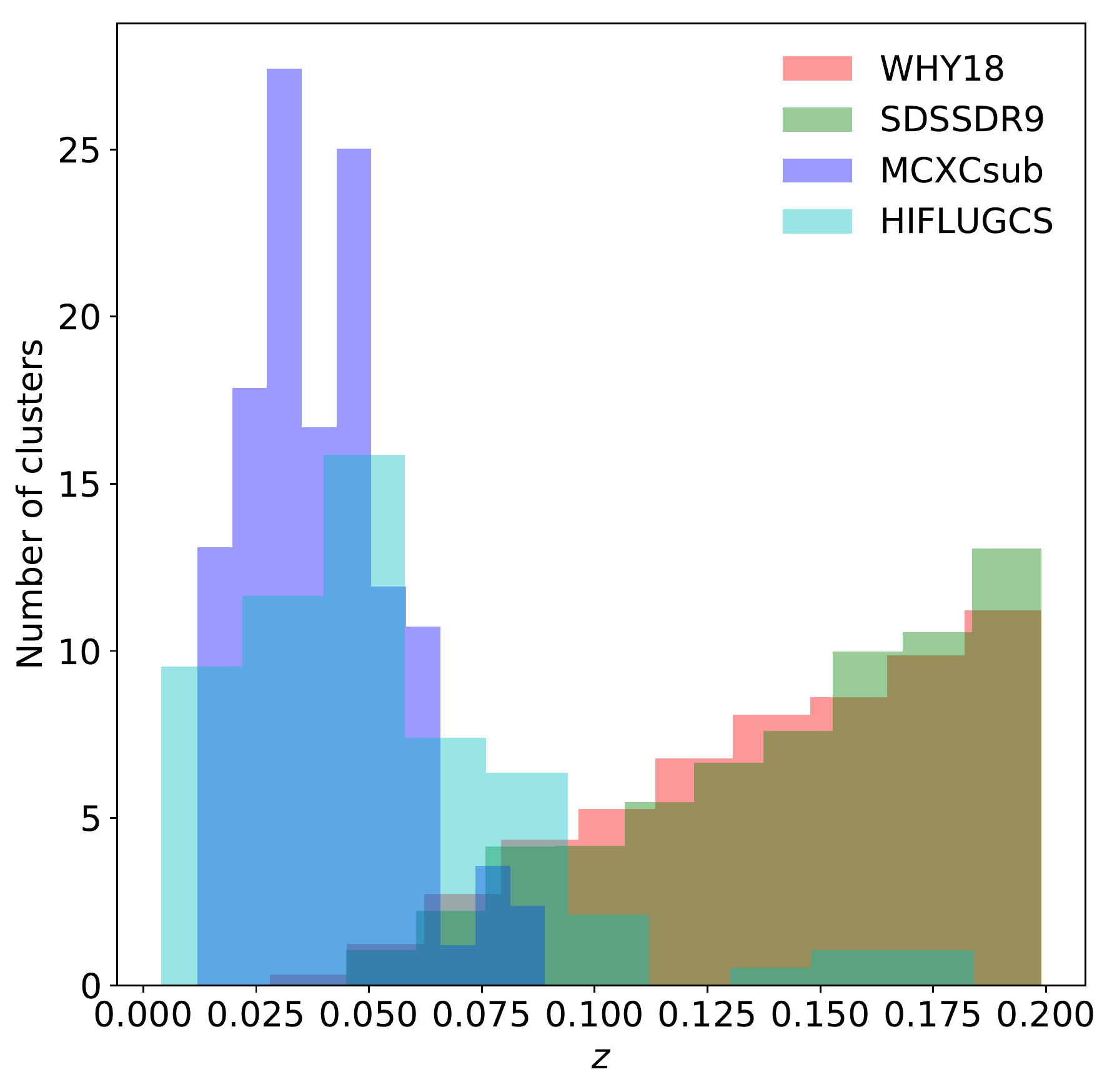}
    \caption{Redshift distribution of the galaxy cluster catalogues used in our analysis: WHY18 (limited to richness larger than 9), SDSSDR9, MCXCsub and HIFLUGCS.}
    \label{fig:zdistSelec}
\end{figure*}

Starting from the catalogues described above, we perform a selection by considering only clusters at redshift smaller than 0.2 and with mass  $\M500 > 10^{13} \Msolar$, since we aim at investigating clusters having an angular size in the sky at least comparable with the \fermi\ PSF. Moreover, in order to consider only clusters with robust identification, we applied a further cut on WHY18 by retaining only clusters with richness larger than 5. Table \ref{tab:deltaChi2FinalFull} reports the number of clusters selected for our analysis in each catalogue.
It is worth to mention that not all the original cluster catalogues report the cluster mass in terms of $\M500$. When a different mass is provided, we have converted it into $\M500$, following the formalism described in the Appendix of \cite{HuEtAll2003}. The final redshift distribution of the cluster catalogues adopted in our analysis is shown in figure \ref{fig:zdistSelec}.

For each of the cluster catalogue, the cross-correlation analysis requires a proper mask that takes into account the portion of sky covered by the catalogue and possible systematic effects due to Galactic contamination or misidentification that derives from high stellar number densities. For the infrared catalogues WHY18 we adopt a mask derived from \cite{InfraredMask2015}; for the optical catalogues SDSSDR9 the mask just refers to removing the portions of sky not covered by the SDSS survey; for the X-ray catalogues MCXCSub and HIFLUGCS the mask is a sharp cut for $|b|>20\degree$ with the additional removal of the Virgo region \cite{XRayMask2017}. For all the catalogues, we assume uniform coverage, at least on the angular scales probed in our correlation analysis (i.e., below a few degrees).

\section{Models}
\label{sec:models}

%In the previous section we have described how to estimate the covariance matrix of the cross-correlation APS. The error matrix is needed as connection between the true signal we can measure cross-correlating the real galaxy cluster and gamma-ray maps and the models we want to test to understand the origin of the signal we are measuring. 
%This procedure, that is the core of all the analysis, is described in the next section. 

In this section we describe how we model the expected cross-correlation APS between the $\g$-ray sky and galaxy clusters at low redshift.

The first expected contribution for the cross APS is provided by $\g$-ray emissions from the center of clusters. This emission can arise from compact sources located inside the cluster like AGNs or from the innermost concentration of the ICM. Given the size of the {\it Fermi}-LAT PSF, such correlation between the $\g$-ray emission and clusters is well represented by correlation at zero angular separation (i.e., the emission comes from a region much smaller than the \fermi\ PSF around the cluster center): this implies that the APS has a flat behaviour in multipole (also called shot-noise term), except for the correction due to the \fermi\ beam window function:
\be
C_{\ell,{\mathrm{FLAT}}}^{c_j \g_i}=\left(\frac{1}{N_{c_j}}\sum_{k=0}^{N_{c_j}}\,F^\gamma_{\Delta E_i,k}\right)W_\ell^{\Delta E_i} \,,
\label{eqn:Cpflux}
\ee
with $N_{c_j}$ being the number of objects in the cluster catalogue $j$ and $F^\gamma_{\Delta E_i,k}$ the $\g$-ray flux coming from the center of the cluster $k$ in the energy bin $\Delta E_i$.
The beam window function $W_\ell^{\Delta E}$ accounts for the finite angular resolution of the \fermi\ instrument that suppresses high multipoles. It is  computed from $W_\ell(E) = 2 \pi \int_{-1}^{1} d \cos\theta \, P_\ell(\cos\theta)  \, {\rm PSF}(\theta, E)$ and then averaging over the energy bin by using an energy spectrum of index $-2.3$, which represents the mean spectral index of the UGRB \citep{Ackermann:2014usa}.

Since gamma-ray sources typically have energy spectra that can be well approximated by a power-law, the energy dependence of the cross APS is modelled as a power law as well, and eq.~\ref{eqn:Cpflux} can be approximated as:
\be
C_{\ell,{\mathrm{FLAT}}}^{c_j \g_i} \ = \ C_P^j \ \bar E_i^{-\alpha_{1,j}} \ \Delta E_i \ W_\ell^{\Delta E_i}\,,
    \label{eqn:flatModel}
\ee
where $\bar E_i=\sqrt{E_i\,E_{i+1}}$ is the geometric mean of the lower ($E_i$)  and upper ($E_{i+1}$) bounds of the energy bin, $\Delta E_i=E_{i+1}-E_i$ is the size of the bin, $C_P^j$ is the normalization of the power-law relation and $\alpha_{1,j}$ is the spectral index. The index $j$ on the parameters refers to the fact that we fit the cross APS separately for each cluster catalogue.

In addition to the shot-noise term a correlation at larger angular separation angles might also be expected. We model this APS following the halo-model approach, where the correlation can be decomposed into the so-called 1-halo and a 2-halo terms (``1h'' and ``2h'', in the notation below). The former refers to the correlation between two points residing in the same physical halo, the latter to the case of two points belonging to two different halos.
Correlations on angular scales larger than the \fermi\ PSF can be due either to a 1-halo term associated to an extended ICM (or DM halo) or to a 2-halo contribution, involving e.g. two AGNs residing in different structures.
For definiteness, since the modeling of $\g$-ray emission from ICM suffers of large uncertainties \citep{ZandanelAndAndo2014}, we limit our modelling of the 2-halo term to the case of the more robust AGN emission.

The general expression defining the theoretical cross APS is:
\begin{equation}
    C_\ell^{c_j \g_i} = \int \frac{d\chi}{\chi^2} \ W_{c_j}(\chi) W_{\g_i}(\chi) \ P_{c_j \g_i}(k=\ell/\chi,\chi)\,,
    \label{eqn:Cldef}
\end{equation}
where $W_{c_j}$ and $W_{\g_i}$ are the window functions of the cluster catalogue $j$ and $\g$-ray source population $i$, $P_{c_j \g_i}$ is the cross-correlation 3D power spectrum, and $\chi$ is the comoving radial distance. To model these quantities we follow \citep{BranchiniEtAl2017}.

The window function of cluster catalogues can be written as:
\begin{equation}
    W_{c_j}(z) = \frac{4\pi\chi(z)^2}{N_{c_j}} \ \int \ dM \ \frac{d^2n_{c_j}}{dM \ dV}\,,
\end{equation}
with the number of objects in the cluster catalogue given by
\begin{equation}
    N_{c_j} = \int \ dM \ dV \ \frac{d^2n_{c_j}}{dM \ dV} \ .
\end{equation}
We empirically derive the cluster mass function from the catalogues themselves, considering the estimated redshift and $M_{500}$ masses provided by the cluster catalogues.

The window function of the $\g$-ray emission from a given source population is
\be
    W_{\g_i}(E,z) = \chi(z)^2 \int_{\mathcal{L}_{\rm min}}^{\mathcal{L}_{\rm max}(F_{\rm sens},z)} d\mathcal{L}
   \, \Phi_{\rm S}(\mathcal{L},z,E) \, \frac{dN_{\rm S}}{dE}\left(\mathcal{L},z\right)\times e^{-\tau\left[E(1+z),z\right]}  \ ,
    \label{eq:win_astro}
\ee
where $\mathcal{L}$ is the $\g$-ray rest-frame luminosity in the energy interval $0.1$ to $100\,\mathrm{GeV}$, $\Phi_{\rm S}$ is the $\g$-ray luminosity function of the source class $i$ of astrophysical emitters included in our analysis, and $d N_{\rm S}/d E$ is its observed (unabsorbed) energy spectrum. The upper bound $\mathcal{L}_{\rm max}(F_{\rm sens},z)={\rm min}[\mathcal{L}(F_{\rm sens},z),\hat{\mathcal{L}}_{\rm max}]$ with $F_{sens}$ being the flux above
which an object is resolved in the FL8Y and 3FHL catalogues and consequently masked in our analysis. The minimum (maximum) intrinsic luminosity $\mathcal{L}_{\rm min}$ ($\hat{\mathcal{L}}_{\rm max}$) depends on the properties of the source class under consideration. 
For definiteness, we focus our analysis on misaligned AGNs, which are modeled as in \cite{BranchiniEtAl2017}.

Finally, the last ingredient we need is the three dimensional power spectrum, here decomposed in its 1-halo and 2-halo terms:
\begin{equation}
    P^{1h}_{c_j,\g_j}(k,z) = \int^{\mathcal{L}_{\mathrm{max},i}(z)}_{\mathcal{L}_{\mathrm{min},i}(z)} \ d\mathcal{L} \ \phi_i(\mathcal{L},z) \times \frac{\mathcal{L}}{\lan f_{\g_i}\ran}\frac{\lan N_{c_j}(M(\mathcal{L}))\ran}{\bar{n}_{c_j}}\,
\label{eqn:1halo}
\end{equation}
\begin{equation}
    P^{2h}_{c_j,\g_i}(k,z) = \biggl[\int^{\mathcal{L}_{\mathrm{max},i}(z)}_{\mathcal{L}_{\mathrm{min},i}(z)}\ d\mathcal{L} \ \Phi_i(\mathcal{L},z)b_{h}(\mathcal{M(L)})\frac{\mathcal{L}}{\lan f_{\g_i}\ran}\biggr] \times \biggl[\int_{M_\mathrm{min}}^{M_\mathrm{max}} \ dM \ \frac{dn}{dM}b_h(M)\frac{\lan N_{c_j}\ran}{\bar{n}_{c_j}}\biggr] \ P^{\mathrm{lin}}(k)\,,
\label{eqn:2halo}
\end{equation}
where $\lan f_{\g_i}\ran$ is the mean luminosity density defined as $\lan f_{\g_i}\ran = \int \ d\mathcal{L} \ \mathcal{L}\Phi_i(\mathcal{L},z)$ and $\bar{n}_{c_j} $ the mean cluster number density defined as $\bar{n}_{c_j} = \int \ dM \ \lan N_{c_j} dn / dM \ran$.
The halo mass function $dn/dM$ and bias $b_h$ are derived from \cite{ShethAndTormen1999}, while the halo mass-luminosity relation is taken from \cite{CameraFornasaEtal2015}.

All these ingredients allow us to model a possible large-scale correlation, resulting in a non-flat dependence on the multipole angular scale. Although we use a  model tuned on AGNs, we allow a certain level of flexibility in order to intercept possible large-scale deviations from the specific AGN emission. This is obtained by allowing a free power-law index  for the energy dependence and a free overall normalisation parameter. This second model is therefore defined as:
\be
    C_{\ell,\mathrm{AGN}}^{c_j \g_i} \ = \ C_{\ell,{\mathrm{FLAT}}}^{c_j \g_i}(C_p^j,\alpha_{1,j}) + \ A_j \ \BL(\frac{\bar E_i}{\bar E_0}\BR)^{-\alpha_{2,j}} \ \Delta E_i \ C_\ell^{\Delta E_0} \ W_\ell^{\Delta E_i} \,;
    \label{eqn:AGNModel}
\ee
it contains the expected small-scale shot-noise term already introduced in eq.~\ref{eqn:flatModel} to which we add the AGN-like model discussed above, with the fudging free parameters $\alpha_2$ and $A$. $C_\ell^{\Delta E_0}$ is the theoretical APS of eq.~\ref{eqn:Cldef} calculated in a specific energy bin $\Delta E_0$, that we choose to be the $(1,2)$ GeV energy interval, with the spectral behaviour of the signal then carried by the free spectral index $\alpha_2$ and the size by the free normalisation parameter $A$. $\bar{E}_i$ is the geometric mean of the boundaries of the $i$ energy bin (reported in Table \ref{tab:Ebins}). $\bar{E}_0$ is the same mean energy for the reference $(1,2)$ GeV energy interval. While here we explicitly indicate the index $j$ (which labels the galaxy catalogue), for the sake of brevity in the rest of the paper we will omit the index $j$.

In conclusion, the FLAT model has two free parameters, the spectral index $\alpha_{1}$ and the normalisation factor $C_P$. It is embedded in the more complete physical AGN-like model, which is endowed with 2 additional free parameters, the spectral index $\alpha_2$ and the normalisation factor $A$ of the non-flat large-scale behaviour.

\section{Cross-correlation signal}
\label{sec:signal}

The goal of the analysis is to investigate the presence of a cross-correlation signal between $\g$-rays and low-redshift clusters, and then to study on which scales this signal originates, especially if a large-scale effect can be identified that could be related to the presence of a diffuse emission from the intra-cluster medium. 
The quantity we measure is the APS $C_{\ell}$ of the cross-correlation between the {\it Fermi}-LAT maps and the galaxy cluster catalogues described in Section~\ref{sec:data}. 
The APS is estimated by using Polspice \citep{chon04}, a 
tool to statistically analyse data pixeled on the sphere: it measures the two point auto- or cross-correlation APS, it is based on the fast spherical harmonic transforms allowed by iso-latitude pixelation such as HEALPix and corrects for the effects of the masks. The statistical method then adopted to quantify the presence of a signal against a null hypothesis is the Signal to Noise Ratio (SNR) defined as (see, e.g., \cite{Becker:2015ilr}):
\begin{equation}
    \mathrm{SNR} = \frac{C_{\ell,\mathrm{DATA}} \ \Gamma^{-1}_{\ell\ell'} \ C_{\ell,\mathrm{MODEL}}}{\sqrt{C_{\ell,\mathrm{MODEL}}\ \Gamma^{-1}_{\ell\ell'} \ C_{\ell,\mathrm{MODEL}}}}\,,
    \label{eqn:SNR}
\end{equation}
where $C_{\ell,\mathrm{DATA}}$ is the measured cross APS, C$_{\ell,\mathrm{MODEL}}$ is a model that grabs the physical features expected for the cross-correlation signal and $\Gamma^{-1}_{\ell,\ell'}$ is the inverse of the cross APS covariance matrix. Theoretical models are described in the previous section, and the estimation of the covariance matrix is described below in a dedicated section. The model employed to assess the significance of the presence of a signal is $C_{\ell,\mathrm{MODEL}} = C_{\ell,\mathrm{FLAT}}$, where $C_{\ell,\mathrm{FLAT}}$ is defined in eq. \ref{eqn:flatModel}, i.e., a model independent of the multipole $\ell$ (as expected from the shot-noise of a population of $\gamma$-ray sources) with an energy dependence similar to the one measured for the UGRB spectrum. The model parameters entering the computation of the SNR are determined as the best-fit parameters obtained as discussed below. The covariance matrix is carefully estimated using mocks, as described in detail in Section \ref{sec:mocks} and Appendix \ref{sec:appendixMocks}.

As a second analysis we investigate whether a more refined and complete model for the cross APS is preferred over the simpler FLAT case. The model we use refers to the case where the $\gamma$-ray emission originates dominantly from AGNs. The model is outlined in Section \ref{sec:models} and its statistical preference over the FLAT case is determined by means of a $\chi^2$ analysis, where the 
$\chi^2$ function is defined as:
\begin{equation}
    \chi^2 = \sum_{i=1}^{N_{\rm bin}}\sum_{\ell=\ell_{\rm min}}^{\ell_{\rm max}}\sum_{\ell'=\ell_{\rm min}}^{\ell_{\rm max}} \ \BL[(C_{\ell,\mathrm{DATA}})^i - (C_{\ell,\mathrm{MODEL}})^i\BR] \ (\textbf{$\Gamma$}_{\ell\ell'}^{-1})^i \ \BL[(C_{\ell',\mathrm{DATA}})^i - (C_{\ell',\mathrm{MODEL}})^i\BR] \,,
    \label{eqn:chi2}
\end{equation}
where index $i$ denotes the energy bins and the sum over the multipole $\ell$ is limited to a range $(\ell_{\mathrm{min}},\ell_{\mathrm{max}})$ whose lower bound is chosen in order to exclude possible galactic-foreground residual contamination, while the upper bound is driven by the energy-dependent \fermi  PSF, as discussed in \cite{AckermannEtAl2018}. The values for 
$\ell_{\mathrm{min}}$ and $\ell_{\mathrm{max}}$ for the different energy bins are reported in Table \ref{tab:Ebins}. 
The best fit FLAT and AGN models are determined by maximixing the Gaussian likelihood $\mathcal{L}$ defined as:
\be
-2 \ln{\mathcal{L}} = \chi^2 .
\label{eq:like} 
\ee
The preference for the AGN model over the FLAT case is then determined by means the $\Delta \chi^2=\chi_{\mathrm{FLAT}}^2-\chi^2_{\mathrm{AGN}}$. This new quantity behaves approximately like a $\chi^2$ distribution with a number of degree of freedom (DOF) defined by the difference of the DOF between the two models that enter in the comparison. For the models used in our analysis and described in Section \ref{sec:models}, we have DOF = 2.

We will also look at the Akaike information criterion (AIC), \cite{akaike1974new}, defined by the following expression:
\begin{equation}
    \mathrm{AIC} = 2k - 2 \ \mathrm{ln} \ \mathcal{L} \,,
    \label{eqn:AICeq}   
\end{equation}
where $k$ is the number of parameters of the model and $\mathcal{L}$ is the likelihood. With this criterion one can estimate the relative quality of models in terms of information lost by the given model: the smaller is the loss of information, the better  the model performs in reproducing the data.

\section{Signal covariance through mocks}
\label{sec:mocks}

In order to determine the presence of a signal and its significance, we need a faithful determination of the covariance matrix, which is then used also to build a robust likelihood function adopted to constrain the models. To this aim, we investigated and compared different options and determined the  optimal choice for our analysis. We performed an extensive and detailed investigation of the sources of covariance in the measurement of the cross APS between galaxy cluster catalogues and unresolved $\gamma$ ray emission. The main result we found is that the full covariance, along the two directions of multipole $\ell$ and energy, can be well approximated by a Gaussian estimate, diagonal in both dimensions, especially when the study is performed on binned (in $\ell$ and energy) data. While we retain in our analysis the full covariance for $\ell$, we can conclude that a Gaussian estimate of the error budget for this type of cross APS is a good approximation. We will substantiate these findings below. At the same time, we derived a general method under the approximation, valid in this case, that the cross-correlation contribution is smaller than the product of the auto-correlations of galaxies and gamma-rays, to provide and compare different estimates of the cross APS covariances, that can be implemented and adopted in any analysis of this kind. Details are provided in Appendices \ref{sec:appendixMocks} and \ref{sec:appendixTheory}.

The method adopted in this paper to build a reliable covariance matrix makes use of mock maps, obtained by randomising the true cluster and $\gamma$-ray data. The idea is to generate a large number of independent mock maps endowed with the same statistical properties of the original true maps. From these originated maps we then derive the covariance matrix in a direct way. The complete set of techniques investigated to produce mock maps, for both clusters and $\gamma$ rays, is described in Appendix \ref{sec:appendixMocks}. For the final analysis, we decided to use the phase randomisation method to produce the mock $\gamma$ ray maps and the \flask log-normal simulation for producing synthetic galaxy cluster catalogues. Even though the full description of these two methods is given in the Appendix, it is worth to provide some details here. 

The concept behind the phase randomisation method is that every field defined on a sphere (like the $\gamma$ ray intensity maps measured by the {\it Fermi}-LAT) can be linearly decomposed in terms of spherical harmonics. The spherical harmonics are weighted by a set of coefficients from which the APS is defined. The APS of the map is invariant under rotations (i.e., under the phase shift on the coefficients described in eq. \ref{eqn:coeffRot}). This means that starting from the measured power spectrum of the true map, we can build mock maps by arbitrarily changing the phases of the spherical harmonic coefficients. All these synthetic maps conserve the APS, but they will give a non zero covariance. 

For what concerns \flask, all the information regarding the code and its implementation can be found in \cite{XavierEtal2016}. \flask\ produces log-normal realisations of maps, starting from an input power spectrum. We feed \flask with the measured APS of the cluster catalogues. All the synthetic maps generated by \flask\ possess the same power spectrum, but are otherwise randomised with respect to the original map that provides the input power spectrum.  We stress that the code allows the user to adopt either a Gaussian or a log-normal probability distribution function: in our case we adopted a log-normal distribution, in order to include effects due to non-gaussianity in the covariance matrix.

To estimate the cross correlation covariance matrix, we produce 2000 realisations of the $\gamma$ maps in each of the 9 energy bin listed in Table \ref{tab:Ebins} and 2000 mocks maps for each of the cluster catalogues. The large number of realisations is required to have numerical control on the off-diagonal terms of the covariance matrix: from our tests, we found that 2000 is a good compromise between statistics and computing time. We performed a large number of tests, which are summarised here in their main features, and discussed in the Appendix.

From these mock realisations, we can then construct the full covariance between different multipoles (for each energy bin, labelled by index $i$), obtained as:
\begin{equation}
    \Gamma_{\ell\ell'}^{c\gamma_i}  \equiv \textrm{cov}[C_\ell^{c\g_i}, C_{\ell'}^{c\g_i}] =  \sum_{k=1}^{N_\gamma}\sum_{j=1}^{N_c} \ (C_\ell^{k,c\g_i} -  \bar{C_\ell}^{c\g_i})(C_{\ell'}^{j,c\g_i} -  \bar{C_{\ell'}}^{c\g_i})\,, 
    \label{eqn:CovEstimator}
\end{equation}
where $N_\gamma (N_c)$ is the total number of $\g$-ray (cluster) mocks, $C_\ell^{a,c\g_i}$ is the APS measurement performed on the $a$-th mock realisations and:
\begin{equation}
    \bar{C_\ell}^{c\g_i} = \frac{1}{N} \sum_{a=1}^{\rm N} C_\ell^{a,c\g_i}
\end{equation}
is the mean of the cross APS over all the realisations. In Appendix \ref{sec:appendixTheory} we demonstrate that the covariance matrix can be obtained by averaging two estimates: $\Gamma_{\ell\ell'}^{\hat{c}\gamma_i}$ obtained by correlating the $N_\gamma$ realisations of the $\gamma$ ray mocks with the true cluster map, and $\Gamma_{\ell\ell'}^{c\hat{\gamma_i}}$ obtained by correlating the $N_c$ realisations of cluster mocks with the true $\gamma$ ray map:
\begin{equation}
    \Gamma_{\ell\ell'}^{c\gamma_i} = \frac{1}{2} \BL(\Gamma_{\ell\ell'}^{\hat{c}\gamma_i} + \Gamma_{\ell\ell'}^{c\hat{\gamma_i}}\BR)\,.
\end{equation}
% 
% is the mean of the cross APS over all the realisations. In Appendix \ref{sec:appendixTheory} we demonstrate that the covariance matrix can be obtained by averaging two estimates: $\textrm{cov}[C_\ell^{\hat{c}\g_i}, C_{\ell'}^{\hat{c}\g_i}] $ obtained by correlating the $N$ realisations of the $\gamma$ ray mocks with the true cluster map, and $\textrm{cov}[C_\ell^{c\hat{\g_i}}, C_{\ell'}^{c\hat{\g_i}}]$ obtained by correlating the $N$ realisations of cluster mocks with the true $\gamma$ ray map:
%
% \begin{equation}
%     \textrm{cov}[C_\ell^{c\g_i}, C_{\ell'}^{c\g_i}] = \frac{1}{2} \BL(\textrm{cov}[C_\ell^{\hat{c}\g_i}, C_{\ell'}^{\hat{c}\g_i}] + \textrm{cov}[C_\ell^{c\hat{\g_i}}, C_{\ell'}^{c\hat{\g_i}}]\BR)\,.
% \end{equation}
%
Each of the ``half'' covariance is obtained by using eq. \ref{eqn:CovEstimator}. This technique allows us to reduce significantly the computing time, since we need $2N$ combinations instead of $N^2$.

\begin{figure*}
    \centering
    \includegraphics[width=0.35\textwidth]{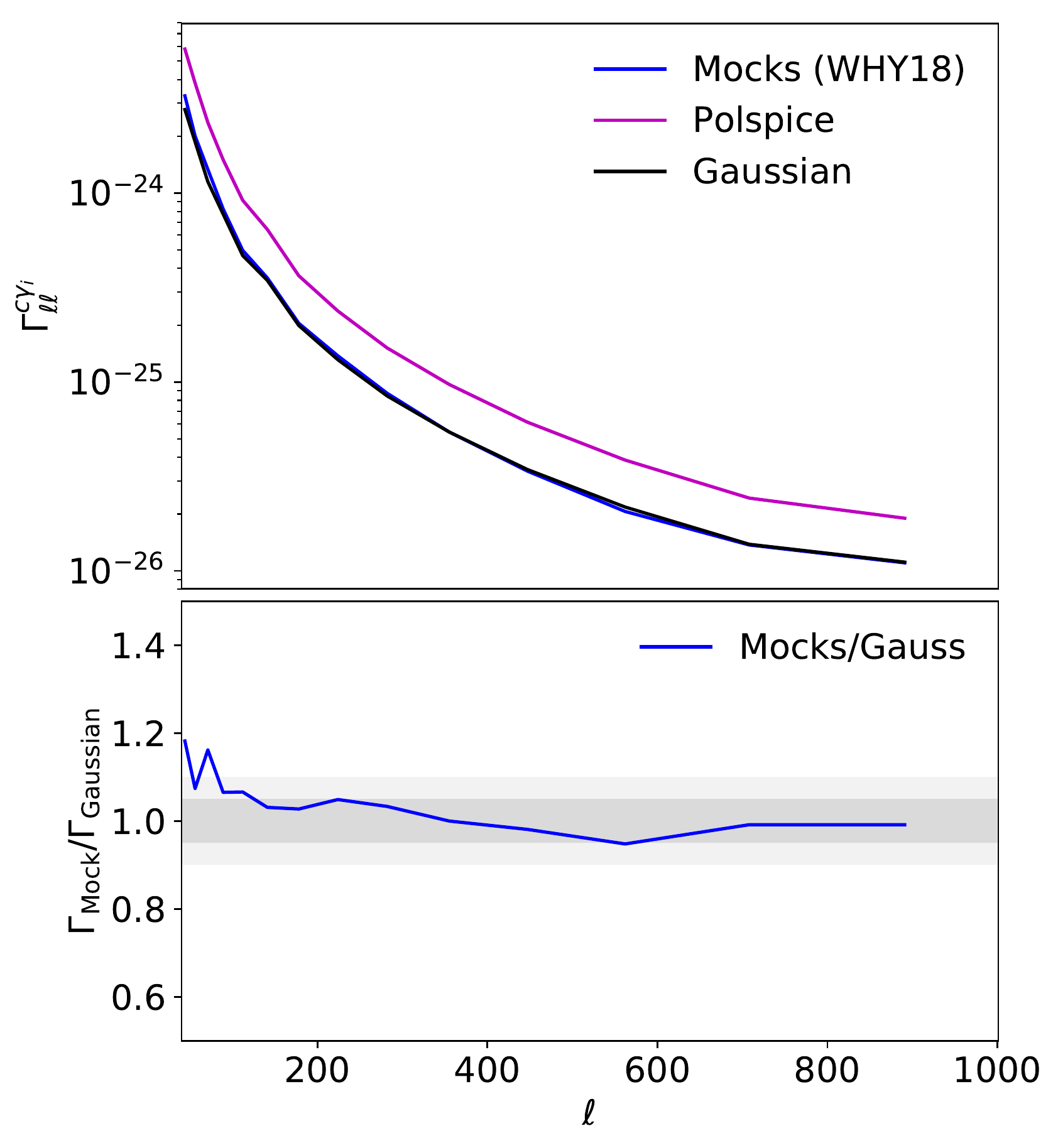}
    \includegraphics[width=0.35\textwidth]{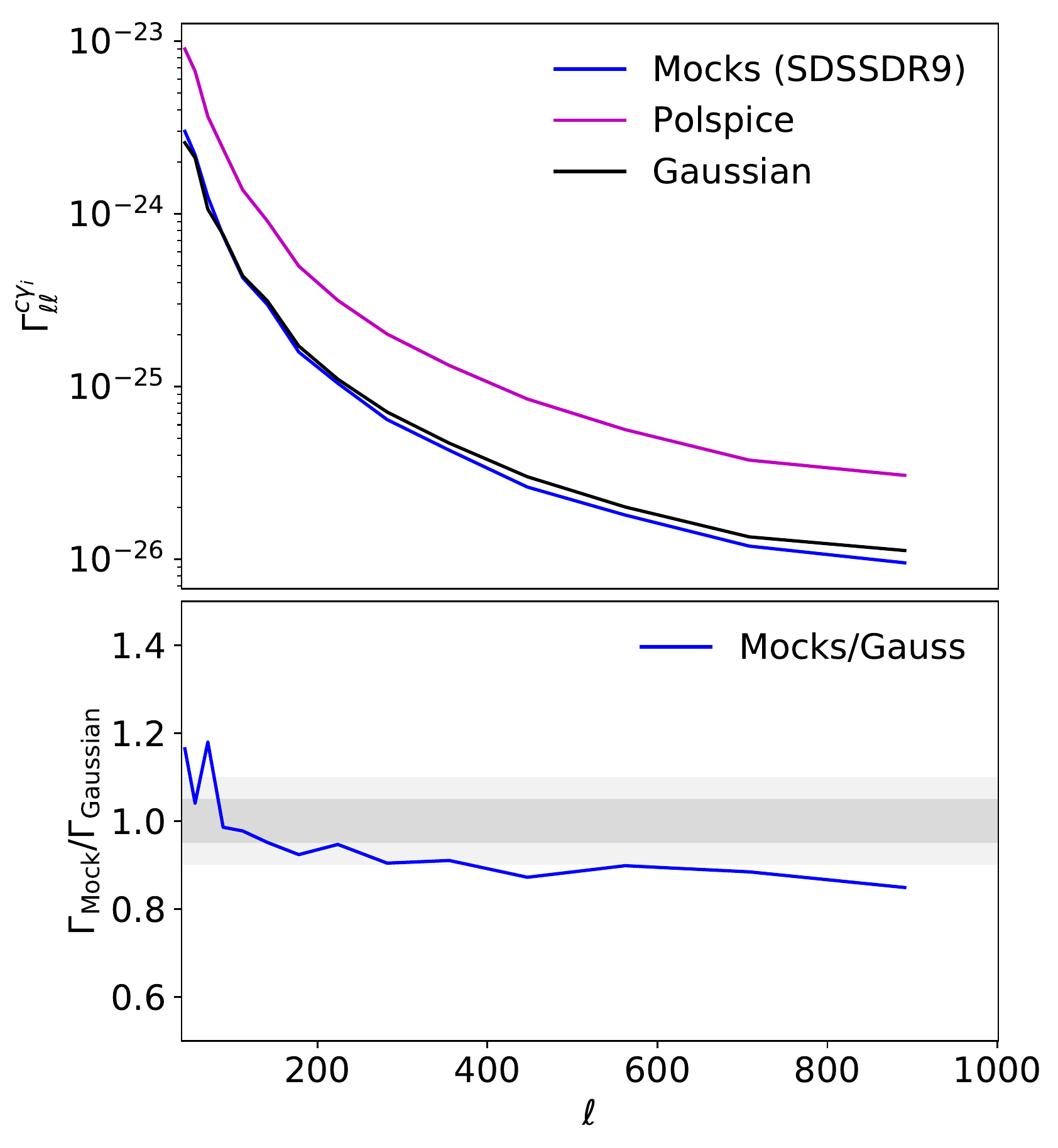}\\
    \includegraphics[width=0.35\textwidth]{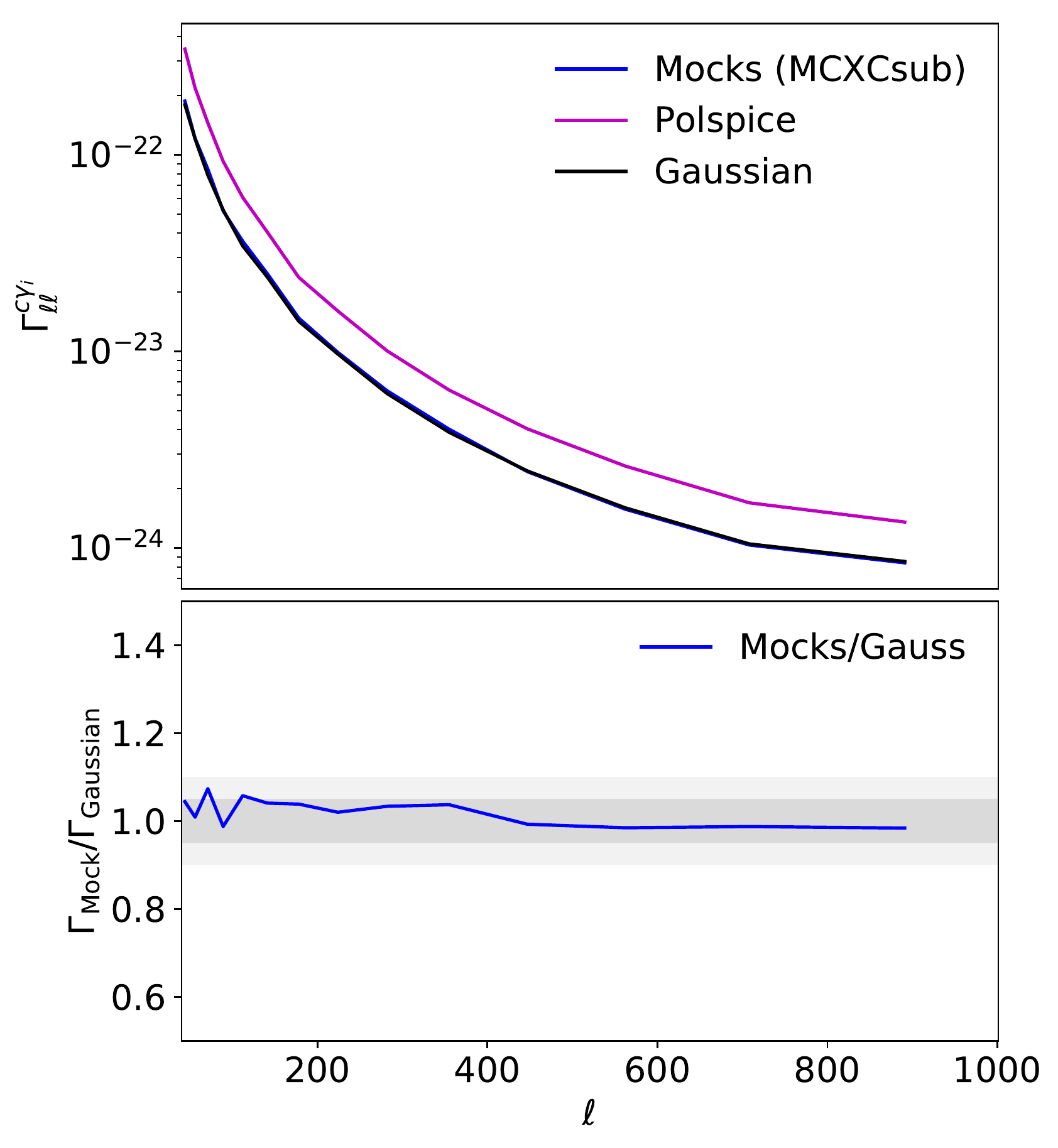}
    \includegraphics[width=0.35\textwidth]{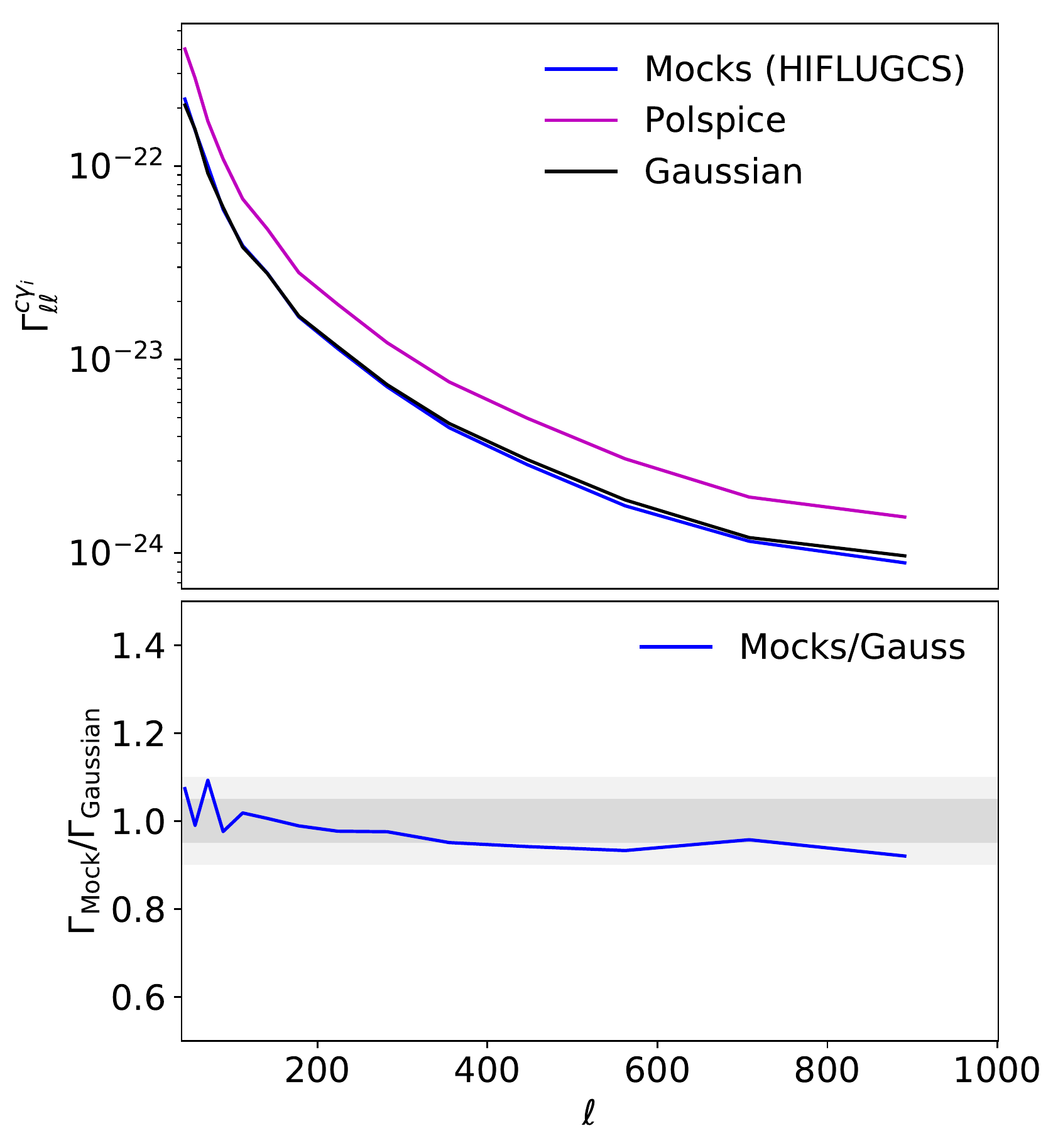}
    \caption{Comparison of the binned cross-APS variance estimated from mocks (blue) with the variance computed with Polspice (magenta) or estimated through the Gaussian prediction (black), in the third energy bin of Table \ref{tab:Ebins}. The upper panel shows the results for the WHY18 and SDSSDR9 catalogues while the lower panel stands for the MCXCsub and HIGLUCS catalogues. For each panel, the ratio between the result obtained with the mock catalogues and the Gaussian prediction is shown. The  shaded areas highlight the levels of 5\% (dark gray) and 10\% (light gray) deviations.}
    \label{fig:varCl}
\end{figure*}

\begin{figure*}
    \centering
    \includegraphics[width=0.45\textwidth]{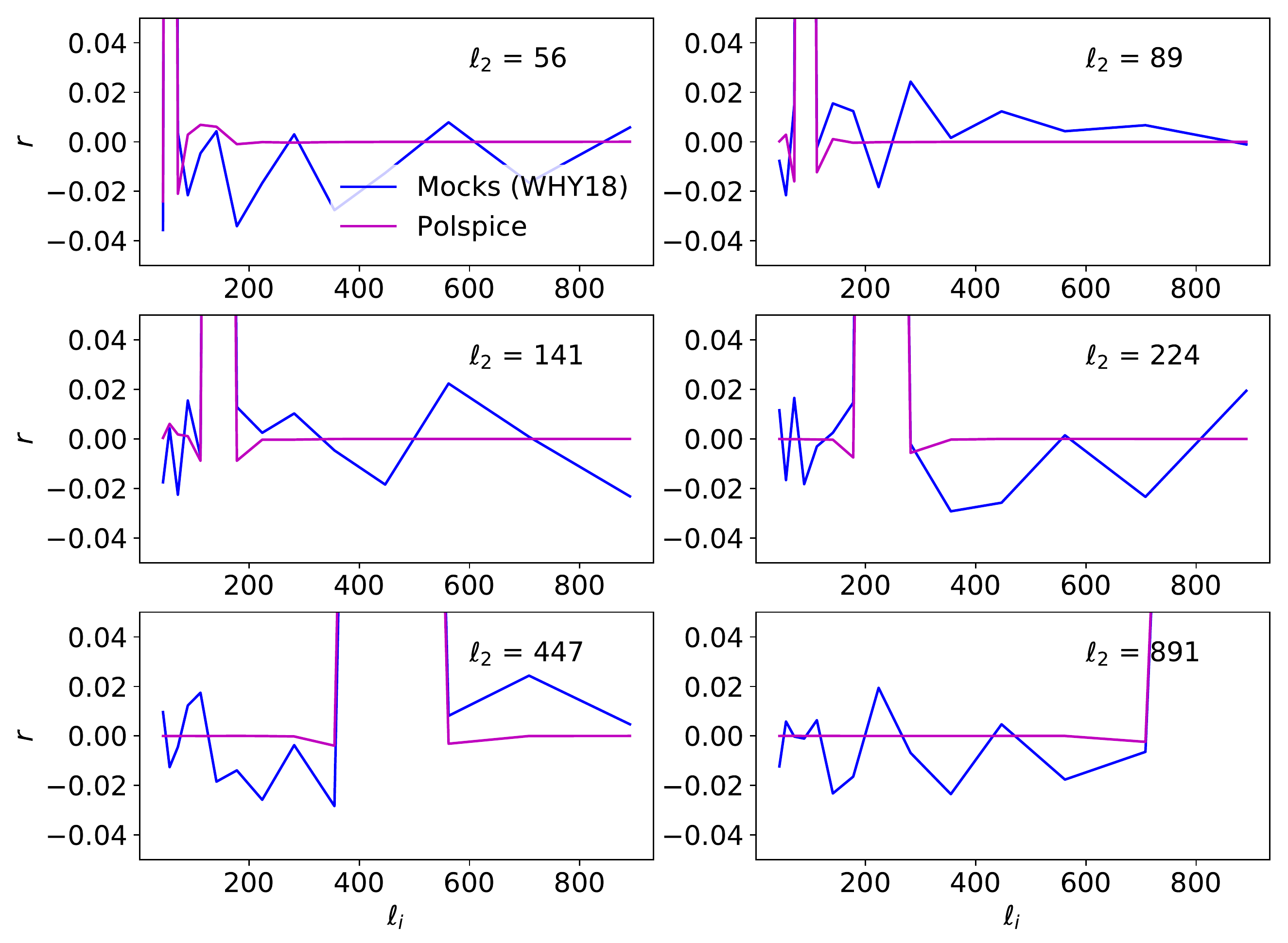}
    \includegraphics[width=0.45\textwidth]{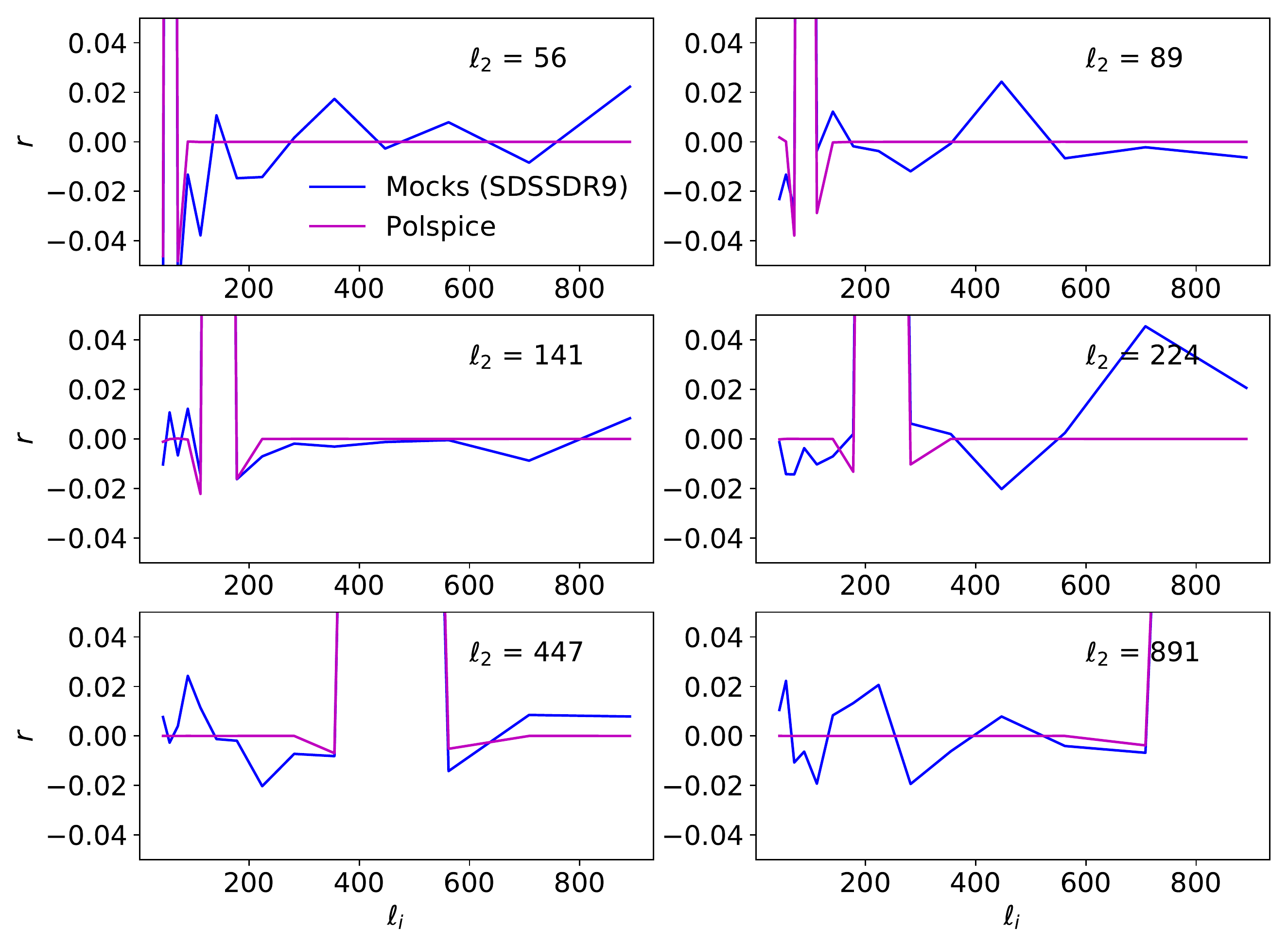}\\
    \includegraphics[width=0.45\textwidth]{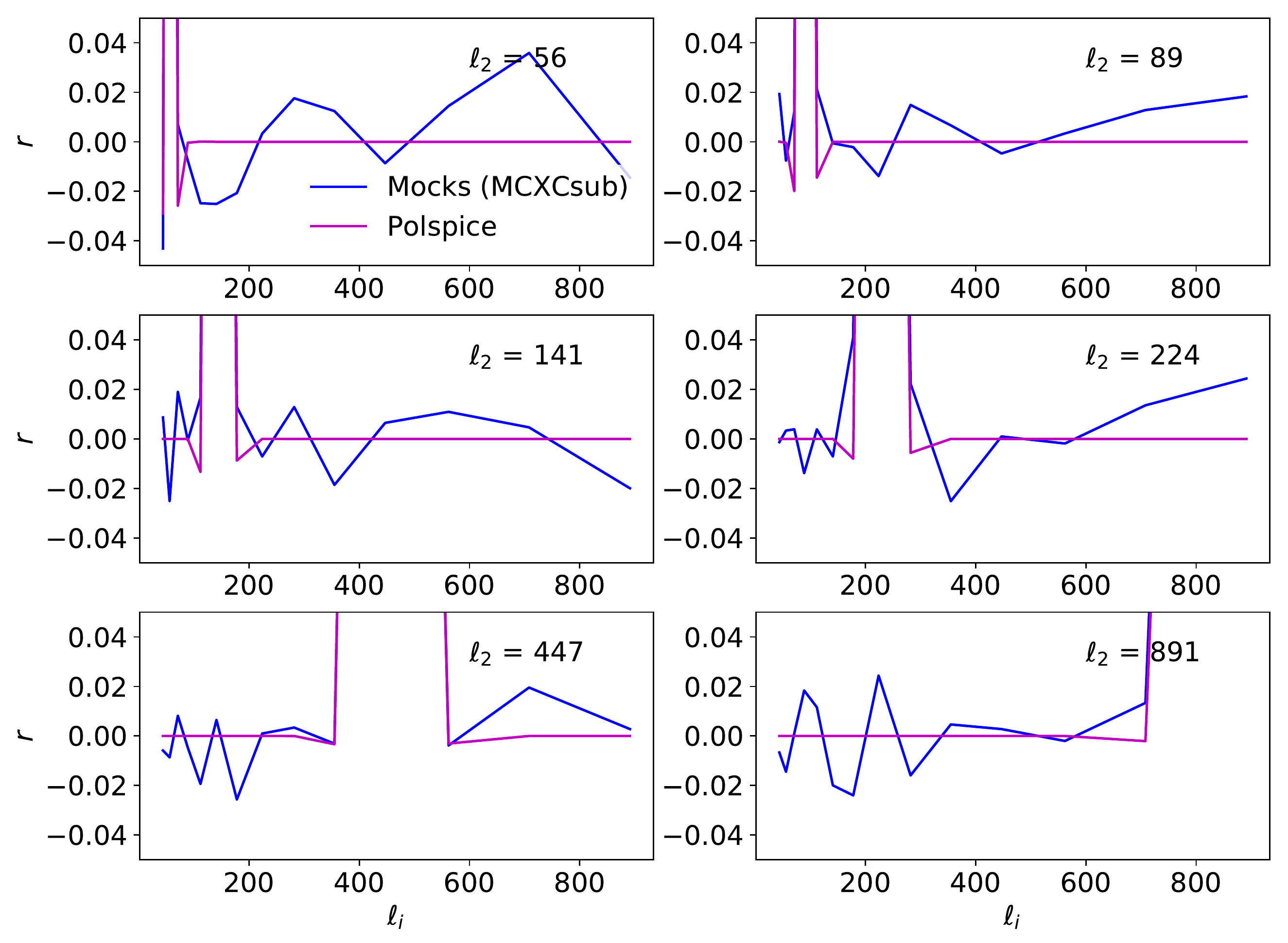}
    \includegraphics[width=0.45\textwidth]{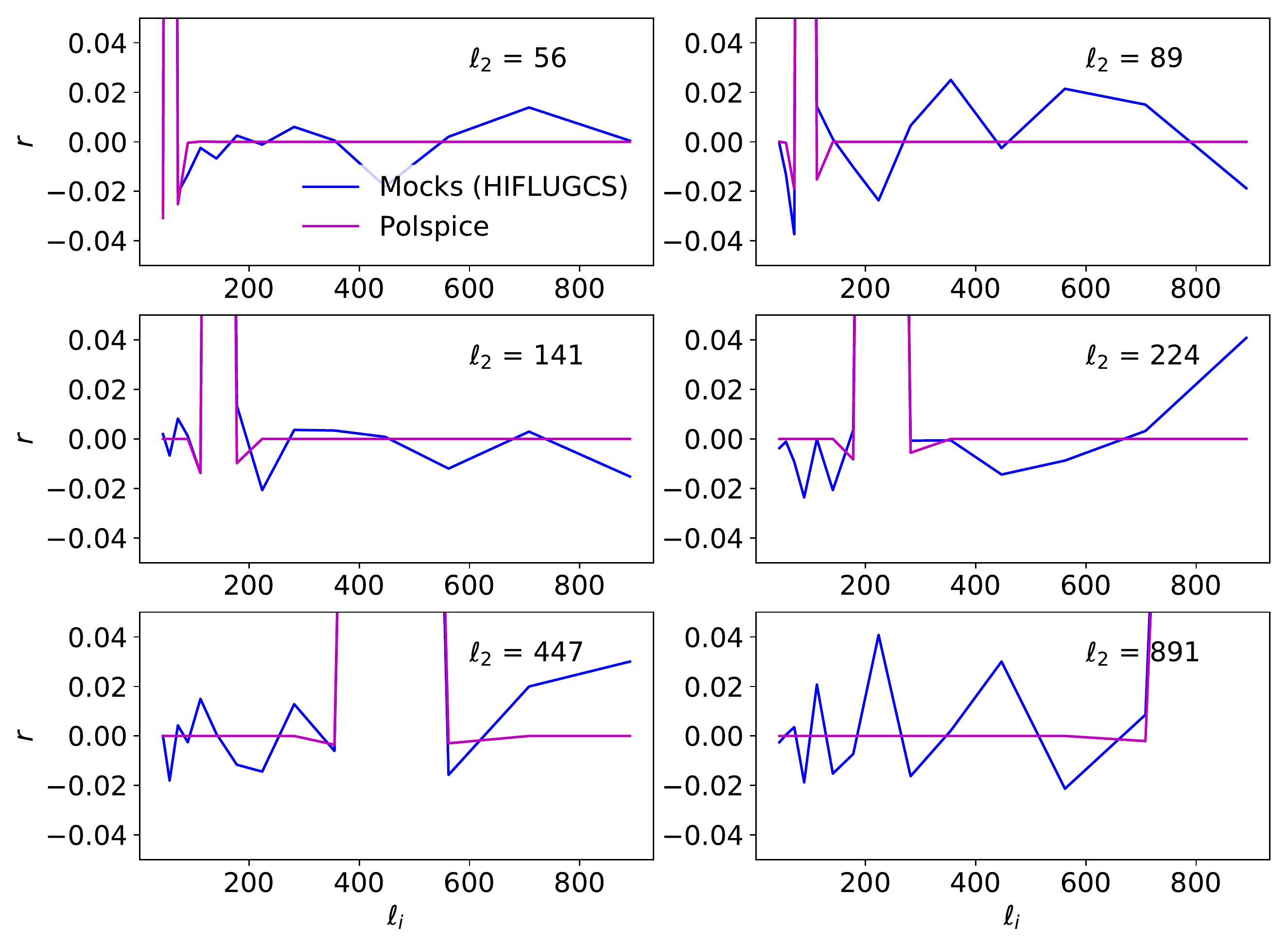}
    \caption{Cross-correlation coefficient $r$, as defined in Eq. (\ref{eqn:corrCl}), that shows the size of the off-diagonal terms of the covariance matrix for the multipole dimension. Top left (right) panels show the WHY18 (SDSSDR9) case, bottom left (right) the MCXCsub (HIFLUGCS) case. Each panel refers to a specific multipole value $\ell_2$ and shows the coefficient $r$ as a function of a different multipole $\ell_j$, reported on the horizontal axis. The blue lines stand for the analysis done with the mock catalogues, while the magenta lines refer to the covariance obtained with Polspice. The relative size of the off-diagonal terms of the covariance matrix as compared to the diagonal ones is therefore always below the few percent level. The peaks occur when the coefficient $r$ sits on the diagonal ($\ell_j = \ell_2$), where $r = 1$, by definition. The various panels show results for some representative multipoles $\ell_2=56,89,141,224,447,981$, for each of the four cluster catalogues.}
    \label{fig:rCl}
\end{figure*}

A selection of the performed tests is shown in Fig. \ref{fig:varCl} and Fig. \ref{fig:rCl}, where results are reported in terms of the binned covariance (the variance and covariance are binned over intervals of size $\Delta l = 60$ with $\ell_\textrm{min} = 40$ and $\ell_\textrm{max} = 1000$ \footnote{It is worth to mention the fact that binning the covariance matrix means to take a block sub-matrix in which the value in each block is given by the average over the covariance values in that block: this means that the diagonal of the binned covariance matrix includes some effect from the off-diagonal contribution of the full covariance matrix.}
). Fig. \ref{fig:varCl} compares the results obtained through the mocks with the covariance estimate provided by Polspice and with the theoretical Gaussian prediction for a given energy bin $i$:
\begin{equation}
     \Gamma_{\ell\ell'}^{c\gamma_i} = \frac{ C_\ell^{cc} C_\ell^{\gamma_i\gamma_i} + (C_\ell^{c\gamma_i})^2}{(2\ell+1) \, \Delta \ell \; f_{\mathrm{sky,i}}} \delta_{ll'}^{\rm K} \,,
    \label{eqn:covCrossE}
\end{equation}
%
% %
% \begin{equation}
%     {\mathrm{cov}}[C_\ell^{c\g_i},C_{\ell'}^{c\g_i}] = \frac{ C_\ell^{cc} C_\ell^{\gamma_i\gamma_i} + (C_\ell^{c\gamma_i})^2}{(2\ell+1) \, \Delta \ell \; f_{\mathrm{sky,i}}} \delta_{ll'}^{\rm K} \,,
%     \label{eqn:covCrossE}
% \end{equation}
% %
where $C_\ell^{cc}$ and $C_\ell^{\gamma_i\gamma_i}$ denote the cluster and $\gamma$-ray autocorrelations, $f_{\rm sky, i}$ is the fraction of the sky probed by the survey in the energy bin $i$, $\Delta \ell$ is the multipole bin width and $\delta^{\rm K}_{ll'}$ is the Kronecker symbol (the gaussian covariance is in fact diagonal, for which reason Fig. \ref{fig:varCl} shows the comparison for the variance at each multipole $\ell$). Polspice is a non minimum-variance 
estimator, while instead the theoretical estimate is not valid in presence of non-gaussianities, in which case it represents an underestimate of the true variance.
For the estimation of the Polspice covariance matrix we refer to \cite{covPol2004} where the procedure to compute the so called pseudo-$C_l$ covariance matrix is described in detail. Here we just list the main steps. The code: (i) computes the $C_l$ from the auto-correlation function; (ii) corrects the auto-correlation function for the effect of the mask; (iii) computes the $C_l$ back; (iv) computes the $V$-matrix (see eq. 15a of \cite{covPol2004}); (v) estimates the final covariance matrix, corrected for the mask, beam and pixel effects, as in eq. 17 of \cite{covPol2004}.
\noindent\\\\
From Figure \ref{fig:varCl} we notice, as expected, that the Polspice variance over-estimates the gaussian prediction, as well as the variance from mocks. At the same time, we find that the variance obtained using the mocks is quite close to the Gaussian prediction. The differences between the two are of the order of $\sim 10\%$ at very large scales ($\ell<100$) for WHY18 and SDSSDR9, and always smaller than 10\% for MCXCsub and HIFLUGCS; at smaller scales ($\ell>400$) they are slightly larger than $10\%$ for SDSSDR9, while smaller for all the other cluster catalogues; on intermediate scales ($100<\ell<300$) we observe differences between 5\% and 10\% for all the catalogues.

The ``gaussianity" of the covariance matrix was not an expected result as it was not expected a large over-estimation of the variance from Polspice. We can confirm that the estimated covariance matrix is nearly (although not exactly) Gaussian by looking at the off-diagonal terms of the covariance matrix, that are small compared with the diagonal ones. In figure \ref{fig:rCl} we show the cross-correlation coefficient defined as:
\begin{equation}
    r = \frac{\textbf{$\Gamma$}_{\ell\ell'}^{c\gamma_i}}{\sqrt{\textbf{$\Gamma$}_{\ell\ell}^{c\gamma_i} \ \textbf{$\Gamma$}_{\ell'\ell'}^{c\gamma_i}}}\,.
    \label{eqn:corrCl}
\end{equation}
One can note that the off-diagonal terms of the mocks covariance matrix are always smaller than 5\% with respect to the diagonal terms. The Polspice correlation coefficient, shown for comparison, is even more diagonal, as expected. Even though we obtain that the covariance matrix is significantly diagonal, nevertheless in our analyses we adopt the full (binned) covariance matrix.

\begin{figure}
    \centering
    \includegraphics[width=0.7\textwidth]{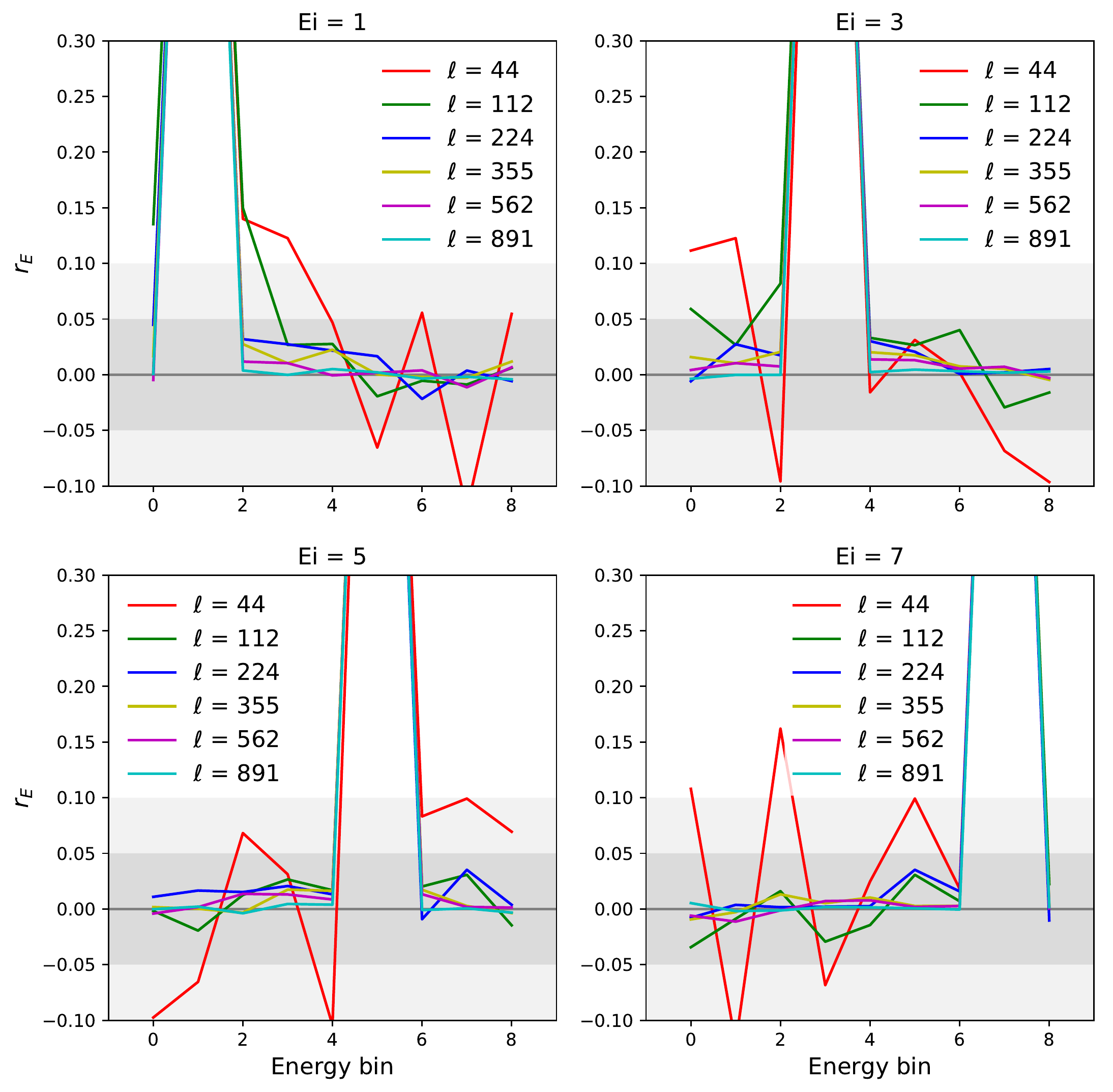}
    \caption{Cross-correlation coefficient $r_E$, as defined in Eq. (\ref{eqn:rE}), that shows the size of the off-diagonal terms of the covariance matrix for the energy dimension. The four panels refer to four arbitrary energy slices of the covariance matrix, representative of the general behaviour. Each line, identified by a different colour, stands for a different multipole $l$, while the horizontal scale refers to the integer index labelling the energy bins. The labels $E_i=1,3,5,7$ on top of each panel refer again to the the energy bin. The relative size of the off-diagonal terms of the covariance matrix as compared to the diagonal ones is therefore almost always below the 10\% level. The peaks occur when the  coefficient $r_E$ sits on the diagonal ($i=1$ in the first panel, and so on), where $r_E = 1$, by definition.}
\label{fig:rCrossE}
\end{figure}

Concerning the covariance between different energy bins, we again find that the cross APS between galaxy catalogues and $\gamma$ rays is quite diagonal, especially for energy bins of the size of those adopted in our analysis (reported in Table \ref{tab:Ebins}). This can be seen by evaluating the Gaussian estimator for the covariance in energy (at fixed multipole, for convenience):
\begin{equation}
    \Gamma_{\ell\ell}^{c\gamma_i\gamma_j}  \equiv \mathrm{cov}[C_\ell^{c\g_i},C_\ell^{c\g_j}] = \frac{C_\ell^{cc} C_\ell^{\gamma_j\gamma_i} + C_\ell^{c\gamma_i}C_\ell^{c\gamma_j}}{(2\ell+1) f_{\mathrm{sky}}}\,,
\end{equation}
where $i$ and $j$ identify the different energy bins and by determining the corresponding correlation coefficient:
\begin{equation}
    r_\mathrm{E} = \frac{\Gamma_{\ell\ell}^{c\gamma_i\gamma_j}}{\sqrt{\Gamma_{\ell\ell}^{c\gamma_i\gamma_i}\Gamma_{\ell\ell}^{c\gamma_j\gamma_j}}}
    \label{eqn:rE}   
\end{equation}
%
% %
% \begin{equation}
%     r_\mathrm{E} = \frac{\mathrm{cov}[C_\ell^{c\g_i},C_\ell^{c\g_j}]}{\sqrt{\mathrm{cov}[C_\ell^{c\g_i},C_\ell^{c\g_i}]\mathrm{cov}[C_\ell^{c\g_j},C_\ell^{c\g_j}]}}
%     \label{eqn:rE}   
% \end{equation}
% %
which is analogous to the coefficient defined in Eq. \ref{eqn:corrCl} to investigate the off-diagonal terms of the covariance matrix for what concerns the multipole. Fig. \ref{fig:rCrossE} shows $r_E$ for some selected angular scales and for MCXCsub. For most of the angular scales, the off-diagonal elements of the covariance matrix are well below a 5\% deviation from the diagonal elements. Only for smaller angular scales the effect reaches deviations of the order of 10\%. Results are similar at different angular scales and for the other cluster catalogues.

\section{Results}
\label{sec:results}

\begin{figure*}
    \centering
    \includegraphics[width=0.45\textwidth]{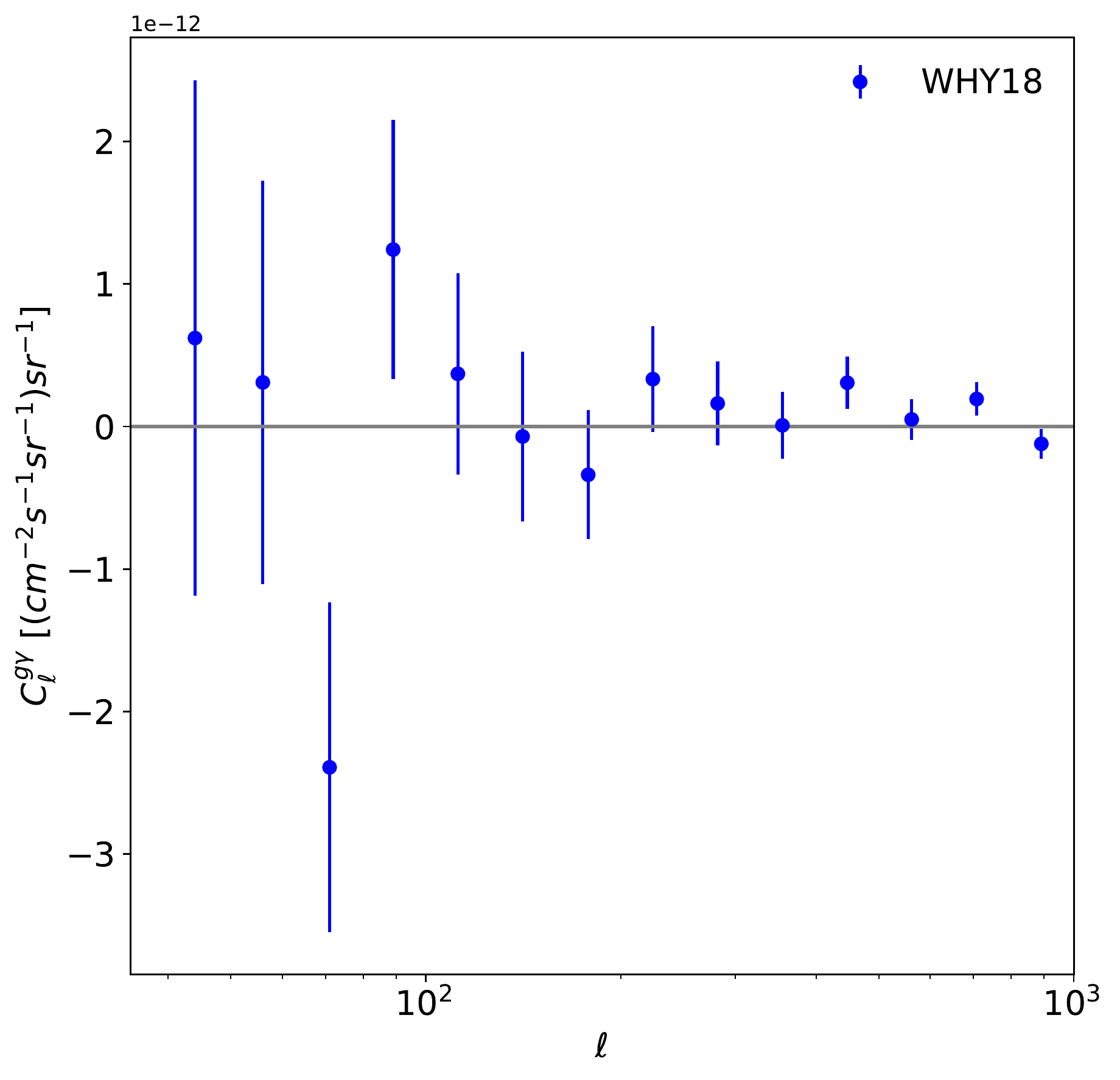}
    \includegraphics[width=0.45\textwidth]{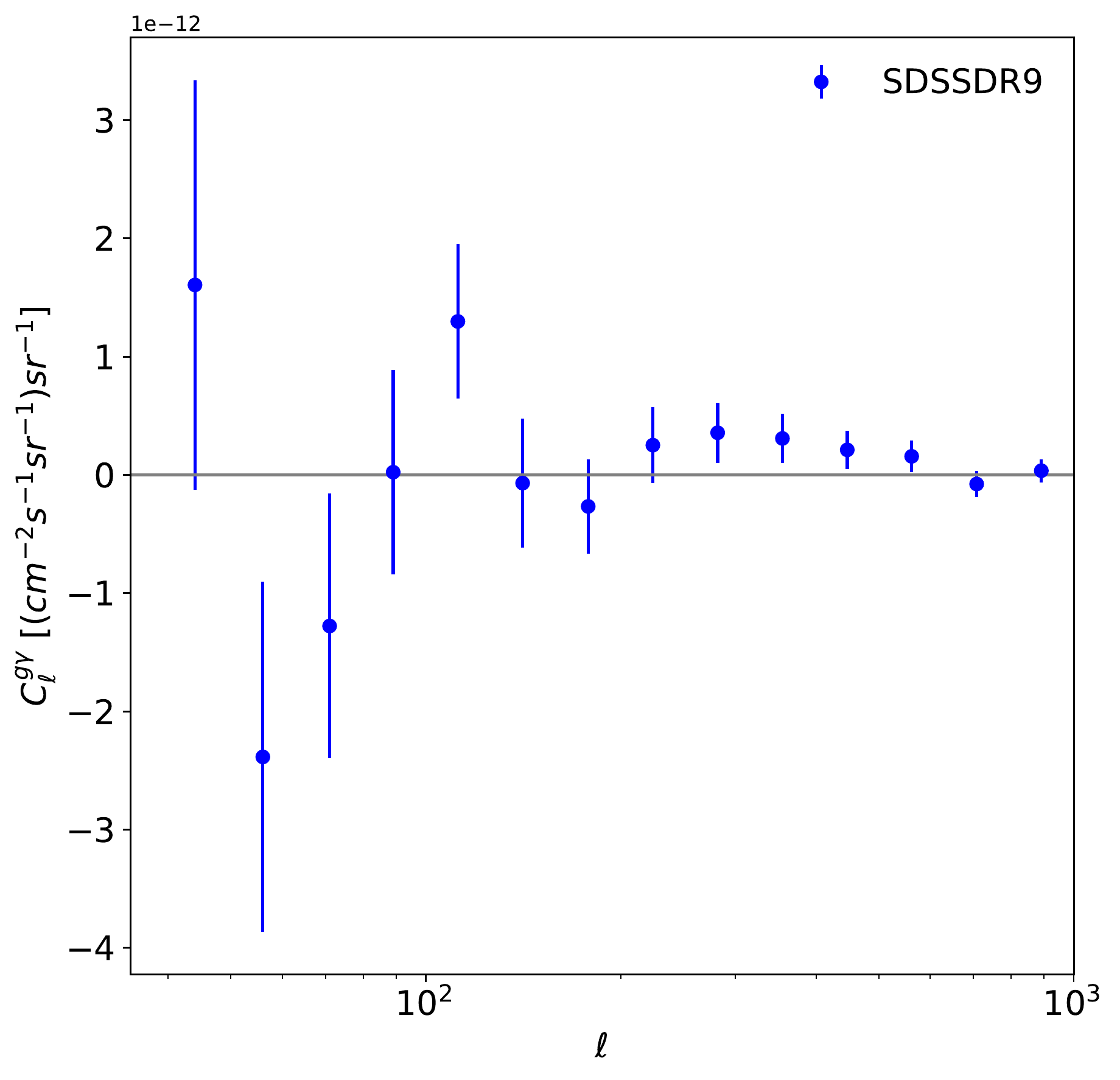}\\
    \includegraphics[width=0.45\textwidth]{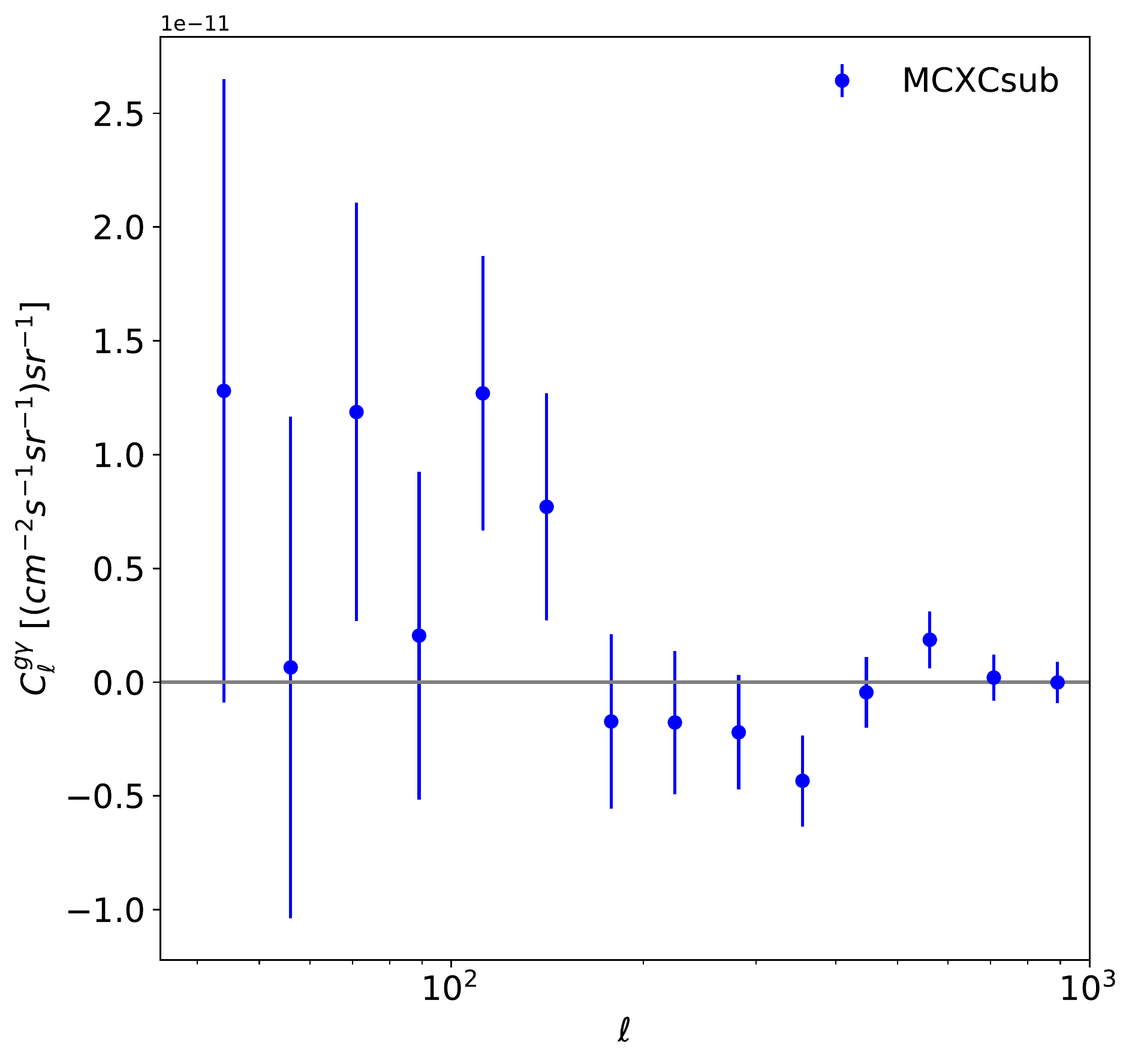} 
    \includegraphics[width=0.45\textwidth]{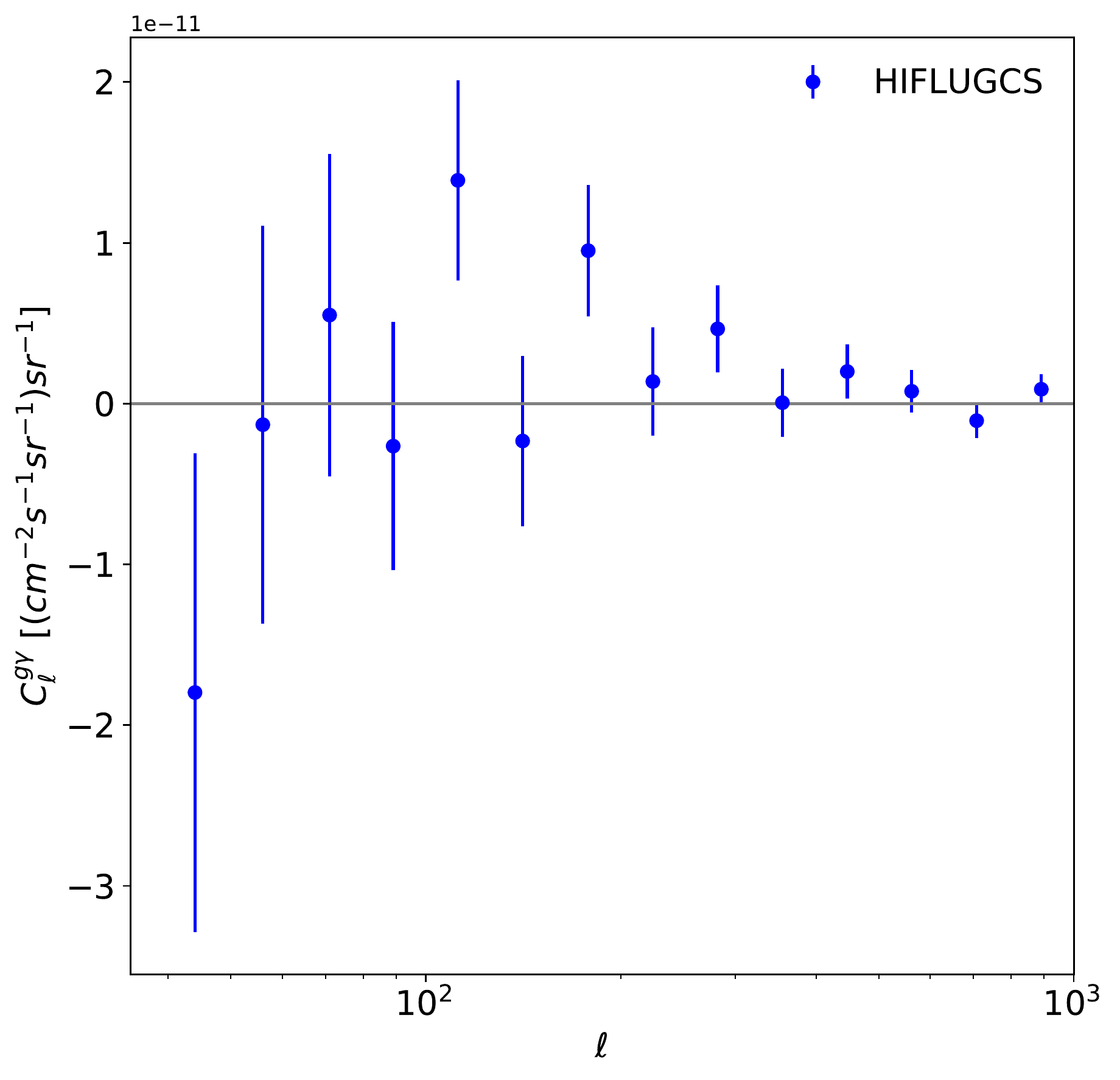}
    \caption{Binned angular power spectrum of the $\gamma$-ray--cluster cross-correlation for the four cluster catalogues adopted in our analysis and in the third $\gamma$-ray energy bin (chosen as a representative example). The error bars are obtained from the binned angular-power-spectrum covariance matrix estimated from 2000 mocks.}
    \label{fig:Clsignal}
\end{figure*}

\begin{figure*}
    \centering
    \includegraphics[width=0.45\textwidth]{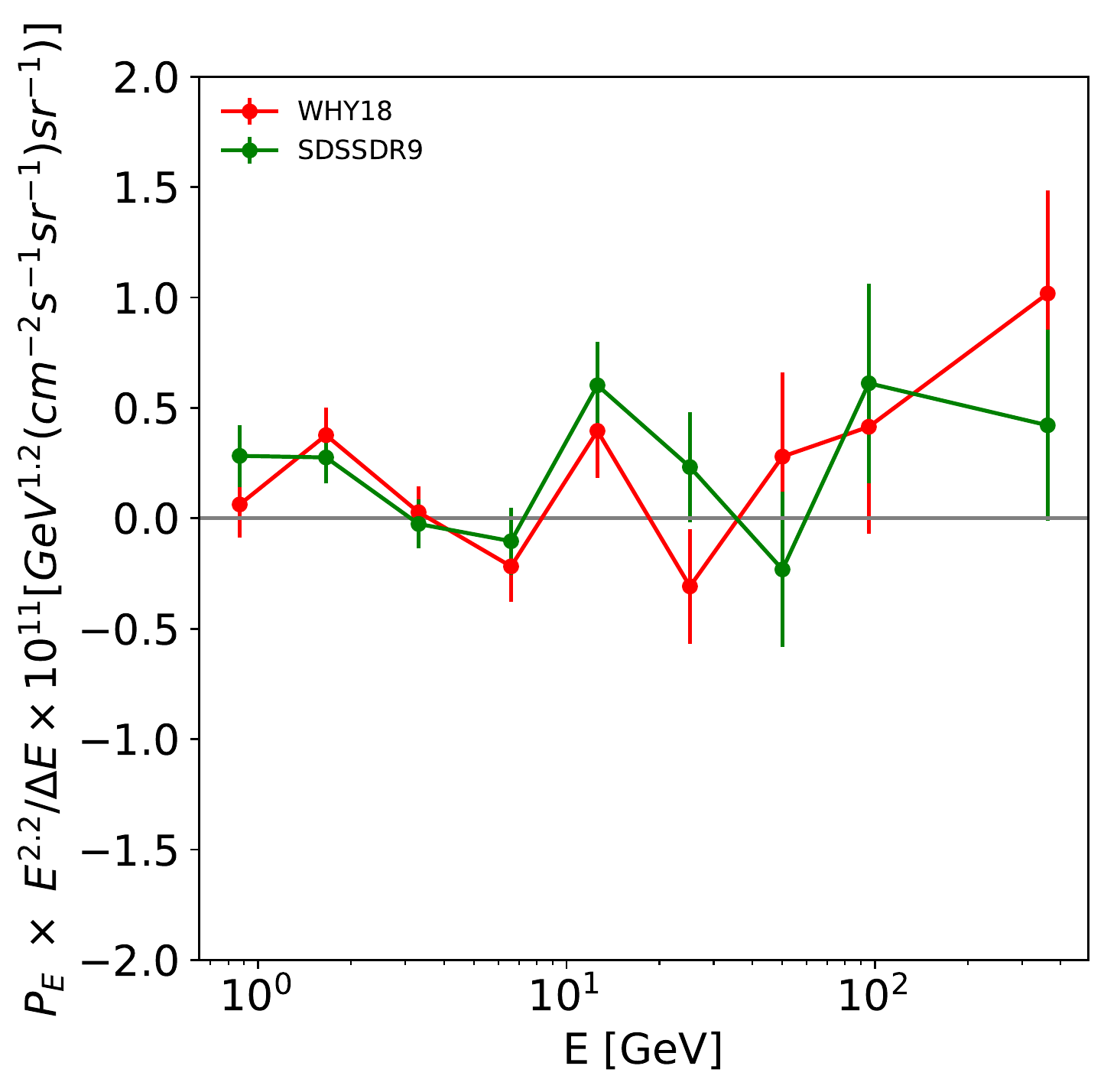}
    \includegraphics[width=0.45\textwidth]{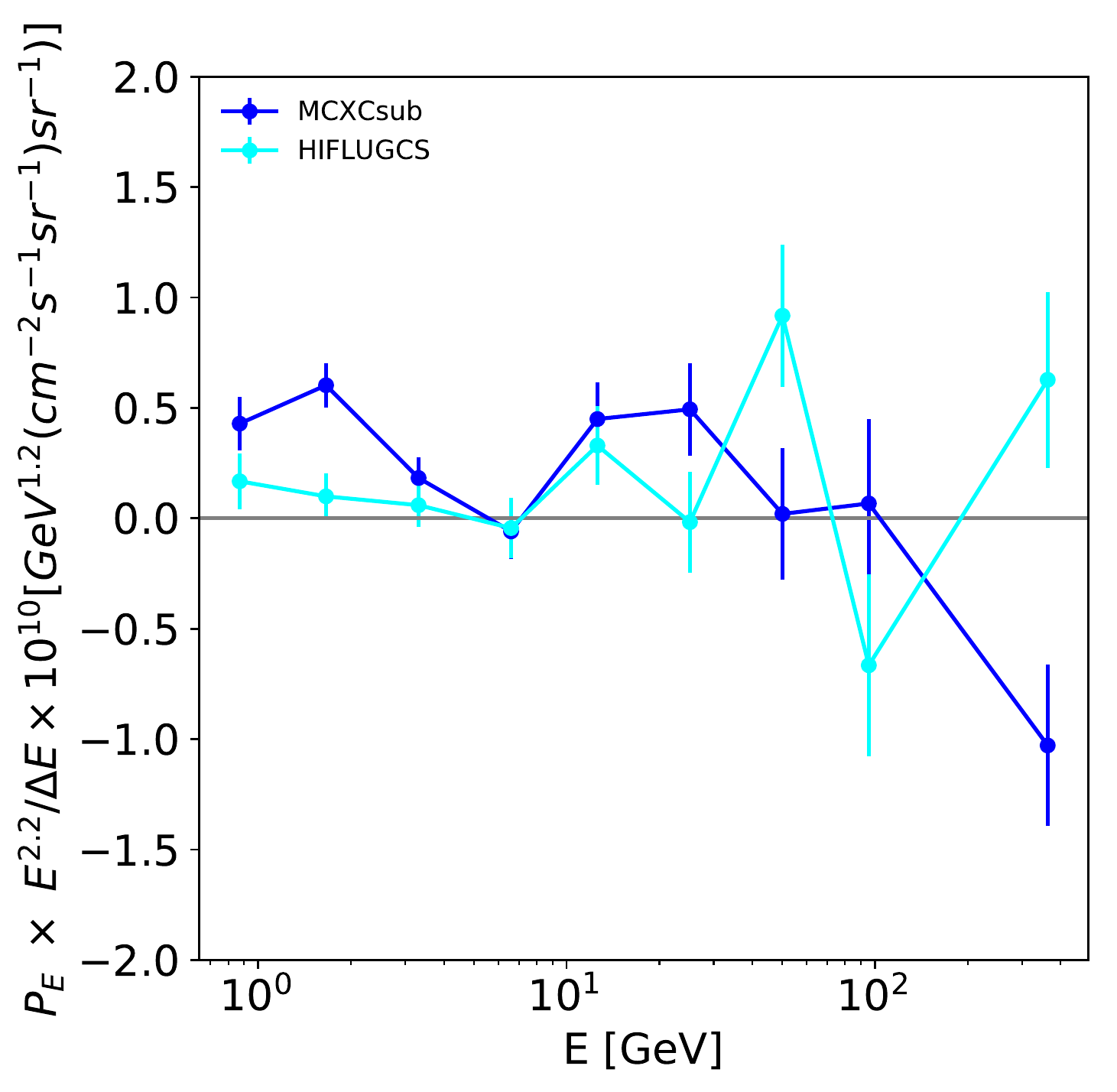}
    \caption{Energy spectrum $P_E$ of the cross-correlation angular power spectrum, for each of the four cluster catalogues. The plot shows $P_E$ rescaled by $E^{2.2}/\Delta E$, where $\Delta E$ is the width of the corresponding energy bin. The error bars are obtained from the angular-power-spectrum covariance matrix estimated from 2000 mocks.}
    \label{fig:Espectrum}
\end{figure*}

The measured cross-correlation APS between the galaxy clusters and the unresolved $\gamma$-ray intensity are shown in Fig.  
 \ref{fig:Clsignal} and Fig. \ref{fig:Espectrum}. The cross APS have been obtained by means of the Polspice estimator and the (co)variances have been derived as discussed in Section \ref{sec:mocks}. 
   
Fig. \ref{fig:Clsignal} shows a representative case of the the binned angular power spectrum $C_\ell^{c \g_i}$ as a function of the multipole $\ell$ (third energy bin of Table \ref{tab:Ebins}),
 for each of the four cluster catalogues. The error bars are the diagonal entries of the covariance matrix obtained from the mock analysis. Fig. \ref{fig:Espectrum} instead shows the energy dependence of the mean cross APS, defined as the average with respect to $\ell$ of the $C_\ell^{c \g_i}$ in each energy bin:
\begin{equation}
    P_{E_i} = \frac{1}{\Delta \ell}\sum_{\ell_{\rm min}}^{\ell_{\rm max}} C_\ell^{c\g_i}\,,
    \label{eqn:energySpectrum}
\end{equation}
where $\ell_{\rm min}$ and $\ell_{\rm max}$ are shown in Table \ref{tab:Ebins} and $\Delta \ell = \ell_{\rm max}-\ell_{\rm min}$. The errors on $P_{E}$ are defined as:
\begin{equation}
    \sigma^2_{P_{E_i}} = \frac{1}{\Delta \ell} \BL(\frac{1}{\Delta \ell}\sum_{\ell_{\rm min}}^{\ell_{\rm max}} \sigma^2_{C_\ell^{c\g_i}}\BR)\,,  % \BL(\frac{\bar{\sigma}_{C_\ell}}{\sqrt{\Delta \ell}}\BR)_{E_i}
    \label{eqn:sigmaPE}
\end{equation}
where $\sigma^2_{C_\ell^{c\g_i}} = \Gamma_{\ell\ell}^{c\gamma_i}$. To easy the visualization in the plot, the data  of eq. \eqref{eqn:energySpectrum} are multiplied by $E^{2.2}$ (expected behaviour of the UGRB \citep{Ackermann:2014usa}) and divided by $\Delta E$ (the width of the energy bin). 

In order to determine the presence of a positive cross-correlation signal, we adopt the SNR defined in Eq. \ref{eqn:SNR}. C$_{\ell,\mathrm{MODEL}}$ is set at the best fit obtained with the featureless FLAT model
defined in Section \ref{sec:models}. We perform all our fits by employing a MCMC technique to determine the likelihood of Eq. \ref{eq:like}. We specifically adopt  a pure-Python implementation of Goodman and Weareas Affine Invariant Markov chain Monte Carlo Ensemble sampler (EMCEE) \cite{EMCEE2013}. Once the best fit C$_{\ell,\mathrm{MODEL}}$ model is obtained, we determine the SNR, whose results are reported in Table \ref{tab:deltaChi2FinalFull}. The SNR analysis shows that the clusters in the WHY18 and SDSSDR9 catalogues exhibit a mild preference for a positive cross-correlation signal, while those in MCXCSub and HIFLUGCS provide a larger SNR, in excess of 3. Therefore, although not large, an evidence of $\gamma$-ray emission from those clusters appears to be present. 

In order to look for a possible large scale contribution, we then fit the measured cross APS by adopting this time a physical model which follows the features of an AGN-like $\gamma$-ray emission, as described in Section \ref{sec:models}. This model, in fact, possesses a large-scale 2-halo term. We test whether the AGN-like model is preferred over the FLAT model by mean of a $\Delta \chi^{2}=\chi_{\mathrm{FLAT}}^2-\chi_{\mathrm{AGN}}^2$ test. Table \ref{tab:deltaChi2FinalFull} shows that for MCXCsub a large-scale contribution is preferred at the $2.1\sigma$ level, whilst the other catalogues do not show a preference for the $\mathrm{AGN}$ model over the $\mathrm{FLAT}$ one.

Thus MCXCsub, with a SNR of 3.5 and a $\Delta \chi^{2} = 4.5$ pointing to a large-scale contribution, turns out to be the most interesting catalogue. 
We would expect HIFLUGCS and MCXCsub samples to provide similar results, since the clusters in the two catalogues share similar mass function and flux distributions. On the other hand, a statistical difference of less than $2\sigma$ between two different samples of (possibly) the same population is nevertheless plausible.

The AIC test confirms the preference for a large-scale contribution in MCXCsub: in particular $\mathrm{AIC}_{\mathrm{FLAT}} = 152.05$ against $\mathrm{AIC}_{\mathrm{AGN}} = 151.35$ is the indication that the AGN model allows for a smaller loss of information, even though with a somewhat smaller evidence than in the $\Delta \chi^{2}$ analysis. All the other three catalogues show instead a preference for the FLAT model, namely $\mathrm{AIC}_{\mathrm{FLAT}} < \mathrm{AIC}_{\mathrm{AGN}}$ . 

\begin{figure*}
    \centering
    \includegraphics[width=0.9\textwidth]{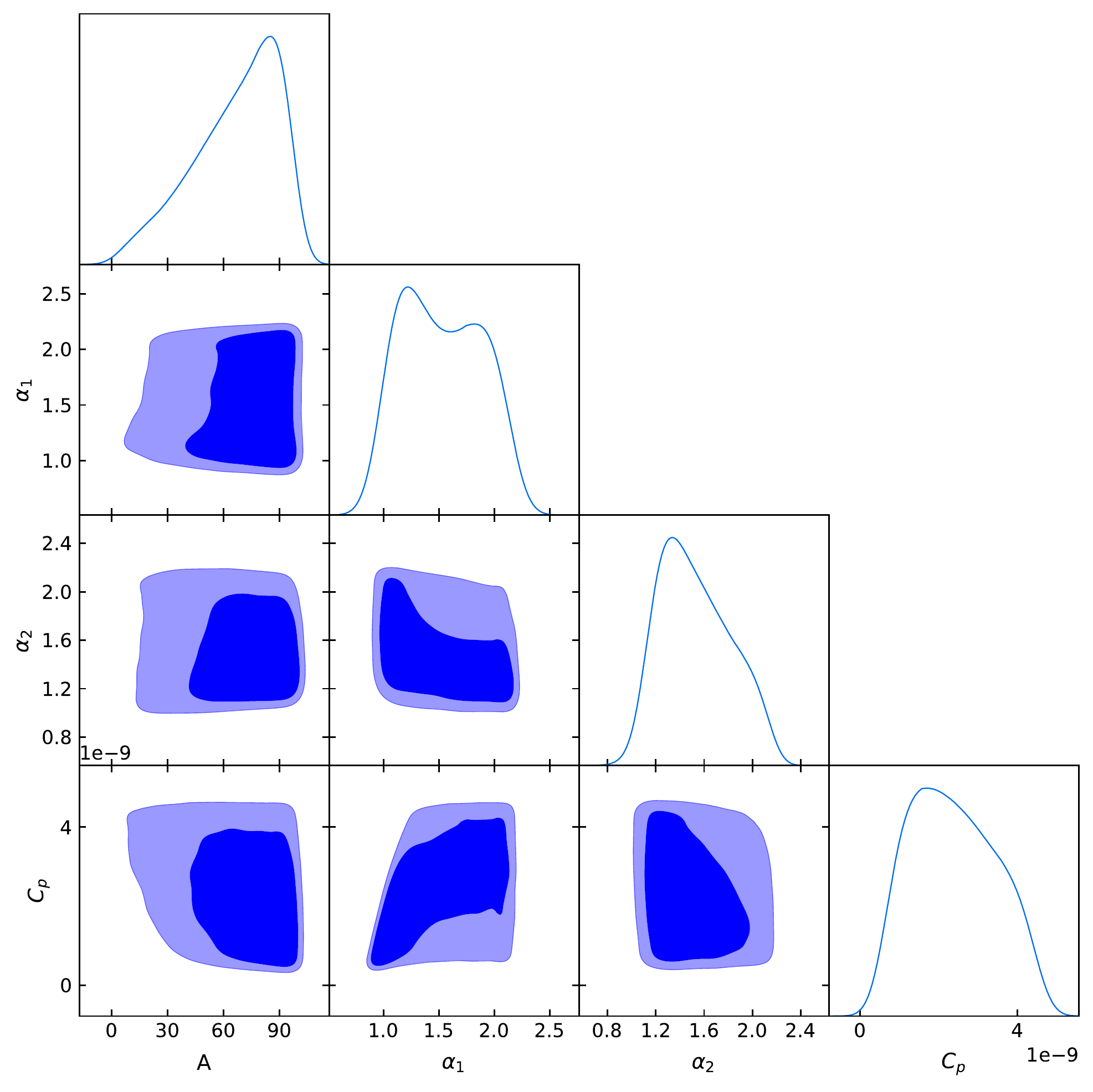}
    \caption{Triangular plot for the bounds on the AGN model parameters obtained from the the fit to the cross-correlation with the MCXCsub catalogue. The contours refer to the 68\% (dark blue) and 95\% (light blue) confidence levels.}
    \label{fig:2d_distro_AGN_mcxcsub}
\end{figure*}

In Fig. \ref{fig:2d_distro_AGN_mcxcsub}, we show the allowed regions obtained for the model parameters in the MCXCsub case. The contours and shaded areas refer to the 2-dimensional allowed regions at 68\% (dark blue) and 95\% (light blue) confidence levels. From Fig. \ref{fig:2d_distro_AGN_mcxcsub} we see that the constraints are somewhat loose, but consistent with an AGN-like model, with the spectral indices close the mean blazar $\gamma$ ray emission 
(\cite{Fermi-LAT:2019pir}), but notably somewhat smaller. A spectral index lower than $\sim 2$ is indicative of a hardening of the spectrum for the unresolved population of blazars, a result compatible with the findings of Ref. \citep{AckermannEtAl2018}. The AGN normalisation, instead, turns out to be much larger than expected, even though with a sizeable uncertainty.
The model adopted in our analysis is normalised such that the integral of the window function over the redshift provides approximately the measured UGRB intensity (see \citep{AmmazzalorsoEtAl2018} for details).
The value of the normalisation we obtain here from the MCMC is $A = 71.5^{+ 19.9}_{- 29.9}$. 
We verified that such conclusion is independent on the specific model of AGN or blazars adopted in eq.~\ref{eqn:AGNModel}.

A value this large could therefore exceed significantly the UGRB intensity, if due to a 2-halo emission: in the halo model, it is the 2-halo term which is directly related to the total $\gamma$ ray emission, while instead the 1-halo term can be large without necessarily inducing an exceedingly large total emission.
Indeed, both the gamma-ray intensity and the two halo term are set by the window function. This can be seen from their definitions: $I_\gamma=\int d\chi\,W_\gamma(\chi)$, and 
$C_\ell^{c\gamma,2h}=\int d\chi/\chi^2\,W_c(\chi)\,W_\gamma(\chi) \langle b_c(\chi)\rangle\,\langle b_\gamma(\chi)\rangle \,P^{lin}(k=\ell/\chi,\chi)$, 
where $\langle b_i\rangle$ is the bias of source $i$ with respect to matter. For a generic population emitting gamma-rays, $\langle b_\gamma \rangle \sim 1$ at low-$z$ (i.e., in the range we are considering).
Therefore a normalization different from one for the cross-correlation APS could only be re-absorbed in the window function, thus affecting the intensity in the same way.
For what concerns the one-halo term, there is instead an additional ingredient, that is poorly constrained, and can significantly change the strength of the correlation without affecting $W_\gamma$ and in turn $I_\gamma$, which is the average gamma-ray luminosity from a cluster of a given mass and redshift. Making this function steeply increasing with the cluster mass can boost the one-halo term.

The result we found might thus indicate that the model we implemented is able to effectively capture a large-scale contribution, but such correlation is not due to a 2-halo term involving $\g$-rays from AGNs (or other galactic sources) in 2 different halos at large physical distances. On the contrary it might be seen as a potential indication in favour of a diffuse emission from the ICM (which would instead be a 1-halo term). Indeed, a relevant 1-halo term providing correlation on scales around $0.5-1$ degree and provided by $\g$-rays from the ICM can be obtained~\citep{BranchiniEtAl2017,ReissEtAl2017}, with no obvious violation of other existing bounds.
%
%In other words from this results it seems that we are detecting a clusters diffuse emission without imposing directly a model for the ICM. 

\begin{figure*}
    \centering
    \includegraphics[width=0.5\textwidth]{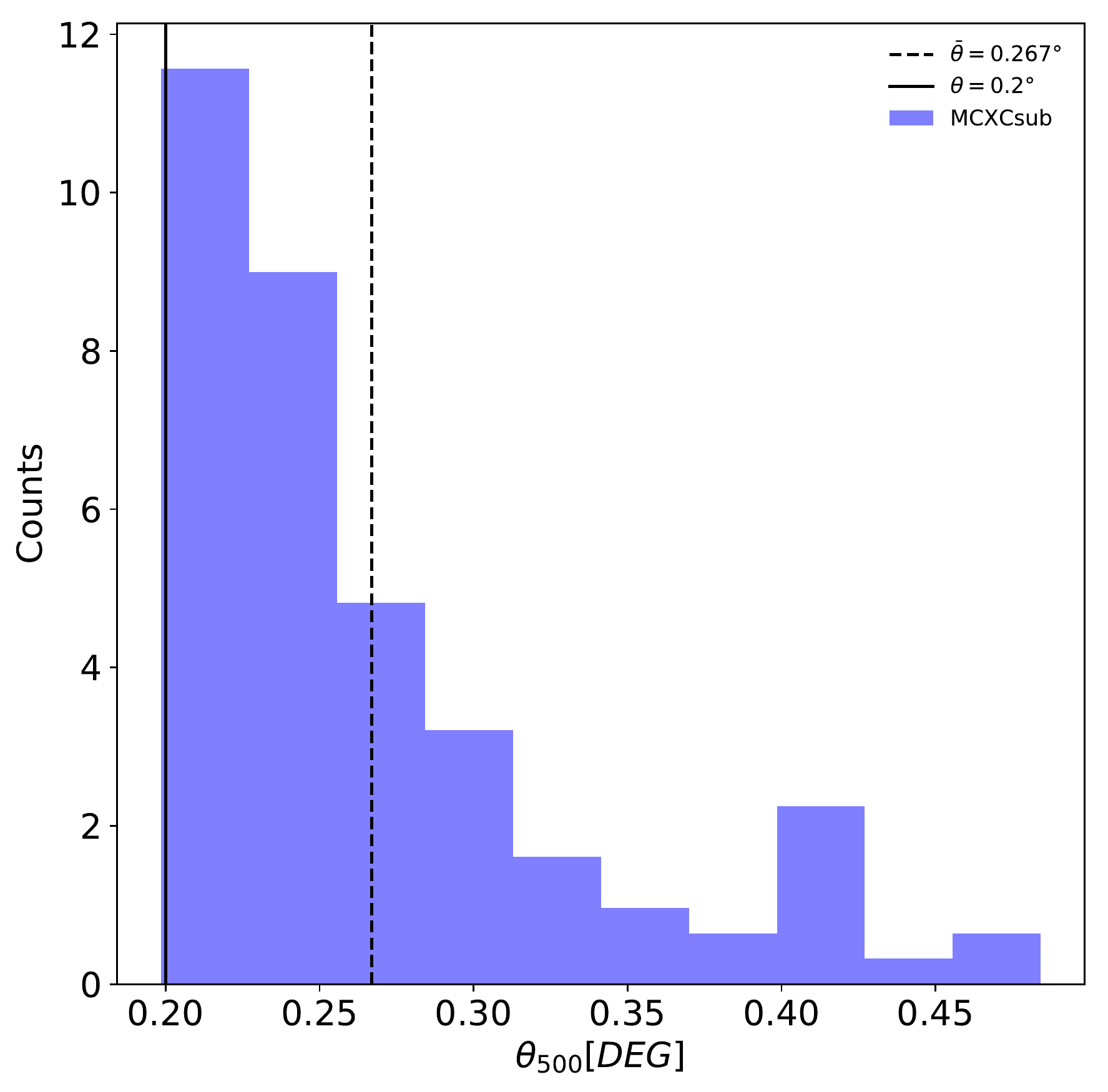}
    \caption{Distribution of $\theta_{500}$ for MCXCsub. The vertical solid line is set at $\theta_{500}=0.2\degree$, as an indication of the average size of the {\it Fermi}-LAT PSF (the values of the 68\% containment angles in the different energy bins are reported in Table \ref{tab:Ebins}), while the dashed vertical line shows the average angular size $\bar{\theta} = 0.267$ of the clusters in the MCXCsub catalogue.}
    \label{fig:thetaDistro}
\end{figure*}

Clearly, if such a signal is present, it must be provided by the clusters with a size of their diffuse emission significantly larger than the \fermi\ PSF. The 1-halo signal from the clusters with an angular dimension below/around the \fermi\ PSF would instead be described by a featureless APS, like in the FLAT case. In order to investigate more deeply this issue, we subdivided the MCXCsub cluster catalogues in two sub-catalogues by looking at the angular size of the clusters. The selection is done according to the angular dimension of the clusters $\theta_{500} = r_{500}/d_A(z)$, where $r_{500}$ is the virial radius relative to an overdensity of 500 (as defined in footnote \ref{foot:virial}) and $d_A(z)$ is the angular diameter distance, which depends on redshift $z$. We compute the average $\bar{\theta}_{500}$ over all the MCXCsub clusters and we focus this new analysis on all those clusters with $\theta_{500} > \bar{\theta}_{500}$. They are expected to be the main contributors to the extended 1-halo correlation if the hypothesis of ICM emission is correct. We have that $\bar{\theta}_{500} = 0.267$ and the total number of clusters with angular size smaller or larger than $\bar{\theta}_{500}$ are 72 and 37, respectively. The angular size distribution of the MCXCsub is shown in Figure \ref{fig:thetaDistro}. We can see that most of the MCXCsub clusters have a size larger than the {\it Fermi}-LAT PSF in most energy bins. The latter is reported in Table \ref{tab:Ebins}. The vertical solid line in the figure refers to an angle of 0.2 deg, which is an approximate illustration of the {\it Fermi}-LAT beam.

%For this check we are interested in MCXCsub, but for completeness we look at the angular distribution of all the cluster catalogues to see also the dimension of those clusters that are not given any kind of hint in the direction of an ICM diffuse emission.
%

\begin{figure}
    \centering
    \includegraphics[width=0.9\textwidth]{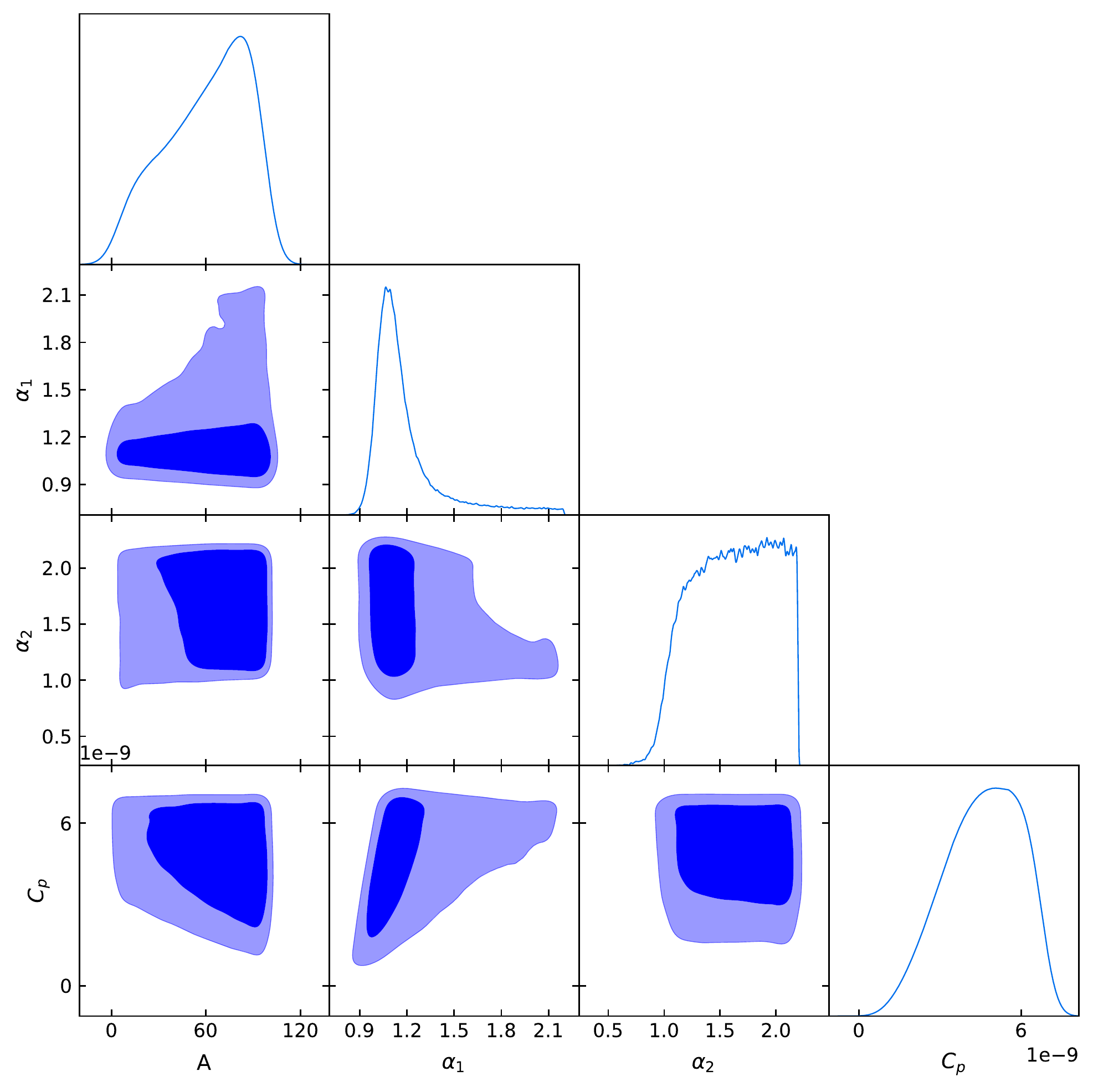}
    \caption{The same as in Fig. \ref{fig:2d_distro_AGN_mcxcsub} but for the MCXCsub clusters selected according the criterion $\theta_{500}>\bar{\theta}_{500}$.}
    \label{fig:2d_distro_AGN_mcxcsubGT}
\end{figure}

We then perform the fit of only the MCXCsub which are larger than the average $\bar{\theta}_{500}$. This analysis  requires to build a new set of cluster mocks (2000) produced in the same way as discussed above, from which 
we determine the APS covariance matrix. The best fit results are shown in \ref{fig:2d_distro_AGN_mcxcsubGT} and the results for the $\DeltaC2_{AGN-FLAT}$ turns out to be only 2.1, which corresponds to $0.96\sigma$. 
Contrary to the expectations for an ICM signal, the statistical significance does not increase when only the largest clusters are considered: instead, it rather decreases as compared to the full MCXCsub sample. This significantly weakens a possible ICM interpretation, and leaves open the alternative between an unresolved blazar population (with a slightly harder energy spectrum as compared to the resolved ones) and a diffuse emission from the cluster itself, like in the case of the intra-cluster medium emission. The value of the best fit parameters are similar to what is found for the full MCXCSub case, with a rather large normalisation parameter: $A = 65.0^{+24.8}_{-35.7}$. This time, the parameter is consistent at $1.8\sigma$ with an interpretation in terms of LSS from AGN emission, although the large uncertainty does not allow to make firm conclusions.

\begin{figure*}
    \centering
    \includegraphics[width=0.5\textwidth]{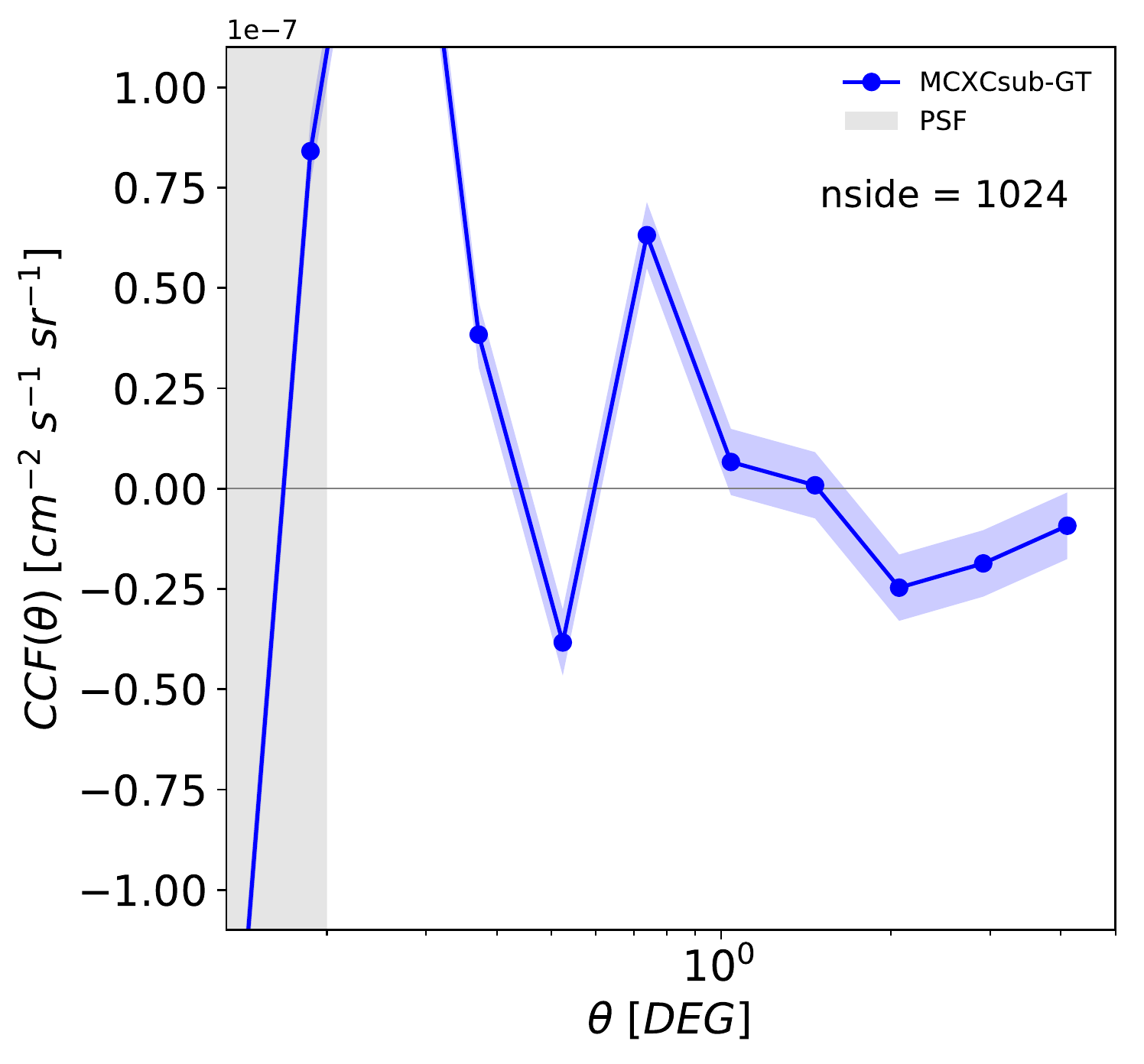}
    \caption{Angular two-point cross-correlation function for the most extended MCXCsub clusters (those with $\theta_{500}>\bar{\theta}_{500}$) and $\gamma$-ray energies in the range $(1,10)$ GeV. The shaded blue area is an estimate of the error obtained from the diagonal of the Polspice covariance matrix; the grey vertical indicates the region where the Fermi PSF effects are not negligible.}
    \label{fig:ccfMCXCsub}
\end{figure*}

For illustrative purposes, in order to visualize the angular scales at which this excess occurs, we show the cross correlation function (CCF) also in configuration space. The CCF for the subset of MCXCsub including most extended clusters (those with $\theta_{500}>\bar{\theta}_{500}$) is reported in Fig. \ref{fig:ccfMCXCsub}. The plot refers to the energy range 1-10 GeV, where the photon count statistics is large and the angular resolution not too poor. The grey area indicates the size of {\it Fermi}-LAT PSF. 

The CCF exhibits a significant "noise" term at small angular scales, compatible with unresolved AGN point-like emission. A peak is present on angular scales of the order of $0.7\degree$, although not so statistically significant to determine a preference for a large-scale term. This excess occurs at similar scales as those obtained in Ref. \cite{mcxcsub2018} by means of a stacking analysis of the $\gamma$-ray emission around galaxy clusters and is there interpreted as due to the presence of virial shocks in the clusters. 

\section{Conclusions}
\label{sec:conclusion}

We analysed the cross-correlation angular power spectrum between the unresolved extra-galactic $\gamma$-ray background measured by the {\it Fermi}-LAT and the large scale structure of the Universe at low redshift traced by four galaxy clusters identified in three different bands: WHY18 (infrared band), SDSSDR9 (optical band), MCXCsub and HIFLUGCS (X-ray band). The main motivation was to investigate whether the cross-correlation technique could identify the presence of an extended $\gamma$-ray emission possibly compatible with an intra-cluster medium emission.

For all the four catalogues, the analysis confirmed that the unresolved $\gamma$-ray emission observed by {\it Fermi}-LAT correlates with the large-scale clustering in the Universe as observed by \cite{FornengoEtAl2015} with the LSS tracer given by the CMB lensing, by \cite{Xia:2015wka} with galaxies and by \cite{BranchiniEtAl2017} specifically with clusters. We found that the largest significance occurs for the galaxy clusters identified in the X-ray band, i.e. MCXCsub and HIFLUGCS, for which the SNR is 3.5 and 3.2, respectively. When compared with a theoretical model which contains an explicit term referring to a large-scale $\gamma$-ray emission, MCXCsub exhibits a clear preference for this type of emission as compared to the a model containing only ``shot-noise'' emission from unresolved point sources, like sub-threshold AGNs. The energy spectrum of this latter component is found to be slightly harder that the mean spectral behaviour of resolved blazars, possibly indicating differences between the resolved and unresolved components of the AGN population, as observed also in \citep{AckermannEtAl2018}. Further investigation of the extended emission could not disentangle between the two options offered by a large-scale 1-halo term, possibly linked to an intra-cluster medium emission, and a large-scale 2-halo contribution, like it would occur if the correlation is due to the large-scale distribution of point-like AGNs.

However, the analysis in angular space shows a peak in the correlation function on angular scales of the order of $0.7\degree$, which appears compatible with the results of \cite{mcxcsub2018} obtained by means of a stacking analysis and where the peak is associated to the $\gamma$-ray emission in virial shocks. In our analysis, we confirm the presence of a fluctuation on similar angular scales, although we do not have the sensitivity to determine whether the peak in the correlation function has a physical origin or instead just reflects a statistical fluctuation.

%\textbf{OLD: However, the analysis in angular space shows a peak in the correlation function on angular scales of the order of $0.7\degree$, which is compatible with the results of Ref. \cite{mcxcsub2018} obtained by means of a stacking analysis. Therefore we confirm the presence of a fluctuation, although we cannot determine whether it has a physical origin, being possibly a hint of $\gamma$-ray emission in virial shocks in the clusters, or a statistical origin.}

In developing the analysis, we also derived and tested technical tools specifically designed to determine reliable multidimensional covariance matrices, which are a key ingredient for the study of cross correlation signals.
These methods are summarised in the Appendices and refer to development, test and comparison of general numerical techniques for the massive and efficient production of mock realisations of the sky for cross-correlation studies, like the correlation of catalogues of galaxies or clusters with $\gamma$-ray maps. We then developed a semi-analytic framework that allowed us to properly join the information coming from the two pieces of the covariance matrix (galaxies/clusters catalogues from one side, and $\gamma$-rays from the other side) without overestimating the error matrix: this is clearly important for the estimation of the significance of the presence of a signal and for the inference of model parameters. These techniques are general enough such that they can be used for any distribution of objects and can be adapted easily to different statistical and astrophysical analyses. 

\section*{Acknowledgements}

We warmly thank Enzo Branchini for very useful and interesting discussions.
This work is supported by the following grants:  {\sl Departments of Excellence} (L. 232/2016), awarded by the Italian Ministry of Education, University and Research (MIUR); {\sl The Anisotropic Dark Universe}, Number CSTO161409, funded by Compagnia di Sanpaolo and University of Torino; {\sl TAsP (Theoretical Astroparticle Physics)} project, funded by the Istituto Nazionale di Fisica Nucleare (INFN); PRIN 2017 project (Progetti di ricerca di Rilevante Interesse Nazionale) {\sl The Dark Universe: A Synergic Multimessenger Approach}, Number 2017X7X85K, funded by MIUR. 
MR acknowledges support by ``Deciphering the high-energy sky via cross correlation'' funded by Accordo Attuativo ASI-INAF n. 2017-14-H.0.
J.-Q.X. is supported by the National Science Foundation of China under grant No.11633001 and No.11690023.

\bibliographystyle{mnras} 
\bibliography{biblio}

\appendix  \label{sec:appendix}
\section*{Appendix}

As pointed out in the main section of the paper, an accurate estimate of the covariance matrix of the cross-correlation angular power spectrum along its different dimensions (multipole, energy, redshift) is necessary to infer the statistical significance of our results. The method
we are using to derive the cross-correlation APS is based on Polspice, which is a non-minimum variance algorithm. We therefore investigated methods based on the production of mock realisations of the two maps which enter the cross-correlation analyses, from which we derive estimates of the covariances of the cross-APS, in order to determine which combination for the generation of the two maps is more suitable for cross-correlation analyses, and to verify if the results are stable across different techniques for the generation of mocks. The method we devise starts from the true maps under investigation, from which the mocks are produced, and is quite general: this makes it directly adaptable to derive the covariance also along those directions (like energy and redshift) for which Polspice cannot be used. Since cross-correlations deal with two observables, the generation of a suitably large number $N$ of maps for each set of observables implies the crossing of $N \times N$ maps: this can make the computation of the covariance quite demanding from the computational point of view, since $N$ needs to be large enough to make all the procedure reliable. We therefore devised, tested and validated a method which allows the computing resources to grow linearly with $N$ rather than quadratically: this is obtained by deriving two estimates of the covariance by correlating the $N$ mocks of the first observable with the true map of the second observable, and then performing the opposite: we demonstrate below that the total covariance can then just be obtained as the average of these two ``half'' covariances. The results thus obtained are a faithful estimate of the global $N \times N$ crossing.

In the reminder of this Appendix we discuss the various techniques adopted to generate mock maps for both galaxy/cluster catalogs and for $\gamma$ ray maps. In Appendix \ref{sec:appendixTheory} we derive a method which allows to drastically reduce the computing power necessary for the mock analysis and in Appendix \ref{sec:appendixComparison} we show results of the methods, applied to some specific cases based, for definiteness, on the 2MPZ galaxy catalog and the {\it Fermi}-LAT $\gamma$ ray maps.

\section{Generation of mocks}
\label{sec:appendixMocks}

The methods we use to generate mock maps starting from a tue map are the following:
\begin{itemize}
  \item  Bootstrap
  \item  Jackknife
  \item  Phase Randomization
  \item  Gaussian Realizations (synfast)
  \item  Lognormal Realizations (Flask)
\end{itemize}
We can group these five methods in two categories: 
\emph{resampling} procedures (bootstrap and jackknife), that allow us to build mocks just with a reorganisation of the
original data sample (galaxes/clusters or $\gamma$-ray map); 
\emph{generated fields} procedures (phase randomization, synfast and
Flask), that use the statistical distribution of the
original data sample to build mocks. 

We pre-process our data sets (either galaxies/clusters or $\gamma$ rays) in \heal format: this allows us to adopt the same procedure for both type of observables. Galaxies/cluster maps are produced in terms of number counts per pixels, $\gamma$ rays maps in terms of photon intensity per pixel. We adopt a \heal pixelation format with resolution parameter
$N_{\rm side} = 1024$, which correspond to a total number of pixels $\Np = 12,582,912$ and mean
spacing of $\sim 0.06^\circ$. Being interested in the unresolved component of the $\gamma$ ray emission, we apply masks for resolved point sources and galactic emission, as described in Section \ref{subsec:fermi}. Masks may apply also to the galaxies/clusters catalogs. The presence of (typically quite different) masks for the two fields make the determination of the covariance matrix quite complex and involved, for which the mock techniques becomes especially useful. 

For each method used to produce the mock realisations of both the galaxies/clusters and $\gamma$ ray maps, we test that the ensuing auto-correlation APS is recovered from the mocks: we measure the auto APS for each mock map and verify that the average of these APS recover the APS of the corresponding data maps.

\subsection{Bootstrap}

A map in \heal format is an array of pixels where each element represents the
intensity of the specific pixel. To make one bootstrap realisation we
follow these steps:
\begin{itemize}
  \item We divided the full array in $\Nsub$ sub-arrays, so
    that each of them has $\Np/\Nsub$ pixels;
  \item We label each of the sub-array;
  \item We randomly pick $\Nsub$ sub-arrays with replacements and form a new resampled \heal map
\end{itemize}
Reiterating these three steps would produce $\Nr$ bootstrap realisations. The new \heal map originated in this way is characterised by the same number of
pixels of the original data set, but each of the sub-array can be
selected more than one times or not selected at all; for this reason
we have to weight each sub-array by the number of times it is
selected. This procedures is called \emph{bootstrap with replacements}
\citep{NorbergEtal2009}.

In this case the estimator of the APS covariance matrix is given by:
\be
\widehat{\Gamma_{\rm B}} = \frac{1}{N_r-1} \ \sum_{k=1}^{N_k} \ (x^k_i - \bar{x}_i)(x^k_j - \bar{x}_j)\,
\label{eqn:covB}
\ee
where $x_i$ is the i-th bootstrap realisation and $\Nr$ is the number of realisations and,
\be
\bar{x}_i = \frac{1}{N_r} \ \sum_{k=1}^{N_r} x_i^k \ .  
\label{eqn:meanx}
\ee

\subsection{Jackknife}

As for the bootstrap technique, also for the jackknife method the map is divided in $\Nsub$ sub-arrays with the
same number of pixels. Each of the sub-arrays is labelled,
but in this case a realisation is obtained by systematically omitting
one of the sub-array in each realisation. The resampling of the
data-set consists of $\Nsub-1$ remaning sub-arrays with volume
$(\Nsub-1)/\Nsub$ times the volume of the original data-set
\citep{NorbergEtal2009}. By definition there are only $\Nr = \Nsub$ different copies of the data set that are created in
this way. In this case the APS covariance matrix estimator reads:
\be
\widehat{\Gamma_{\rm J}} = \frac{\Nr-1}{\Nr} \ \sum_{k=1}^{\Nr} \ (x^k_i - \bar{x}_i)(x^k_j - \bar{x}_j)\,
\label{eqn:covJ}
\ee
where $\bar{x}_i$ is given by eq. \eqref{eqn:meanx}.
The factor $(\Nr-1)$ accounts for the lack of independence between
the $\Nr$ copies of the data set. 

\subsection{Phase randomization}

The implementation of this procedure is described in
\cite{DeDomenicoEtal2012}.
It based on the fact that it is always possible to write an intensity map as a linear
combination of spherical harmonics:
\be
f(\theta,\phi) = \sum_{\ell=0}^\infty\sum_{m=-\ell}^\ell \ \alm \ Y_{\ell m}(\theta,\phi)\,.
\label{eqn:intensitymap}
\ee
from which the angular power spectrum is obtained: 
\be
C_\ell = \frac{1}{2\ell+1} \ \sum_{m=-1}^\ell \ |\alm|^2 \ .
\label{eqn:Cl}
\ee
It is clear that Eq. \eqref{eqn:Cl} is invariant under a phase rotation on the harmonic amplitudes $a_{lm}$: 
\be
\alm \quad \longrightarrow \quad \alm e^{i\varphi_{\ell m}} \qquad \varphi_{\ell m} \in \mathbb{R}  \,;
\label{eqn:coeffRot}
\ee
Taking advantage of this symmetry, we can build independent realisations of the initial intensity map, each sharing the same APS. Since we determine the true-map APS from a masked sky, we need to correct for it, in order to produce a mock map that contains the correct statistical properties of the original map. The procedure we adopt is:
\begin{itemize}
\item Measure the  auto APS ($C_\ell^{\gamma\gamma}$ or $C_\ell^{gg}$ for $\gamma$ rays or galaxies/clusters) from the masked data maps
\item Transform: $\alm \longrightarrow \tilde{a}_{\ell m} = \alm e^{i\varphi_{\ell m}}$
\item Construct a full-sky mock map: $\tilde{f}(\theta,\phi) = \sum_{\ell=0}^\infty\sum_{m=-\ell}^\ell \ \tilde{a}_{\ell m} \ Y_{\ell m}(\theta,\phi)$
\item Correct the mock map for incomplete sky: $\tilde{f}(\theta,\phi) \longrightarrow \tilde{f}(\theta,\phi) \ \times W(\theta,\phi) \ \times f^{-1/2}_{\rm sky}$
\end{itemize}
The $(f_{\rm sky})^{-1/2}$ accounts for the fact that the original map was masked and therefore the obtained harmonic amplitudes has reduced power as compared to the true one. $W(\theta,\phi)$ restores the mask on the mock map.
Let us notice that this method looses information on the shot-noise, and therefore it can produce underestimate of the covariance in situations where the shot-noise is large.

The evaluation of the APS covariance matrix is finally done with eq. \eqref{eqn:covJ}%, where here $\Nsub$ denotes the number of realisations.

\subsection{Gaussian realisations (Synfast)}

Synfast\footnote{\url{https://healpix.jpl.nasa.gov/html/facilitiesnode14.htm}}
is a \heal routine that allows to generate realisations of a Gaussian random fields on a sphere, starting from an input APS. The procedure
 We therefore start from the APS describing the statistical distribution of
the data sample we want to replicate:
\begin{itemize}
\item Measure the  auto APS ($C_l^{\gamma\gamma}$ or $C_l^{gg}$ for $\gamma$ rays or galaxies/clusters) from the masked data maps
\item The obtained APS is fed to synfast, which outputs a full-sky mock map: $\tilde{f}(\theta,\phi)$
\item Mask the mock map: $\tilde{f}(\theta,\phi) \longrightarrow \tilde{f}(\theta,\phi) \ \times W(\theta,\phi)$
\end{itemize}

The evaluation of the APS covariance matrix is finally done with eq. \eqref{eqn:covJ}%, where here $\Nsub$ denotes the number of realisations.

\subsection{Lognormal realisations (FLASK)}

Flask\footnote{\url{http://www.astro.iag.usp.br/~flask}} is a
C++ code, parallelised with OpenMP, based on the work of
\cite{XavierEtal2016} and created to generate mock realisations of galaxy distributions starting from their 3D power spectrum.
Like synfast, it generates multiple correlated
fields on spherical shells, after providing the power spectrum
describing the distribution to be replicated. Differently from synfast, the generated maps are obtained from a lognormal distribution. The tomographic
approach used by Flask slices the three dimensional space into
spherical shells (redshift slices), each one discretized in Healpix
maps. After generating the fields, Flask can apply selection functions
and noise to them. The output can be in the form of a source catalogue
and/or Healpix maps, among others. 

We use Flask to generate $\Nr$ independent realizations. The evaluation of the APS covariance matrix is finally done with eq. \eqref{eqn:covJ}.
Although the code is thought to work for galaxy distributions in
different redshift bins, we tried to use it also for $\gamma$ ray maps, by using the APS instead of the 3D power
spectrum.

%\subsection{$\g$-rays angular power spectrum fit}

%In order to check that the produced $\g$-ray mock maps maintain the
%same two-point statistics as the real ones, we compare the APS of the
%mocks with the true one. The true APS is obtained by computing the
%auto- and cross-power spectrum with Polspice, feeding the algorithm
%with the $\g$ -ray maps and masks. We re-bin the ``raw" results into 15
%logarithminc multipole bins and we perform a noise removal and a beam
%window correction, in order to extract the ``signal" APS:
%
%\be
%C^{ij}_{\ell,signal} = \frac{C^{ij}_{\ell,raw} - C^{ij}_N}{W^i_\ell \, W^j_\ell}\,,
%\ee
%
%where \textit{i} and \textit{j} denote the energy bin, the noise term
%$C^{ij}_N$ is non-zero only when $i=j$ and $W^i_l, W^j_l$ are the
%beam window functions. This procedure is the same as described by \cite{FornasaEtAl2016} and the photon noise term is computed in the same way as equation (5) of section C of the same reference. 

%We fit the resulting APS with the following model:
%
%\be
%C_l = N \ \ell^{-\alpha} + C_p\,,
%\ee
%
%where $N, \alpha$ and $C_p$ are the free parameters. We define a multipole range where the fit is performed, setting minimum and maximum multipole in order to avoid, respectively, possible foreground contamination due to a imperfect removal and overcorrection of the beam effect associated the PSF finite resolution. Values are reported in table \ref{tab:Ebins}. The curves resulting from the fit of the true maps are compared to the mock APS in order to check the consistency of the mock realizations.

\subsection{Covariance estimators' relations}

As we have shown in the previous sections, we can set a different APS
covariance matrix estimator for each of the methods we use to produce
mocks. It can be useful to have a look to the relation between the
different estimators.

Given a data vector $\textbf{X} = \{x_1,x_2,...,x_{\Nr}\}$ we can write
the generic expression for the covariance matrix of $\textbf{X}$ as:
\be
\textrm{cov}[\textbf{X}]  = \frac{1}{\Nr} \ \sum_{k=1}^{\Nr} \ (x_i^k-\bar{x}_i)(x_j^k-\bar{x}_j)\,
\label{eqn:cov}
\ee
with $\bar{x}_i$ is given by eq. \eqref{eqn:meanx}.
Eq. \eqref{eqn:cov} can also be rewritten as:
\be
\textrm{cov}[\textbf{X}] = \frac{1}{\Nr} \ \sum_{k=1}^{\Nr} \ x_i^kx_j^k - \frac{1}{\Nr^2} \ \sum_{k=1}^{\Nr}x_i^k\sum_{k=1}^{\Nr}x_j^k \ .
\ee
The unbiased definition of the sample covariance when the mean is derived from the sample itself is:
\bea
\widehat{\Gamma}_{\rm U} &\equiv& \frac{\Nr}{\Nr-1} \ \times \ \textrm{cov}[\textbf{X}] \nn\\
&=&\frac{1}{\Nr-1} \ \sum_{k=1}^{\Nr} \ x_i^kx_j^k - \frac{1}{\Nr}\frac{1}{\Nr-1} \ \sum_{k=1}^{\Nr}x_i^k\sum_{k=1}^{\Nr}x_j^k\,. \nn \\ 
\label{eqn:covU}
\eea
We can relate the estimator of eq. \eqref{eqn:covU} to the
ones obtained with the jackknife \eqref{eqn:covJ} and bootstrap
\eqref{eqn:covB} techniques as:
\bea
\textrm{Jackknife:} \quad \widehat{\Gamma}_{\rm J} &=& \frac{(N_r-1)^2}{N_r} \ \times \ \widehat{\Gamma}_{\rm U}\\
\textrm{Bootstrap:} \quad \widehat{\Gamma}_{\rm J} &=&  \widehat{\Gamma_{\rm U}}
\eea

\section{Semi-analytic prediction of the cross-correlation covariance}
\label{sec:appendixTheory}

The derivation of the covariance matrix for the cross-correlation APS requires to combine the information arising from the generation of many different set of maps. In order to obtain stable results, the required number of realisations for each of the two observables can be large (in our analysis on the cross-correlations between clusters and $\gamma$ rays, we produced 2000 mocks for each set), and can require to produce maps in several energy bins (for $\gamma$ rays) and redshift bins (for galaxies/clusters). In this section we show that we can obtain a reliable estimate of the full covariance matrix by performing a simpler combination, namely we can construct two partial estimates of the covariance matrix by combining separately: (i) the galaxies/clusters mock with the measured $\gamma$ ray map; (ii) the $\gamma$ ray mocks with the measured galaxies/clusters maps. The final estimate of the covariance is obtained as the average of these two ``half'' covariances. This reduces the number of combinations from $(\Nr)^2 \times (n_E \times n_z)$ to $2(\Nr \times n_E \times n_z)$ (where $\Nr$ denotes the number of mock maps produced for each of the $n_E$ energy bins and $n_z$ redshift bins), which is $\Nr/2$ times faster. We show that this approach is correct in the limit of a large number $(\Nr)$ of realisations, by following a theoretical derivation based on the Gaussian prediction for the APS covariance matrix. We have numerically verified that the result shown here hold in a more general situation where gaussianity is not necessarily present.
The method we are going to describe is valid in the case where the cross-correlation term is small when compared to the product of the auto-correlations for galaxies and gamma-rays.

Let us start with the Gaussian prediction for the APS covariance matrix:
\citep{HuAndJain2004}:
\be
\Gamma_{\ell\ell}^{g\gamma_i} \equiv \textrm{cov}[C_\ell^{g\gamma_i},C_\ell^{g\gamma_i}]= \frac{C_\ell^{gg}C_\ell^{\gamma_i\gamma_i} + (C_\ell^{g\gamma_i})^2}{(2\ell+1)\Delta \ell \ f_{\rm sky}} \delta_{\ell\ell'}^{\rm K} \,,
\label{eqn:covGauss}
\ee
where $\Delta l$ is the bin width, $f_{\rm sky}$ is the fraction of
the sky probed by the surveys and $\delta^{\rm K}$ is the Kronecker
symbol. 

Let us denote with a hat symbol quantities which are measured on the real maps, while quantities obtained from mocks do not have the hat symbol. For instance, $\hat{C}_l^{gg}$ and
$\hat{C}_l^{\gamma\gamma}$ are the galaxies/clusters the $\gamma$ ray auto-correlation APS measured on the true data maps.
Let us define a covariance term obtained by computing the
cross-correlation between the real galaxy distribution and the mock
$\gamma$ rays realisations. From Eq. \eqref{eqn:covGauss} and considering that we construct the covariance from the mock by averaging over the $\Nr$ realisations: 
\be
\Gamma_{\ell\ell'}^{\hat{g}\gamma_i} \equiv \textrm{cov}[C_\ell^{\hat{g}\gamma_i},C_{\ell'}^{\hat{g}\gamma_i}] \ \propto \ \hat{C}_\ell^{gg}\frac{1}{N_r}\sum_{n=1}^{N_r} \ C_\ell^{\gamma_i\gamma_i,n} +\frac{1}{N_r}\sum_{n=1}^{N_r} \ (C_\ell^{\hat{g}\gamma_i,n})^2 \, ;
\label{eqn:covGaussgam}
\ee
Let us then define the corresponding counterpart term obtained by using the mocks for  galaxies/clusters and data
for $\gamma$ rays:
\be
\Gamma_{\ell\ell'}^{g\hat{\gamma_i}} \propto \ \hat{C}_l^{\gamma_i\gamma_i}\frac{1}{N_r}\sum_{n=1}^{N_r} \ C_l^{gg,n} + \frac{1}{N_r}\sum_{n=1}^{N_r} \ (C_l^{g\hat{\gamma_i},n})^2 \,.
\label{eqn:covGaussgal}
\ee
When instead we use only mocks, Eq. \eqref{eqn:covGauss} gives:
\be
\Gamma_{\ell\ell'}^{g\gamma_i} \ \propto \ 
\left(
\frac{1}{N_r}\sum_{n=1}^{N_r} \ C_l^{gg,n}
\right)
\left(
\frac{1}{N_r}\sum_{n=1}^{N_r} \ C_l^{\gamma_i\gamma_i,n}
\right)
+ \frac{1}{N_r^2}\sum_{n=1}^{N^2_r} \ (C_l^{g\gamma_i,n})^2 \,.
\label{eqn:covGaussgalgam}
\ee
If we take the average of the expressions \ref{eqn:covGaussgam} and \ref{eqn:covGaussgal}, we obtain:
\be
(\Gamma_{\ell\ell'}^{g\gamma_i})_{\rm ave}  = \frac{1}{2} \BL(\Gamma_{\ell\ell'}^{\hat{g}\gamma_i}+\Gamma_{\ell\ell'}^{g\hat{\gamma_i}}\BR)
\ \propto \ \hat{C}_l^{\gamma_i\gamma_i}\frac{1}{N_r}\sum_{n=1}^{N_r} \ C_l^{gg,n} +\hat{C}_l^{gg}\frac{1}{N_r}\sum_{n=1}^{N_r} \ C_l^{\gamma_i\gamma_i,n} + \frac{1}{N_r}\sum_{n=1}^{N_r} \ (C_l^{g\gamma_i,n})^2 \,,
\label{eqn:covsum}
\ee
Since the measurements of the APS ($\hat{C}_l^{gg}$ and
$\hat{C}_l^{\gamma\gamma}$) obtained using the real map are well
reproduced by the APS measurements from mock maps:
\bea
\hat{C}_l^{gg} & \simeq & \frac{1}{N_r}\sum_{n=1}^{N_r} \ C_l^{gg,n} \nn \\ 
\hat{C}_l^{\gamma_i\gamma_i} & \simeq & \frac{1}{N_r}\sum_{n=1}^{N_r} \ C_l^{\gamma_i\gamma_i,n} \,. 
\label{eqn:assumption}
\eea
we obtain:
\be
(\Gamma_{\ell\ell'}^{g\gamma_i})_{\rm ave} \ \propto \ 
\left(
\frac{1}{N_r}\sum_{n=1}^{N_r} \ C_l^{gg,n}
\right)
\left(
\frac{1}{N_r}\sum_{n=1}^{N_r} \ C_l^{\gamma_i\gamma_i,n} 
\right)
+ \frac{1}{N^2_r}\sum_{n=1}^{N^2_r} \ (C_l^{g\gamma_i,n})^2 \,.
\label{eqn:covsum2}
\ee
We then observe that in the limit of large $\Nr$:
\be
\frac{1}{N^2_r}\sum_{n=1}^{N^2_r} \ (C_l^{g\g_i,n})^2 \simeq \frac{1}{N_r}\sum_{n=1}^{N_r} \ (C_l^{g\g_i,n})^2\,.
\ee
In this case, Eq. \eqref{eqn:covGaussgalgam} and Eq. \eqref{eqn:covsum2} give the same result. Therefore we can obtain a reliable estimate the covariance by simply averaging over the sum of the two ``half'' contributions:
\be
\Gamma_{\ell\ell'}^{g\gamma_i} = \frac{1}{2} \BL(\Gamma_{\ell\ell'}^{\hat{g}\gamma_i}+\Gamma_{\ell\ell'}^{g\hat{\gamma_i}}\BR)\,.
\label{eq:cov2halves}
\ee

\section{Comparison of the methods to produce map mocks}
\label{sec:appendixComparison}

\begin{figure*}
    \centering
    \includegraphics[width=0.45\textwidth]{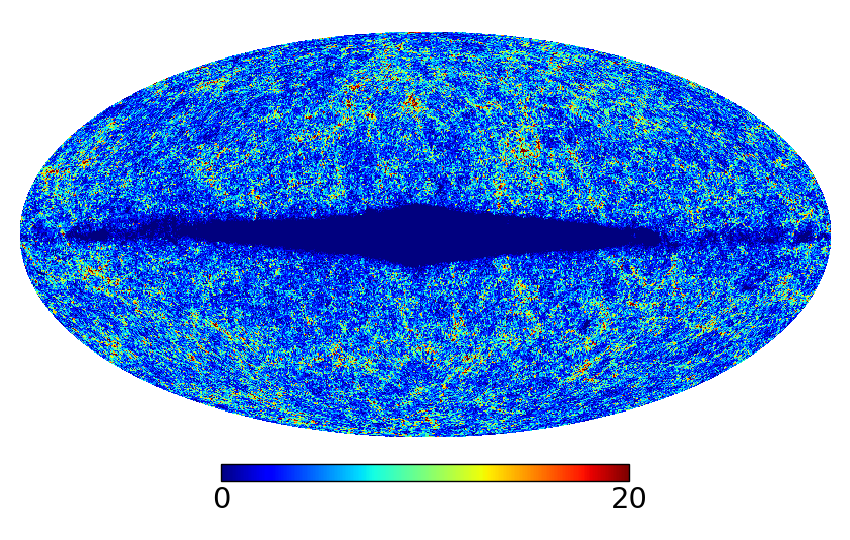}
    \includegraphics[width=0.45\textwidth]{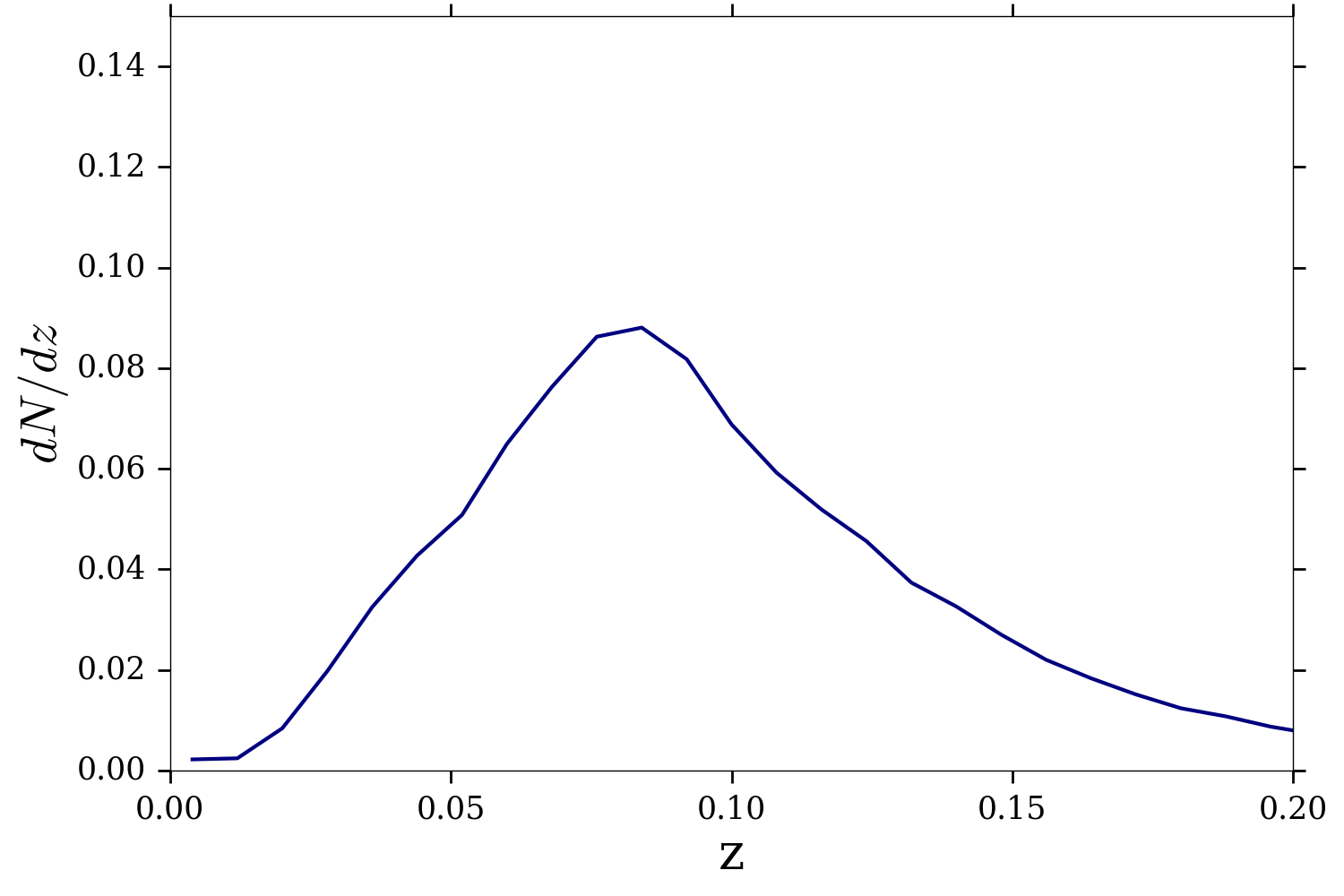}
    \caption{Left panel: All-sky map of the 2MPZ catalogue in \heal projection with $N_{\rm side} = 128$. Right panel: Redshift distribution of the 2MPZ catalogue.}
    \label{fig:2MPZ}
\end{figure*}

In this Appendix we report some results for the determination of the cross-correlation covariance matrix obtained using the methods described in Appendix \ref{sec:appendixMocks} and using the relation shown in Eq. \ref{eq:cov2halves} to join the separate information coming respectively from the $\gamma$ and galaxy mock maps. For definiteness, in this analysis we use the the 2MASS Photometric Redshift catalogue (2MPZ, \cite{BilickiEtAl2014}) that is a galaxy catalogue built by cross-matching 2MASS XSC, WISE and SuperCOSMOS all-sky samples with galaxy photometric redshift reconstructed via an artificial neural network. The employed algorithm is the one described in \cite{CollisterAndLahav2004} and trained on several redshift surveys (2MRS, SDSS, 6dFGS, 2dFGRS and ZCAT). The all-sky accuracy of the redshift reconstructed by the network is close to $\sigma_z = 0.015$ for nearly all the dataset with few outliers. The resulting 2MPZ sample contains almost 1 million galaxies with a median redshift of $z=0.07$. In the left panel of fig. \ref{fig:2MPZ} we show the 2MPZ catalogue in \heal projection and in the right panel we show its redshift distribution.
We use this catalogue to test our methods to exploit the large statistics in terms of number of galaxies (934,175), that allow us to reduce the galaxy shot-noise contribution.

We produce 256 galaxy mock maps starting from the 2MPZ galaxy catalogue and 256 $\gamma$-ray maps in each energy bins used also in the main text analysis (see Table \ref{tab:Ebins}) using the five methods described in Appendix \ref{sec:appendixMocks}. With this relative small number of mocks, per energy bin, we can test the accuracy in the estimation of the diagonal of the cross-correlation covariance matrix and compare it with its Gaussian prediction of Eq. \ref{eqn:covGauss}). For the non diagonal terms, we need to produce a much larger number of mocks, and in the analysis of the main text we use 2000 realizations for each observable.

\begin{figure}
    \centering
    \includegraphics[width=0.3\textwidth]{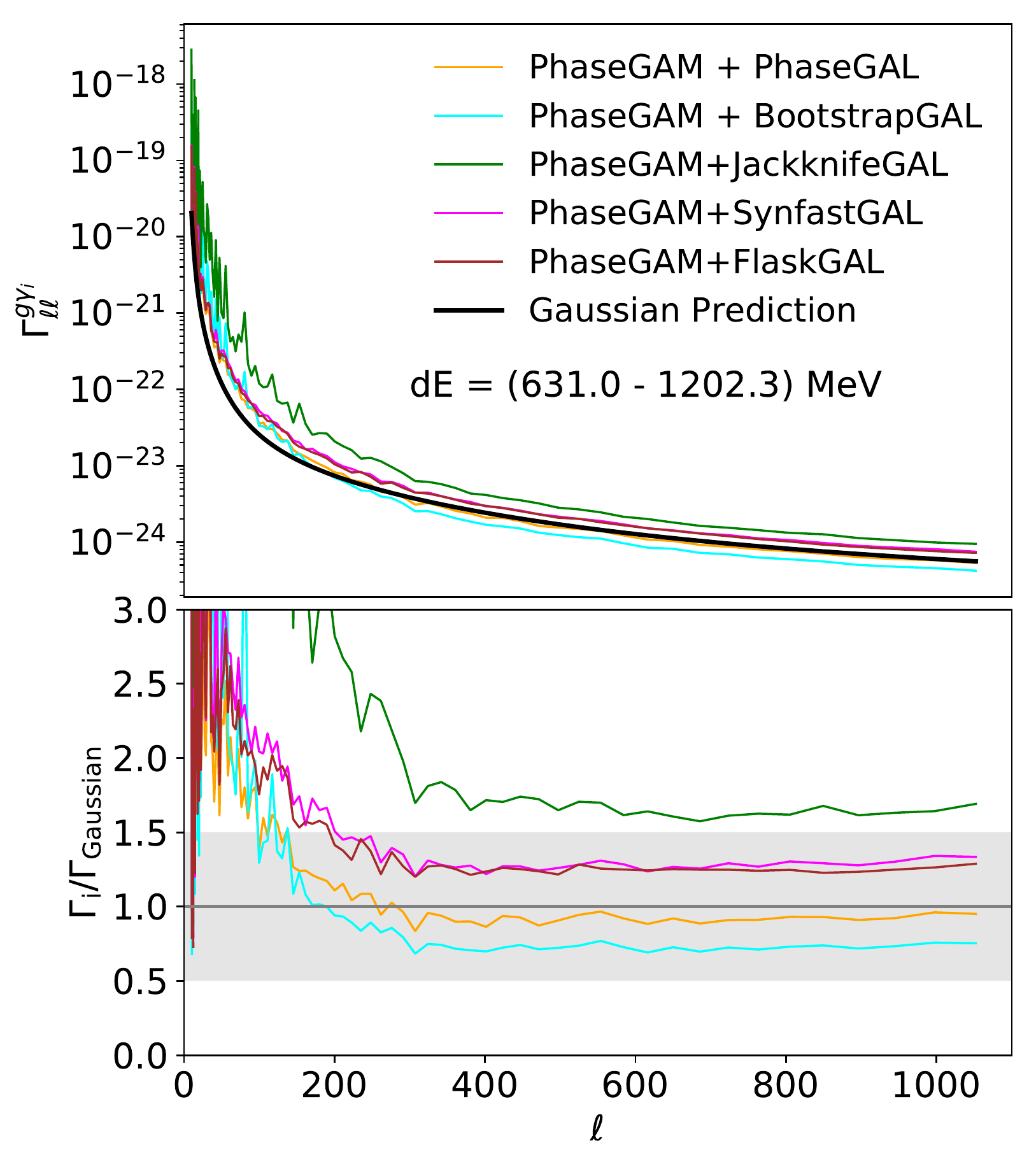}
    \includegraphics[width=0.3\textwidth]{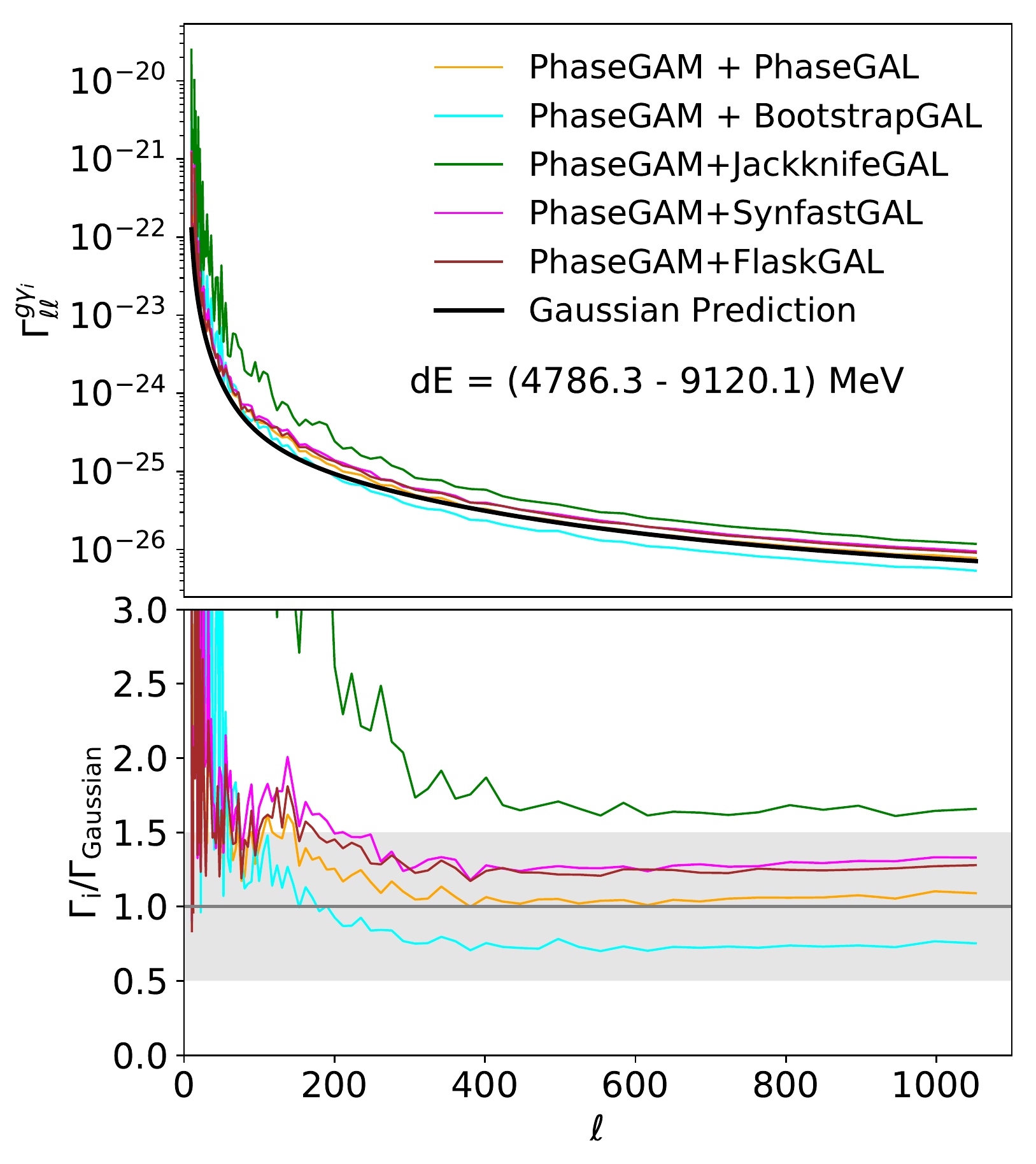}
    \includegraphics[width=0.3\textwidth]{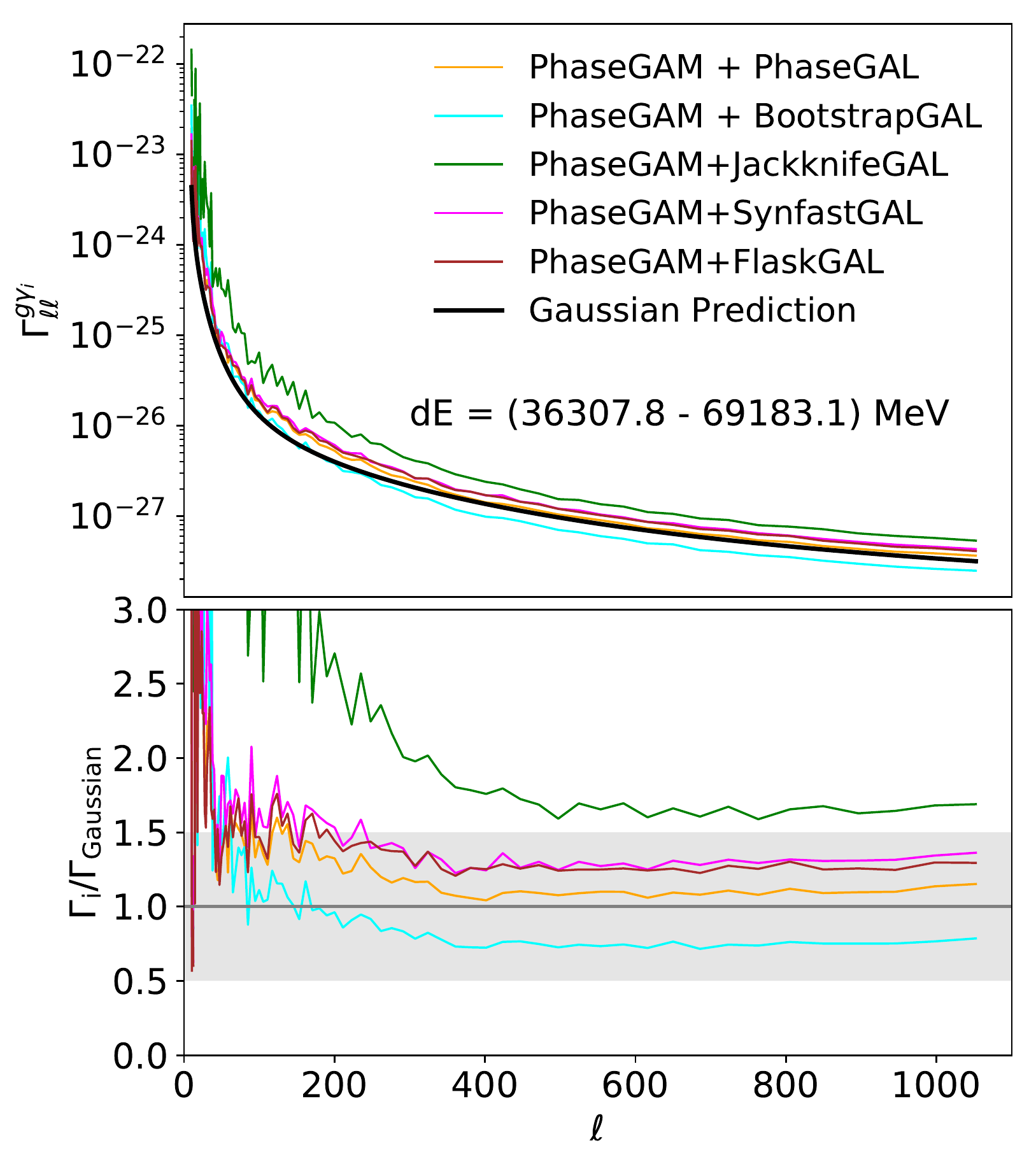}
    \caption{Diagonal of the cross-correlation covariance matrix estimated using 256 mocks; in black the Gaussian predictions; the coloured lines are the variances estimated using the phase randomization method for $\gamma$ maps + all the other methods for galaxy maps. In the lower panels it is shown the ratio between each of the coloured lines and the Gaussian variance; the grey shaded area represents a 50\% interval.}
    \label{fig:PHASEall}
\end{figure}

\begin{figure}
    \centering
    \includegraphics[width=0.3\textwidth]{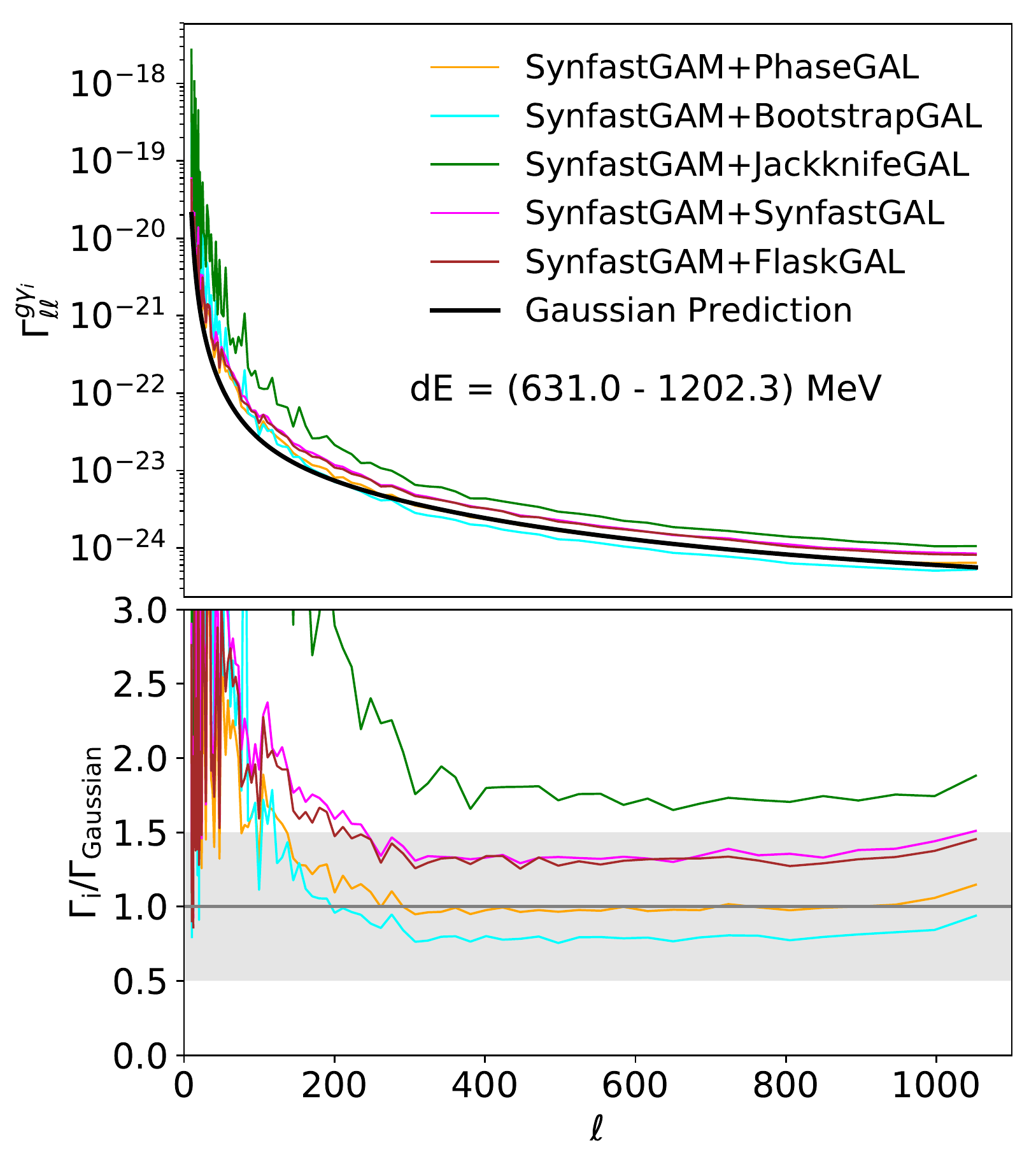}
    \includegraphics[width=0.3\textwidth]{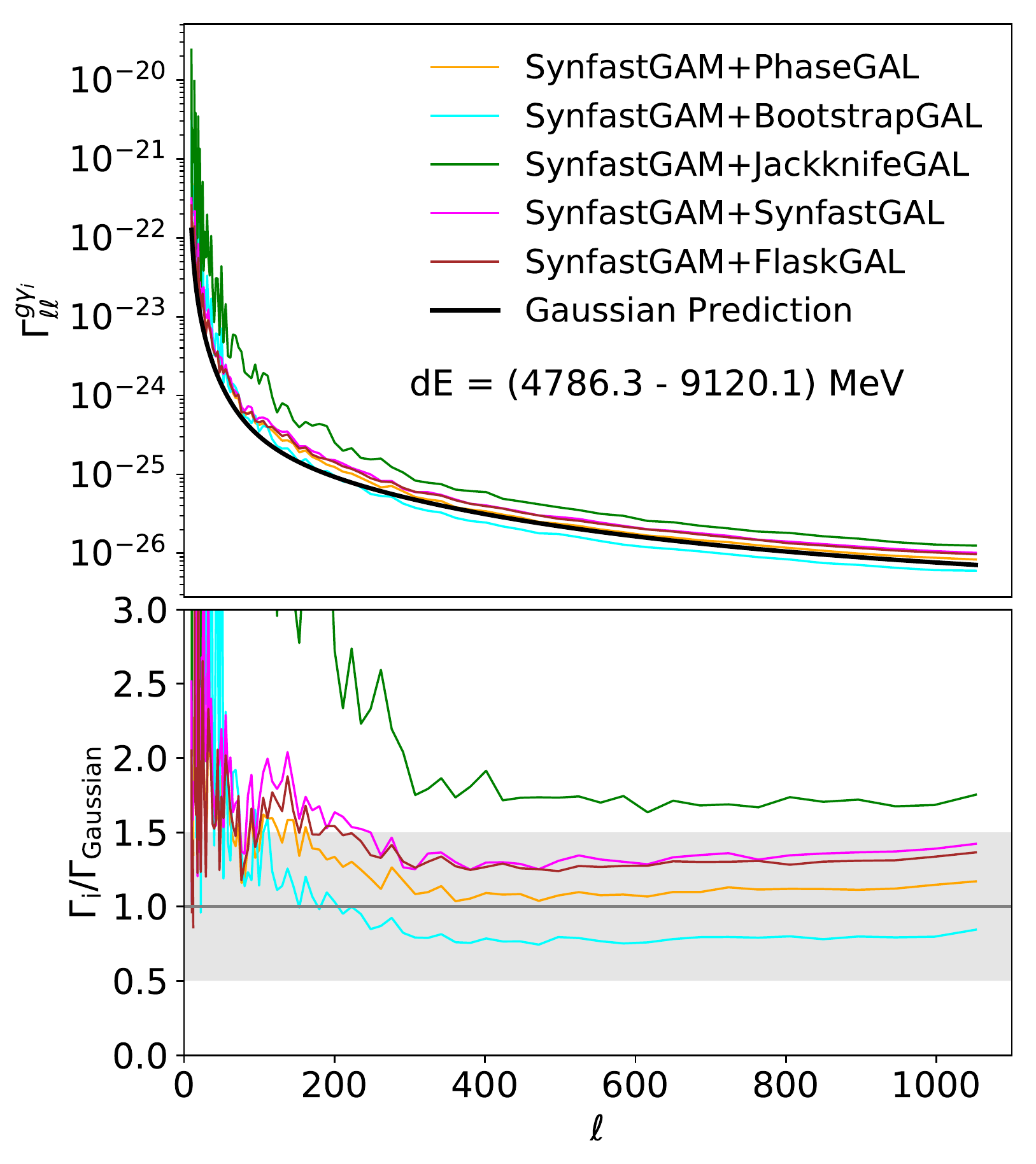}
    \includegraphics[width=0.3\textwidth]{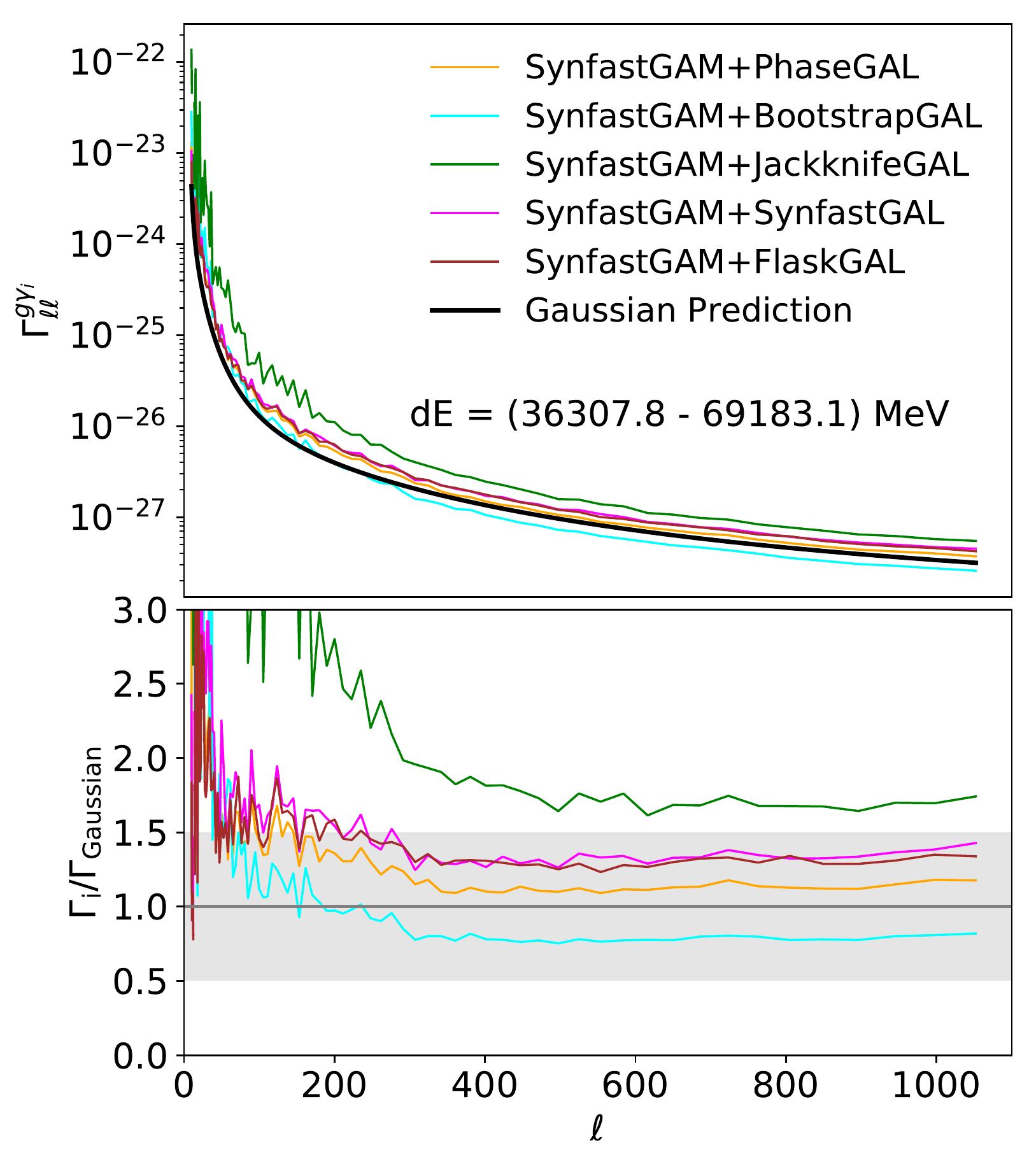}
    \caption{Same as figure \ref{fig:PHASEall} but using the Synfast methods for the $\gamma$ maps.}
    \label{fig:SYNFall}
\end{figure}

\begin{figure}
    \centering
    \includegraphics[width=0.3\textwidth]{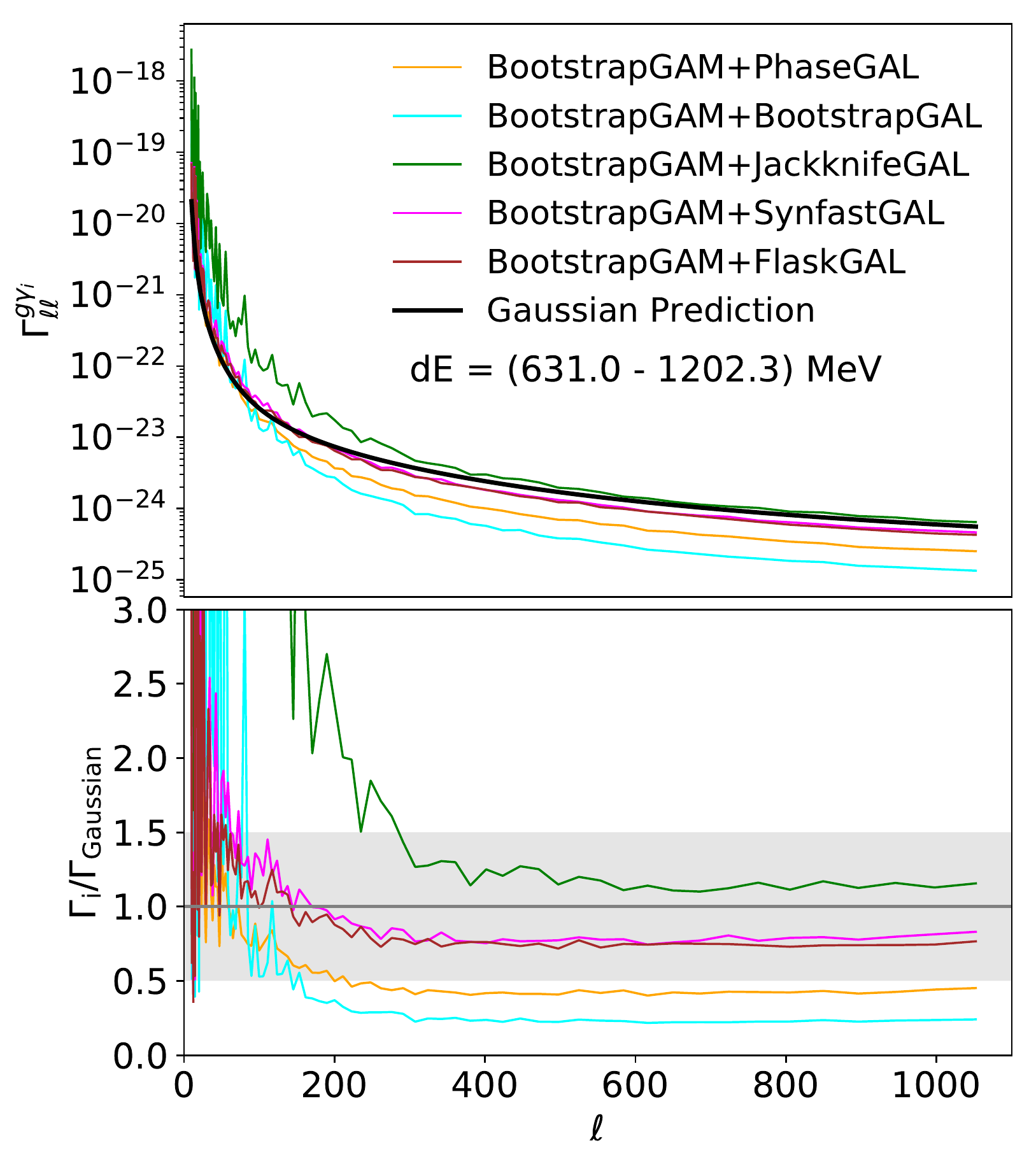}
    \includegraphics[width=0.3\textwidth]{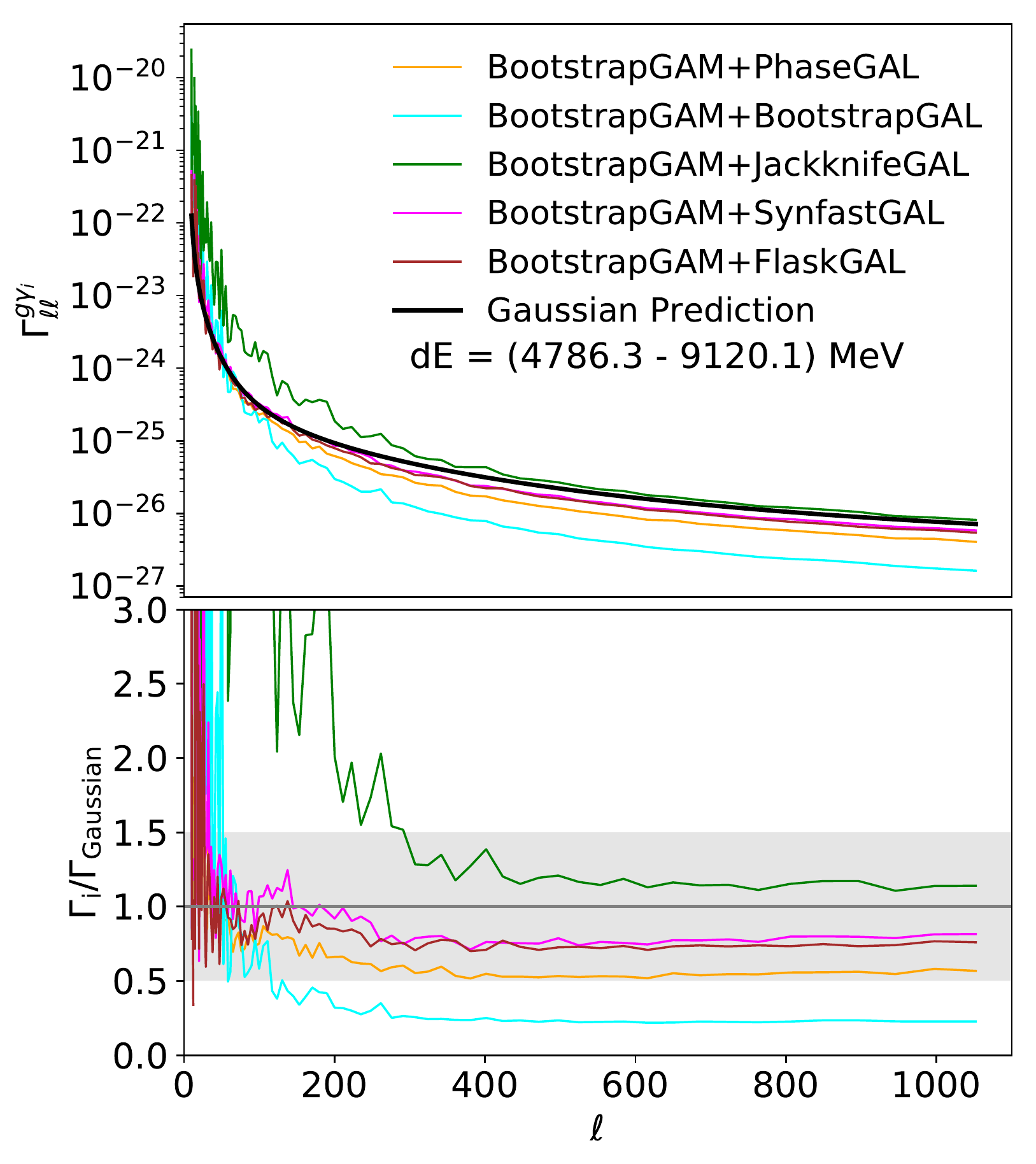}
    \includegraphics[width=0.3\textwidth]{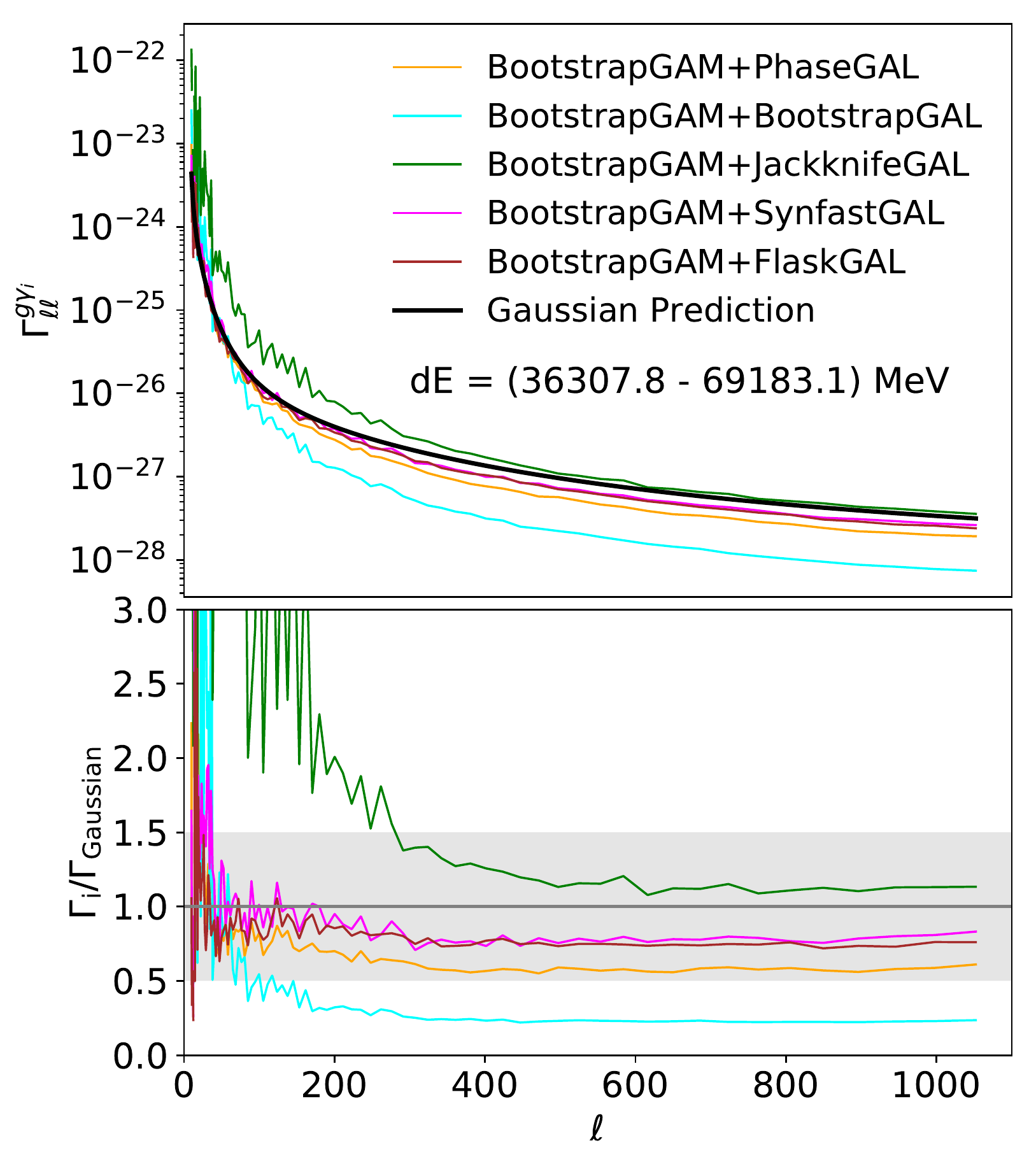}
    \caption{Same as figure \ref{fig:PHASEall} but using the Bootstrap methods for the $\gamma$ maps.}
    \label{fig:BOOTall}
\end{figure}

\begin{figure}
    \centering
    \includegraphics[width=0.3\textwidth]{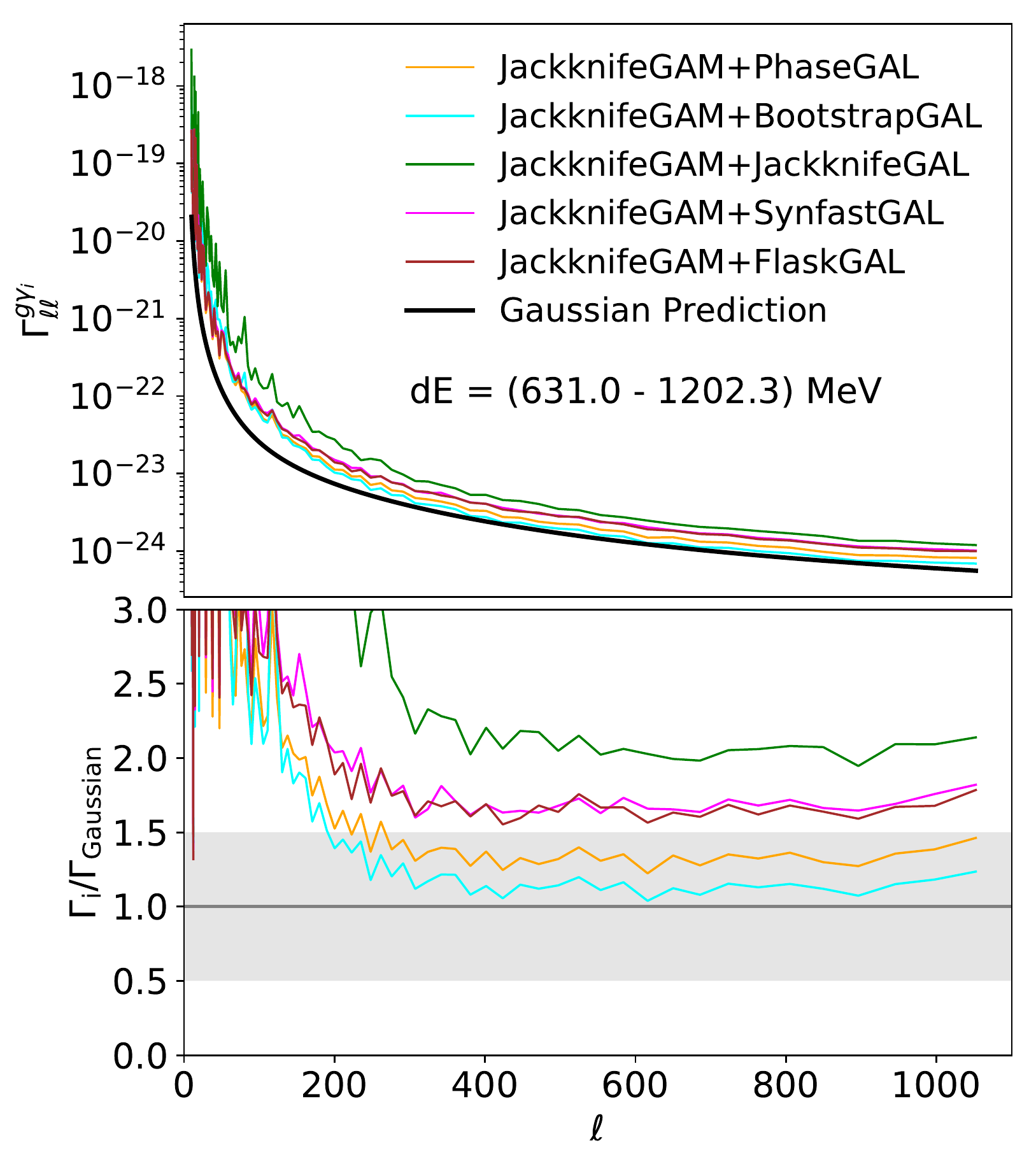}
    \includegraphics[width=0.3\textwidth]{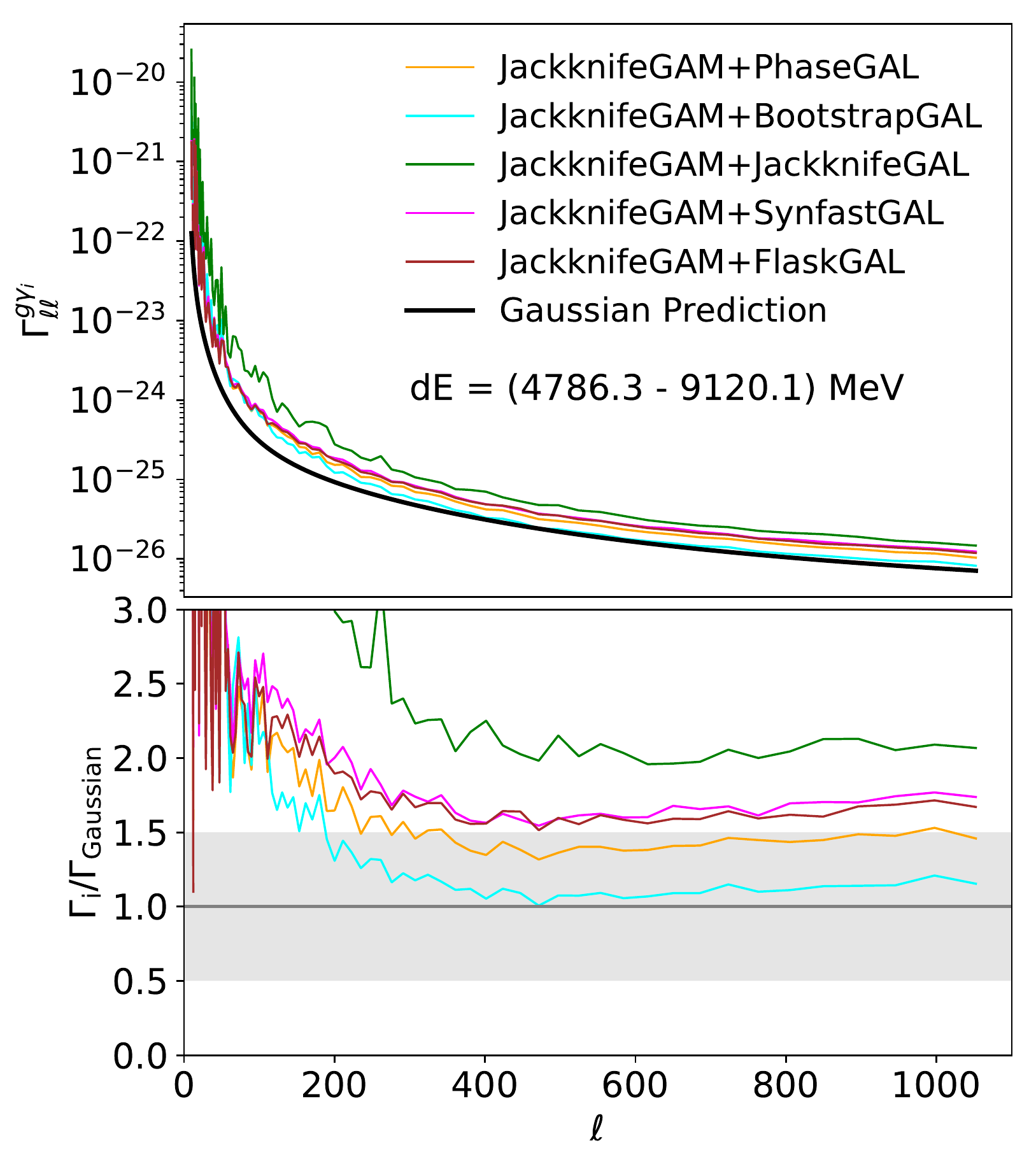}
    \includegraphics[width=0.3\textwidth]{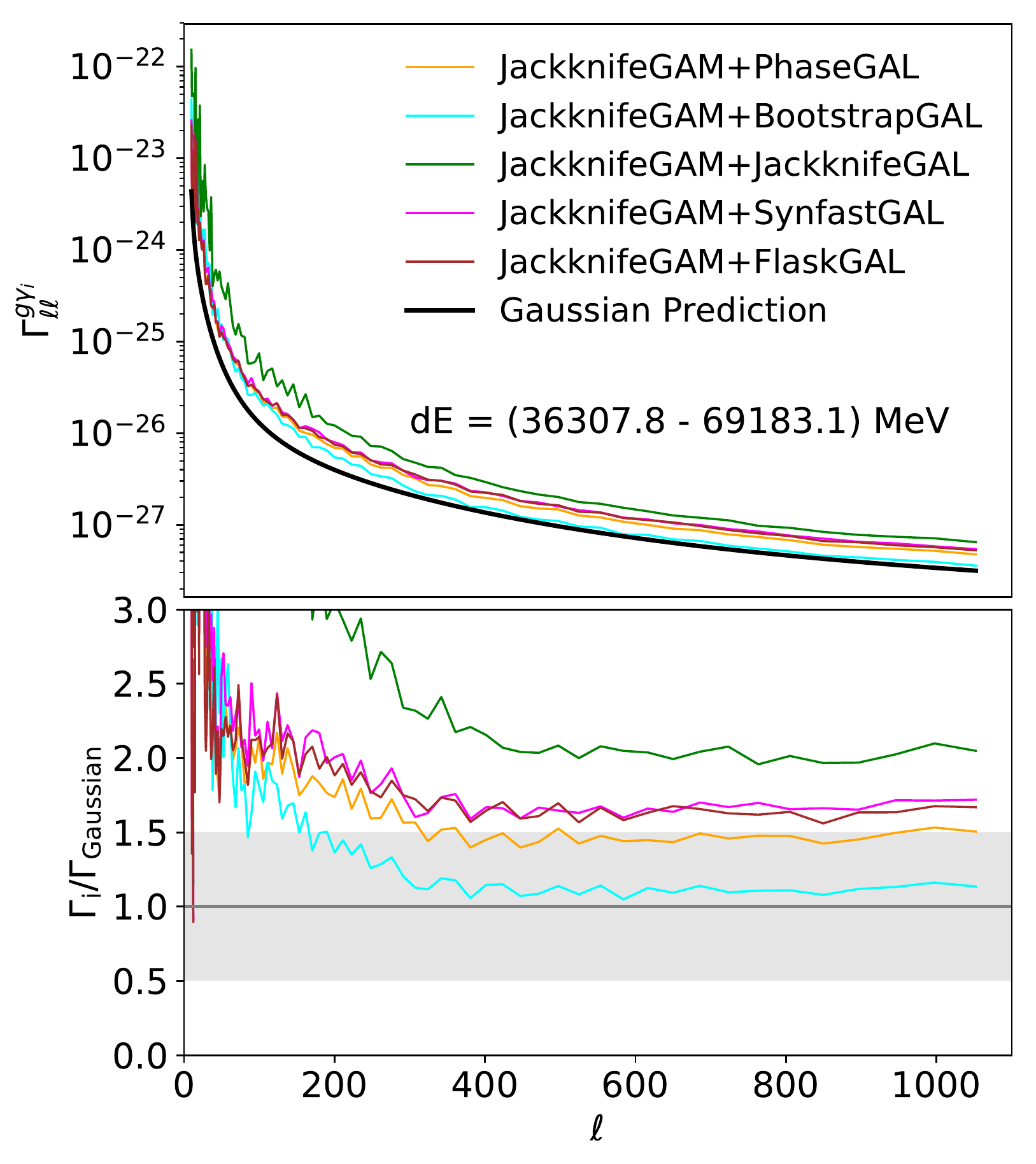}
    \caption{Same as figure \ref{fig:PHASEall} but using the Jackknife methods for the $\gamma$ maps.}
    \label{fig:JACKKall}
\end{figure}

\begin{figure}
    \centering
    \includegraphics[width=0.3\textwidth]{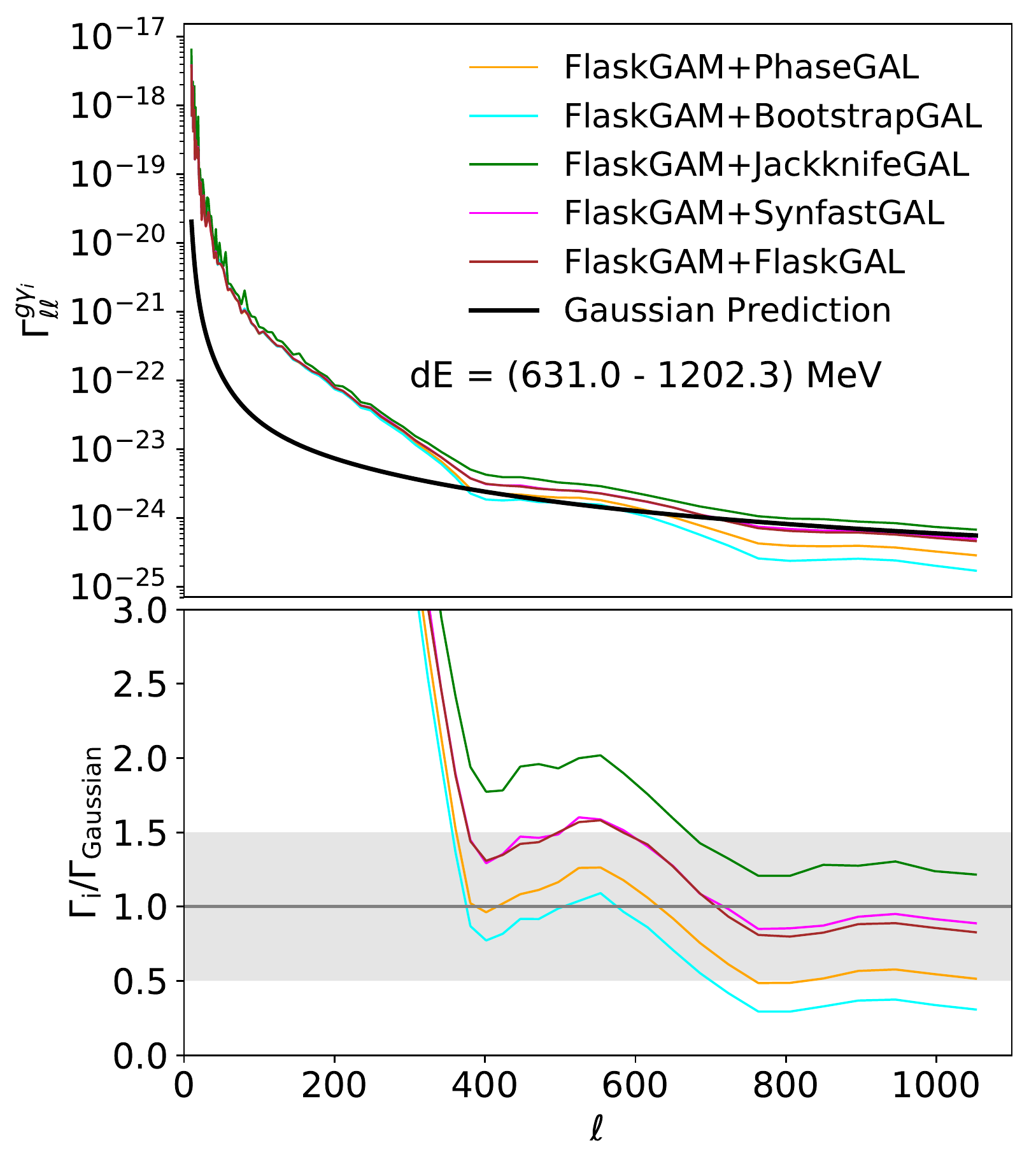}
    \includegraphics[width=0.3\textwidth]{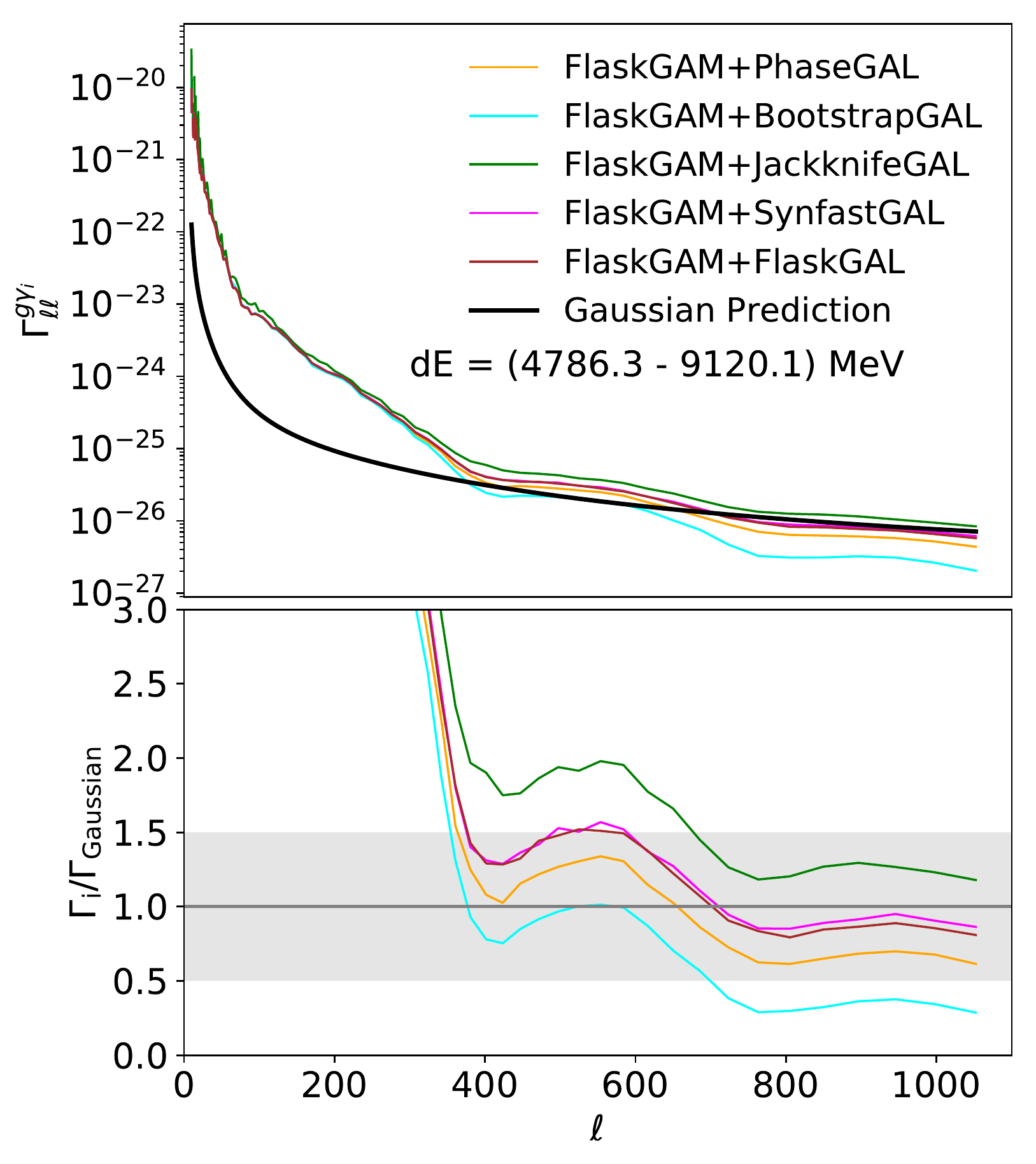}
    \includegraphics[width=0.3\textwidth]{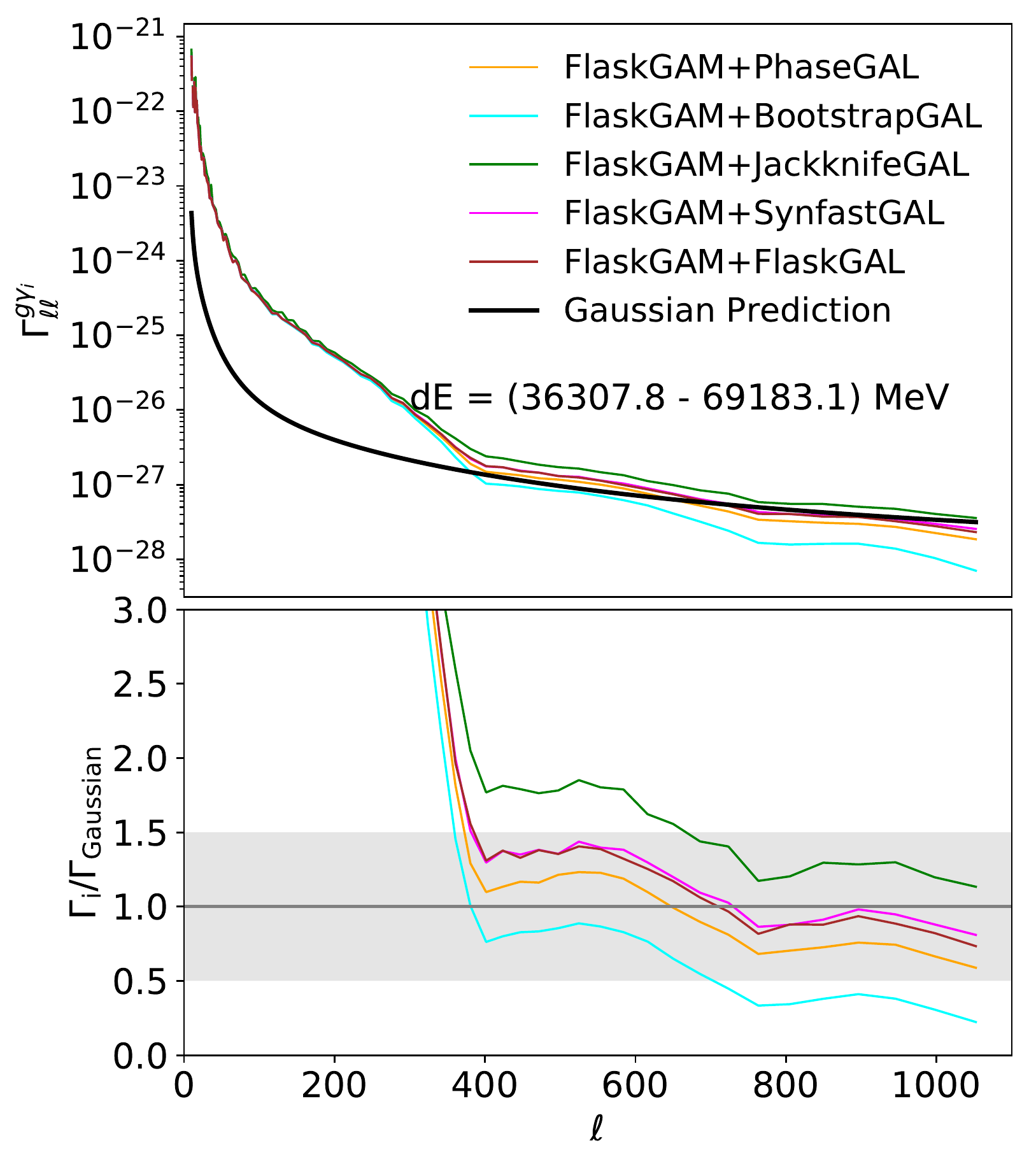}
    \caption{Same as figure \ref{fig:PHASEall} but using the Flask methods for the $\gamma$ maps.}
    \label{fig:FLASKall}
\end{figure}

The results are shown in figures \ref{fig:PHASEall}-\ref{fig:FLASKall}. In each figure we show the variance estimated using Eq. \ref{eqn:covsum2} (coloured lines) for a low (left), intermediate (center) and large (right) energy bin; in all plots, the black line represents the Gaussian prediction. Each plot is accompanied by a lower panels, where we show the ratio between the result for each combination and the Gaussian variance. 

In Fig. \ref{fig:PHASEall} the method to produce $\gamma$ ray mocks is fixed to phase randomisation, and the galaxy mocks methods are rotated among the five options discussed in Appendix \ref{sec:appendixMocks}. We notice that phaseGAM+bootstrapGAL tend to systematically underestimate the covariance, producing results even smaller by $\sim 30\%$ than the Gaussian prediction. This happens for all energy bins. The combination phaseGAM+phaseGAL slightly underestimante by a few percent the Gaussian variance in the first energy bin, but it is almost Gaussian in the other cases; the other combinations produce estimates in excess of the the Gaussian prediction of about $\sim 30\%$ for phaseGAM+phaseGAL/flaskGAL and about $60\%$ for phaseGAM+jackknifeGAL. 

A similar behaviour is observed in Figure \ref{fig:SYNFall}, where synfast is used to produce the $\gamma$ ray mocks. All the combinations show almost the same trend observed in Fig. \ref{fig:PHASEall}. We expected similar results between the phase randomisation and synfast technique, since the two technique are quite similarly implemented in the generation of mock maps.

Figure \ref{fig:BOOTall} shows the result for the bootstrap method applied to $\gamma$ rays, in combination with all methods for the galaxy mocks. In this case we observe that all the combinations except bootstrapGAM+jackknifeGAL underestimate the Gaussian prediction of about $\sim 30\%$ in the case of bootstrapGAM+synfastGAL/flaskGAL, and by more than $50\%$ for bootstrapGAM+phaseGAL/bootstrapGAL.

The jackknife applied to $\gamma$ rays in combination with all methods for galaxies is shown in Fig. \ref{fig:JACKKall}. In this case, all the combinations largely overestimate the gaussian prediction with just jackknifeGAM+bootstrapGAL/phaseGAL remaining below a $50\%$ difference. 

Finally, we show in figure \ref{fig:FLASKall}, the adoption of Flask to produce $\gamma$ ray mocks: it is clear that all the combinations show an inconsistent behaviour, with peculiar fluctuations for multipole scales smaller than 800. This behaviour is likely due to the fact that Flask is built to reproduce a galaxy distribution and is not general enough to be used for $\gamma$ ray maps.  We decided to test its use also to generate $\gamma$ ray mocks: although the APS is well reproduced, the behaviour of the covariance does not give results which look trustable, especially when compared with all the other methods shown above.

In conclusion, from the extensive analysis of the different combinations, we found that FlaskGAL + PhaseGAM represents a good options for estimating the covariance matrix for the cross-correlation APS of galaxies/clusters with $\gamma$ rays. This combination produces covariance in slight excess of the gaussian prediction for almost all situation tested. 

\end{document}